\documentclass{aa}

\usepackage{graphicx}
\usepackage{txfonts}
\def\ie {{\rm i.e.}}

\def\eg {{\rm e.g.}}

\def\HI  {\hbox{H$\textsc{i}$}}

\def\kms {\hbox{${\rm km\,s}^{-1}$}}

\def\Jypb {\hbox{Jy\,beam$^{-1}$}}
\def\Jypbkms {\hbox{${\rm Jy\,beam^{-1}\,km\,s^{-1}}$}}
\def\Jykms {\hbox{${\rm Jy\,km\,s}^{-1}$}}

\def\arcsec {\hbox{$^{\prime\prime}$}}
\def\arcmin {\hbox{$^{\prime}$}}
\def\ffas {\hbox{$\,.\!\!^{\prime\prime}$}}
\def\ffam {\hbox{$\,.\!\!^{\prime}$}}
\def\ffad {\hbox{$\,.\!\!^{\circ}$}}

\def \la{\mathrel{\mathchoice   {\vcenter{\offinterlineskip\halign{\hfil
$\displaystyle##$\hfil\cr<\cr\sim\cr}}}
{\vcenter{\offinterlineskip\halign{\hfil$\textstyle##$\hfil\cr
<\cr\sim\cr}}}
{\vcenter{\offinterlineskip\halign{\hfil$\scriptstyle##$\hfil\cr
<\cr\sim\cr}}}
{\vcenter{\offinterlineskip\halign{\hfil$\scriptscriptstyle##$\hfil\cr
<\cr\sim\cr}}}}}
\def \ga{\mathrel{\mathchoice   {\vcenter{\offinterlineskip\halign{\hfil
$\displaystyle##$\hfil\cr>\cr\sim\cr}}}
{\vcenter{\offinterlineskip\halign{\hfil$\textstyle##$\hfil\cr
>\cr\sim\cr}}}
{\vcenter{\offinterlineskip\halign{\hfil$\scriptstyle##$\hfil\cr
>\cr\sim\cr}}}
{\vcenter{\offinterlineskip\halign{\hfil$\scriptscriptstyle##$\hfil\cr
>\cr\sim\cr}}}}}

\begin{document}

   \title{$\HI$ mapping of the Leo Triplet}
   \subtitle{Morphologies and kinematics of tails and bridges}

   \author{Gang Wu
          \inst{1}\fnmsep\inst{2}
          \and
          David Mart\'{i}nez-Delgado\thanks{Talentia Senior Fellow}
          \inst{3}
          \and
          Christian Henkel
          \inst{1}\fnmsep\inst{4}\fnmsep\inst{2}
          \and
          Pavel Kroupa
          \inst{5,6}
          \and
          Fabian Walter
          \inst{7}
          \and
          Nico Krieger
          \inst{7}
          \and
          Alberto D. Bolatto
          \inst{8}
          \and
          Timothy Robishaw
          \inst{9}
          \and
          Joshua D. Simon
          \inst{10}
          \and
          \'{A}lvaro Ib\'{a}\~{n}ez P\'{e}rez
          \inst{11}
          \and
          Karl M. Menten
          \inst{1}
          \and
          Jarken Esimbek
          \inst{2}
          }

   \institute{Max-Planck-Institut f\"{u}r Radioastronomie, Auf dem H\"{u}gel 69, D-53121 Bonn, Germany\\
   \email{gwu@mpifr-bonn.mpg.de}
   \and Xinjiang Astronomical Observatory, Chinese Academy of Sciences, 830011 Urumqi, Xinjiang, PR China
   \and Instituto de Astrof\'{i}sica de Andaluc\'{i}a, CSIC, E-18080, Granada, Spain
   \and Astronomy Department, King Abdulaziz University, PO Box 80203, Jeddah 21589, Saudi Arabia
   \and Helmholtz-Institut f\"{u}r Strahlen und Kernphysik (HISKP), University of Bonn, Nussallee 14-16, D-53115 Bonn, Germany
   \and Charles University in Prague, Faculty of Mathematics and Physics, Astronomical Institute, V  Hole\v{s}ovi\v{c}k\'ach 2 CZ-18000 Praha Czech Republic
   \and Max Planck Institute for Astronomy, K\"{o}nigstuhl 17, D-69117 Heidelberg, Germany
   \and Department of Astronomy, University of Maryland, College Park, MD 20742, USA
   \and National Research Council Canada, Herzberg Programs in Astronomy \& Astrophysics, Dominion Radio Astrophysical Observatory, P.O. Box 248, Penticton, BC V2A 6J9, Canada
   \and Observatories of the Carnegie Institution for Science, 813 Santa Barbara Street, Pasadena, CA 91101, USA
   \and Asociaci\'{o}n Astron\'{o}mica AstroHenares, 28823 Coslada, Madrid, Spain
   }
   \date{Received September 15, 1996; accepted October 16, 1996}


  \abstract
   {A fully-sampled and hitherto highest resolution and sensitivity observation of neutral hydrogen ($\HI$) in the Leo Triplet (NGC\,3628, M\,65/NGC\,3623, and M\,66/NGC\,3627) reveals six $\HI$ structures beyond the three galaxies. We present detailed results of the morphologies and kinematics of these structures, which can be used for future simulations. In particular, we detect a two-arm structure in the plume of NGC\,3628 for the first time, which can be explained by a tidal interaction model. The optical counterpart of the plume is mainly associated with the southern arm. The connecting part (base) of the plume (directed eastwards) with NGC\,3628  is located at the blueshifted (western) side of NGC\,3628. Two bases appear to be associated with the two arms of the plume.
   A clump with reversed velocity gradient (relative to the velocity gradient of M\,66) and a newly detected tail, $\ie$ M\,66SE, is found in the southeast of M\,66. We suspect that M\,66SE represents gas from NGC\,3628 which was captured by M\,66 in the recent interaction between the two galaxies.
   Meanwhile gas is falling toward  M\,66, resulting in features already previously observed in the southeastern part of M\,66, $\eg$ large line widths and double peaks.
   An upside-down `Y'-shaped $\HI$ gas component (M\,65S) is detected in the south of M\,65 which suggests that M\,65 may also have been involved in the interaction.
   We strongly encourage modern hydrodynamical simulations of this interacting group of galaxies to reveal the origin of the gaseous debris surrounding all three galaxies.
   }
   \keywords{galaxies: individual (NGC\,3628, M\,65/NGC\,3623, and M\,66/NGC\,3627) -- galaxies: interactions -- galaxies: ISM -- galaxies: peculiar -- radio lines: galaxies}
   \maketitle

%

\section{Introduction}

\begin{figure*}
\centering
\includegraphics[width=0.49\hsize]{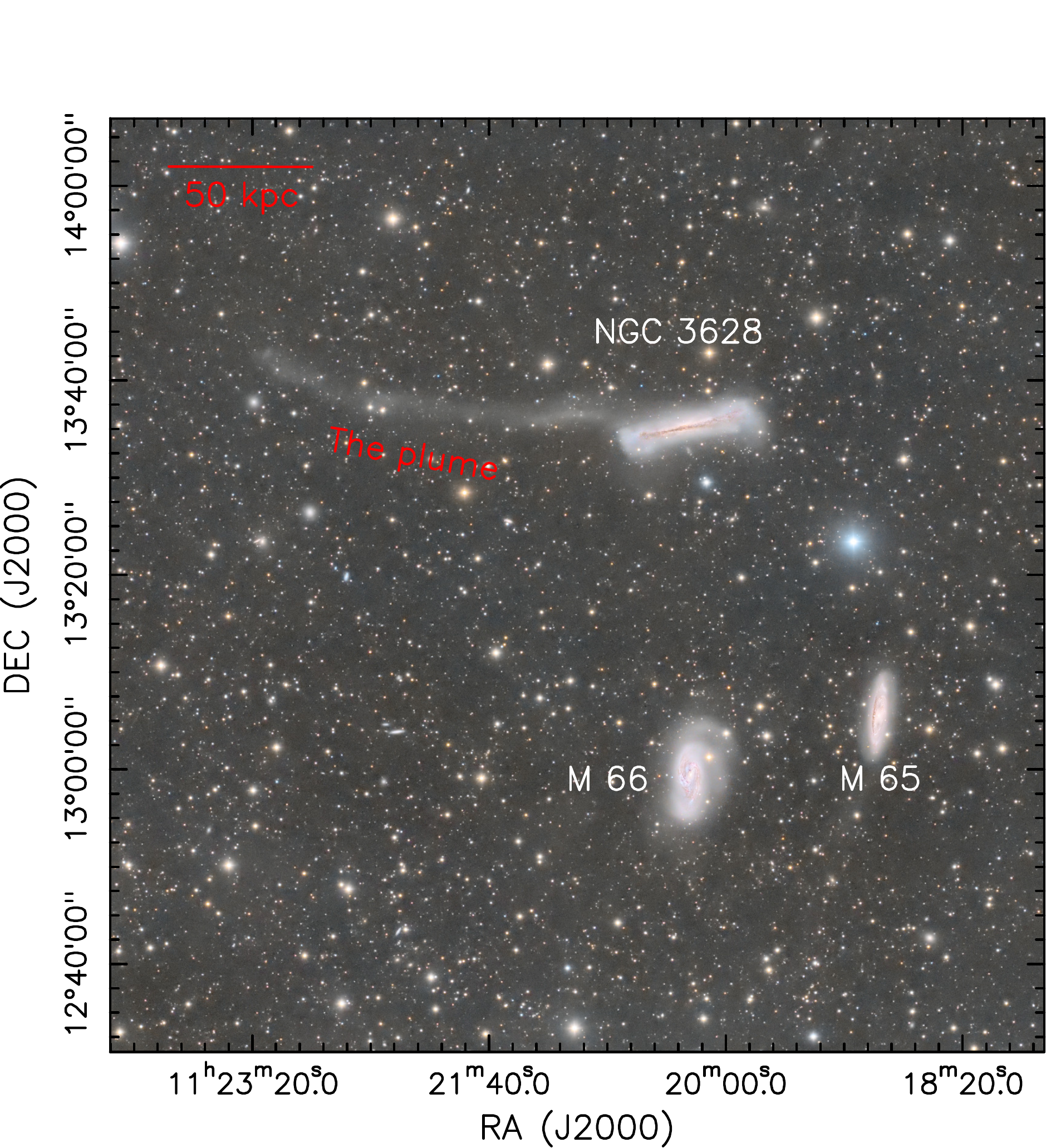}
\includegraphics[width=0.49\hsize]{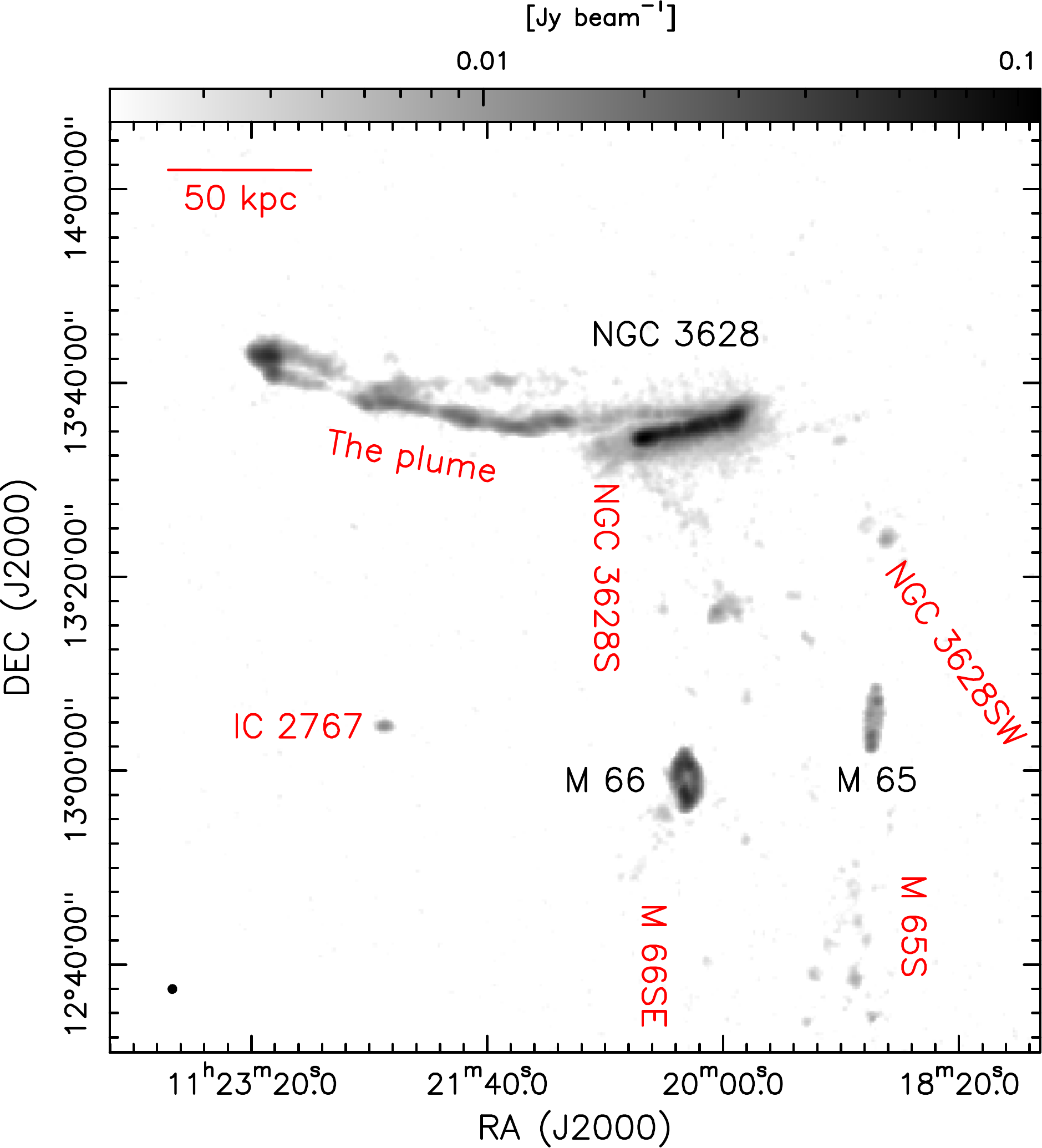}
\caption{Left panel: A full color optical image of the Leo Triplet obtained with a 10.6-cm Takahashi FSQ106EDX apocromatic refractor (see Section \ref{op}).
Right panel: The peak intensity image of the $\HI$ spectra in the Leo Triplet system. The galaxies of NGC 3628, M\,65, and M\,66 are labeled in black and the $\HI$ structures beside the three galaxies are labeled in red.}
\label{fig:op-tpeak}
\end{figure*}

Interactions between galaxies have significant impacts on their participants, leading to asymmetries, warps, exchange of gas and momentum.  These interactions can finally result in non-circular potentials, leading to inflow of gas, enhancement of bars, and also triggering starburst activity in the galaxies.
An excellent example is the Leo Triplet galaxy system (see the left panel of Fig. \ref{fig:op-tpeak}), also known as Arp 317 \citep{1966ApJS...14....1A}, mainly including the SAB(rs)a galaxy M\,65 (NGC\,3623), the SAB(s)b galaxy M\,66 (NGC\,3627), and the SAb pec galaxy NGC 3628.

The most famous feature in this system is probably the spectacular tail (the plume) from NGC\,3628 extending towards the east. The plume was first detected by Zwicky (1956) and \citet{1974AJ.....79..671K} via optical observations. Then \citet{1978AJ.....83..219R} and \citet{1979ApJ...229...83H} reported  neutral hydrogen ($\HI$) observations in the Leo Triplet and revealed a $\approx$150 kpc (rescaled to a distance of 11.3$\pm$0.5 Mpc) long $\HI$ structure\footnote{
\citet{2021MNRAS.501.3621A} recently compiled a catalog of the best available distances of 118 galaxies. Here we adopt the average of the red giant branch distance of M\,65 (11.3$\pm$1.1 Mpc) and the group distances (from the galaxy group and numerical modeling  of their orbits) of M\,66 (11.32$\pm$0.48 Mpc) and NGC\,3628 (11.3$\pm$1.1 Mpc) in \citet{2021MNRAS.501.3621A} as the distance to the Leo Triplet. See \citet{2009AJ....138..332J} and \citet{2021MNRAS.501.3621A} for more details. }
consistent with its optical counterpart. In these $\HI$ observations, an additional bridge-like structure is also detected extruding from NGC\,3628 and likely pointing southwards to M\,66.
\citet{1978AJ.....83..219R} applied the methods of \citet{1972ApJ...178..623T} and basically recovered the morphologies and kinematics of $\HI$ plume and bridge by assuming a recent (8$\times$10$^{8}$ years ago) tidal encounter between NGC\,3628 and M\,66.
However, there remained some tension between the model and the observational data as argued by \citet{1979ApJ...229...83H}.
Moreover, according to the scenarios simulated by \citet{1972ApJ...178..623T}, the bridge seems to be too prominent for two galaxies with similar total mass \citep[$\eg$][]{1978AJ.....83..219R}.

In spite of that, major mergers (involving two galaxies with comparable mass) are the commonly cited mechanism to explain the observations. An alternative is that a non-merger tidal interaction between passing galaxies with comparable masses pulls out tidal tails. Such strong encounters are only possible if dynamical friction on the extensive dark matter halos does not operate \citep{2015CaJPh..93..169K, 2016MNRAS.463.3637R}.
Evidence for interactions either related to M\,66 or NGC\,3628 have also been found in optical, radio continuum, spectroscopic, and polarization observations
\citep[$\eg$,][]{1978AJ.....83..219R, 1979ApJ...229...83H, 1983ApJ...269..136Y,  1992ApJ...385..188B, 1993ApJ...418..100Z, 1993MNRAS.263.1075W,  1996A&A...306..721R, 1998AJ....115.2331C, 2001A&A...378...40S, 2003A&A...405...89C, 2011ASPC..446..111D, 2012A&A...544A.113W}.
However, previous simulations and observations are mainly focused on the interaction between M\,66 and  NGC\,3628. The role of the third member, M\,65, in this triplet system has remained unclear. On the one hand,  M\,65 looks quiescent and no clear distortion is seen in it, which supports the view that M\,65 is not involved in the interaction \citep[$\eg$, ][]{2005A&A...429..825A, 2006AJ....132.1581D}.  On the other hand, simulations of the interaction solely between NGC\,3628 and M\,66 still show some discrepancies with observations \citep[$\eg$][]{1979ApJ...229...83H, 2015ApJ...812L..10J}.

An alternative scenario for the formation of the giant tidal tail associated with NGC 3628 has also been discussed more recently, the minor merger ($\ie$ a merger between a large and a less massive satellite galaxy).
Within the hierarchical galaxy formation framework, minor mergers are expected to be significantly more common than mergers with comparable masses at the current epoch  \citep[$\eg$][]{2000MNRAS.319..168C}.
If the satellite galaxies become tidally disrupted, they should leave behind extended low surface brightness substructures known as tidal streams \citep[$\eg$] []{2010AJ....140..962M, 2015AJ....150..116M}.
Two typical examples in our local volume are the Sagittarius tidal stream surrounding our Milky Way \citep[$\eg$][]{2003ApJ...599.1082M} and the Great Southern stream around the Andromeda galaxy \citep[$\eg$][]{2001Natur.412...49I}. Recently, \citet{2015ApJ...812L..10J} identified and characterized a compact massive star cluster, NGC\,3628-UCD1, embedded in the base of the  plume. It shows similar characteristics as the most massive Milky Way globular cluster, $\omega$ Centauri, which is believed to be a stripped dwarf galaxy remnant \citep[$\eg$][]{2019ApJ...886..121B}.
NGC\,3628-UCD1 (see Section \ref{twoArm} below) is also surrounded by a resolved stellar population showing a typical S-shaped structure, which strengthens the scenario of a young tidal stream.
Therefore, \citet{2015ApJ...812L..10J} suggested the possibility of a minor merger between NGC\,3628 and a dwarf elliptical  galaxy (the progenitor of the star cluster NGC\,3628-UCD1) adding additional complexity to the standard interaction scenario.

The appendages, $\ie$ tails and bridges, provide a fossil record of the interaction history and pose constraints on dynamical models.
To recover the morphologies and kinematics of these appendages provides a principal benchmark for dynamical models \citep[$\eg$][]{1978AJ.....83..219R, 1979ApJ...229...83H}.
$\HI$ observations bring enormous convenience to identifying appendages. For example, the $\HI$ plume \citep[]{1979ApJ...229...83H} is more remarkable than the faint optical plume \citep[][]{1974AJ.....79..671K} and so far no CO has been detected from the plume \citep[][]{1983ApJ...269..136Y}.
Meanwhile, to obtain a full view of the appendages, the observations should include the three galaxies but should also cover a more extended region.
Previous $\HI$ observations covering the entire Leo Triplet system were all conducted with the Arecibo 305 meter telescope with a resolution of about 4$\arcmin$ \citep{1979ApJ...229...83H, 2009AJ....138..338S}. Higher resolution and sensitivity observations are required to further constrain the dynamical model and to also clarify the origin of some interesting features observed in this system. For example, the $\HI$ emission observed along the plume surprisingly shows an almost constant velocity field but contains two velocity regimes \citep{1979ApJ...229...83H}. 
A `peculiar velocity' clump is detected in the southeast of M\,66 which is found to be not corotating with  M\,66 \citep[][]{1979ApJ...229...83H, 1993ApJ...418..100Z}.

In this paper, we report high resolution and sensitivity fully-sampled neutral hydrogen observations of the entire Leo Triplet system by combining VLA and Arecibo observations. We provide detailed morphologies and kinematics of the appendages and seek to understand the interaction history of this system.

\section{Observations}
\label{ob}
\subsection{VLA and Arecibo observations}
\label{radio}
The VLA observations (Project code: AB1074, PI: A. Bolatto) were conducted on 2003 March 29th and 30th in the D configuration.
Eighteen fields were used to completely cover the Leo Triplet and the integration time on each field is $\ga$ 40 minutes.
The details of the observations are summarized in Table \ref{tab:smob}.
Calibration of the VLA observations was performed with the VLA pipeline in AIPS\footnote{http://www.aips.nrao.edu/} and the continuum was subtracted from the visibilities using the task "uvlin" with a linear fit. The visibilities are then imaged and cleaned in CASA \citep{2007ASPC..376..127M}  using a "multiscale" clean in the "tclean" task. Within a clean mask defined by the "auto-multithresh" algorithm, the emission is cleaned down to 1.5 m$\Jypb$ ($\approx$3$\sigma$).
The synthesized beam is 61$\ffas$06 $\times$ 59$\ffas$40 at P.A. $\approx$ \hbox{--65$^{\circ}$} and the pixel size is 15$\arcsec$. The channel spacing is $\Delta V = 20.7\,\kms$ (see Table \ref{tab:smob}).
The rms noise in a single channel is $\sigma_{\rm rms}$ $\la$ 0.43\,m$\Jypb$.

The Arecibo observations (Project code: A1581, PI: Bolatto) were conducted over 11 days from 2002 April 7 through 2002 April 20. A $\approx$ 2$\ffad$2 $\times$ 1$\ffad$5 area of the sky was covered by the Arecibo observations using the on-the-fly mapping technique (see Fig. \ref{fig:arecibo}) with a total observing time of about 27.25 hours.
These data were used to fill the zero spacing of the VLA HI observations.  The L-band Narrow dual-polarization frontend ($T_{\rm sys}\,\approx\,30$\,K) was used in combination with the correlation backend configured for 9-level sampling over a 6.125\,MHz bandwidth with 2048 channels.  This yielded a 1320 km s$^{-1}$ bandpass sampled in 0.64 km s$^{-1}$-wide channels.
At this frequency, the telescope has a full width at half maximum (FWHM) beam width of $\approx$3.3$\arcmin$.
The orthogonal polarizations were combined by weighting by the reciprocal of the squares of their noise levels. The polarization-averaged on-the-fly (OTF) data were sampled onto a regular grid after applying a slant orthographic projection and a Gaussian convolution function with an FWHM of 3$\ffam$3. An effective integrated time per beam is about 70 seconds. The rms noise in a single channel is $\sigma_{\rm rms}$ $\la$ 3.6 m$\Jypb$.
The interferometric and single dish cubes were combined with "immerge" (revision 1.7) in miriad \citep{1995ASPC...77..433S} and the weighting factor between Arecibo and VLA is set to unity.
The rms noise in a single channel of the combined data is $\sigma_{\rm rms}$ $\la$ 0.5 m$\Jypb$.
Integrated over the full velocity range of a single channel, this yields $\sigma_{\rm ch}$ = $\sigma_{\rm rms} \times \Delta V $ $\la$ 0.01 $\Jypbkms$.

Here we also roughly estimate the flux recovered by the combined data. The integrated line fluxes are all obtained by integrating from $V_{\rm HEL}$ =506.7 to 1125.1\,$\kms$ (see Section \ref{results}).
Then the total flux of the combined data is found to be 220.5 $\Jykms$ by spatially integrating the pixels with signal-to-noise ratios (SNRs) larger than 5.
Within the same pixels, the total flux of the VLA data is 185.6 $\Jykms$.
The total flux of the Arecibo data within the pixels with SNRs larger than 5 is 233.1 $\Jykms$.
Therefore, the combined data have recovered most of the flux filtered by the VLA. The total $\HI$ mass of the Leo Triplet system is $6.6\times 10^{9}$ M$_{\odot}$ by using the expression M$_{\rm HI}$ = 2.36 $\times$10$^{5}$ D$^{2}_{\rm Mpc}$$\int$S$_{\nu}$d$\nu$ \citep{1986AJ.....91..791V}, which is consistent with the results reported by \citet{2009AJ....138..338S} ($\approx 9\times 10^{9}$ M$_{\odot}$) and \citet{1979ApJ...229...83H} ($\approx 3\times 10^{9}$ M$_{\odot}$) for the distance of 11.3 Mpc adopted here.

In this paper, if not otherwise noted, $\HI$ maps represent the combined data.
The velocity-integrated emission (zeroth moment), intensity-weighted velocity (first moment), and intensity-weighted velocity dispersion (second moment) maps are all constructed using the routines in the GILDAS software package\footnote{https://www.iram.fr/IRAMFR/GILDAS}.
A 3$\sigma_{\rm ch}$ threshold, where $\sigma_{\rm ch}$ is the rms noise level in a channel, is used to derive the first and second moment maps in order not to emphasize noisy features.

\begin{table*}
\caption{Summary of the VLA Observations.}
\label{tab:smob}
\centering
\begin{tabular}{cccccccc}     
\hline\hline
Obs. Date & Array  & \multicolumn{2}{c}{Freq. Cove. (MHz)}   &  Spec. Res.  & Bandpass   & Gain    &  Flux   \\
\cline{3-4}
  & Configuration & spw 1 & spw 2   & kHz ($\kms$)   & Cal. & Cal.   & Cal.  \\
 (1) & (2)  & (3) & (4)   &  (5)  & (6) & (7)  & (8)  \\
\hline
2003 March 29-30   & D   & 1416.198-1419.226 & 1413.854-1416.882&  97.665 (20.7) & 3C 147  & 1120+143  &  3C 286  \\

\hline
\end{tabular}
\tablefoot{Column 1: observing dates. Column 2: array configuration. Columns 3 and 4: the ranges of rest frequencies covered by the two spectral windows. Column 5: channel width. Columns 6-8: bandpass, time dependent gain, and absolute flux calibrators.}
\end{table*}

\subsection{Optical observations}
\label{op}

An optical wide-field image of the Leo Triplet was obtained using high throughput clear filters with near-IR cut-off, known as luminance filters with a Takahashi FSQ106EDX 106mm F/5 apochromatic refractor at native focal length (530\,mm) operated in Cuenca and Guadalajara (Spain) during 2019. It used an Atik 16200 CCD camera and Astrodon LRGB filters with a
pixel scale of 2$\ffas$34.
The reduction was carried out with the usual steps of bias and dark subtraction. The flat-fielding was done in Pixinsight, using a Geoptik 200-mm flat-field generator. Astrometry was obtained using \texttt{SCAMP} \citep{2006ASPC..351..112B}. A final stacked image (see Fig.1, left panel) was obtained by combining the
25$\times$600-s best images in luminance and the
18$\times$300-s best images in RGB filters, with a total exposure time of 520 minutes (8.40 hours).

\section{Results}
\label{results}

We first present the peak intensity (the intensity of the strongest channel) image of the $\HI$ data to show the overall distribution of all the various $\HI$ components and to highlight the narrow features in the right panel of Fig. \ref{fig:op-tpeak}.
Here, we see that several structures, $\ie$ the plume, NGC\,3628W, NGC\,3628S, NGC\,3628SW, IC\,2767, M\,66SE, and M\,65S (named after their relative positions except the plume and IC\,2767), in addition to NGC\,3628, M\,65, and M\,66, can clearly be identified and also labeled.
We should note that the peak intensity map in the right panel of Fig. \ref{fig:op-tpeak} biases the $\HI$ emission against that of galaxies with broad and irregular profiles.

\begin{figure*}
\centering
\includegraphics[width=0.45\hsize]{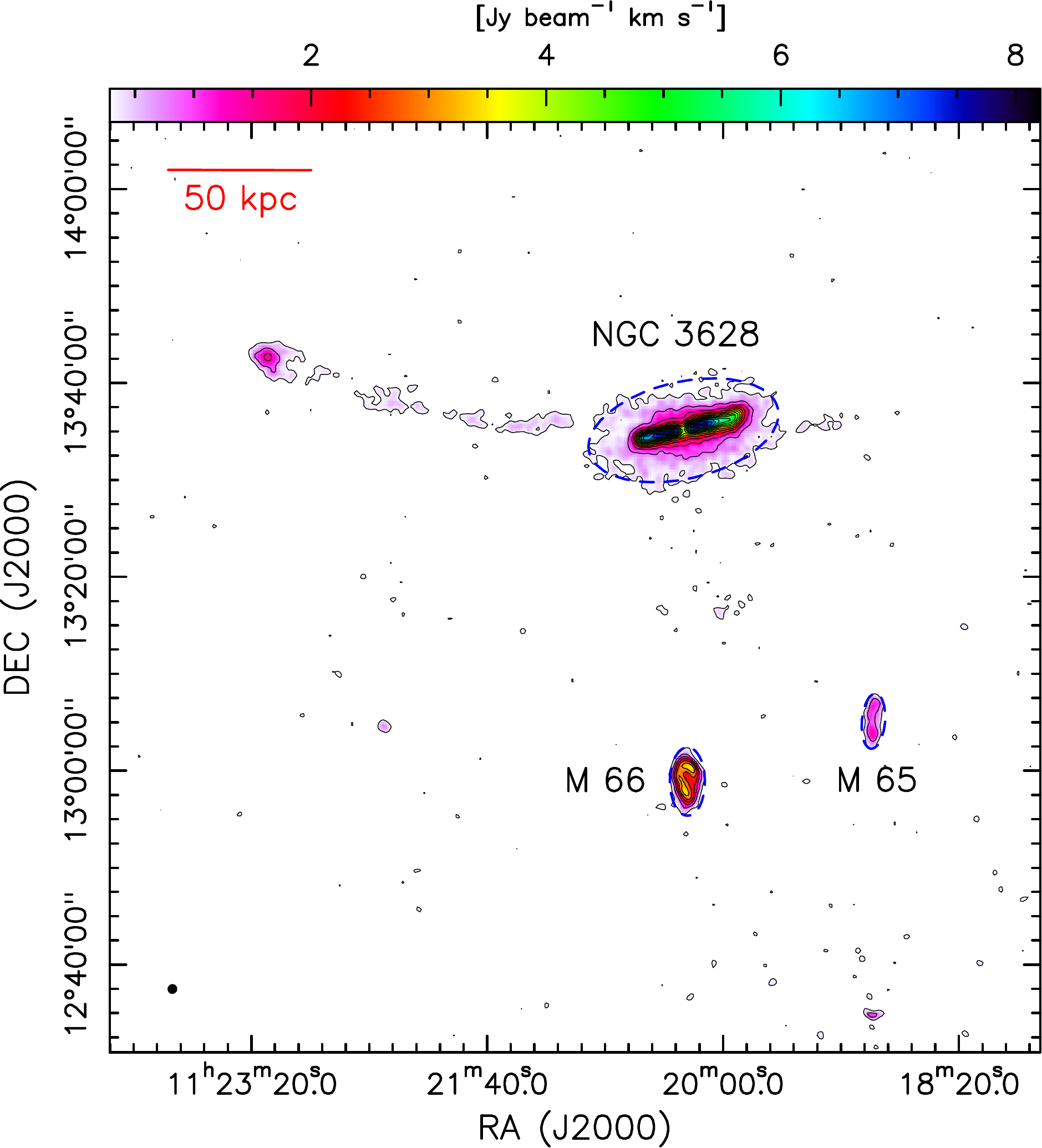}
\includegraphics[width=0.45\hsize]{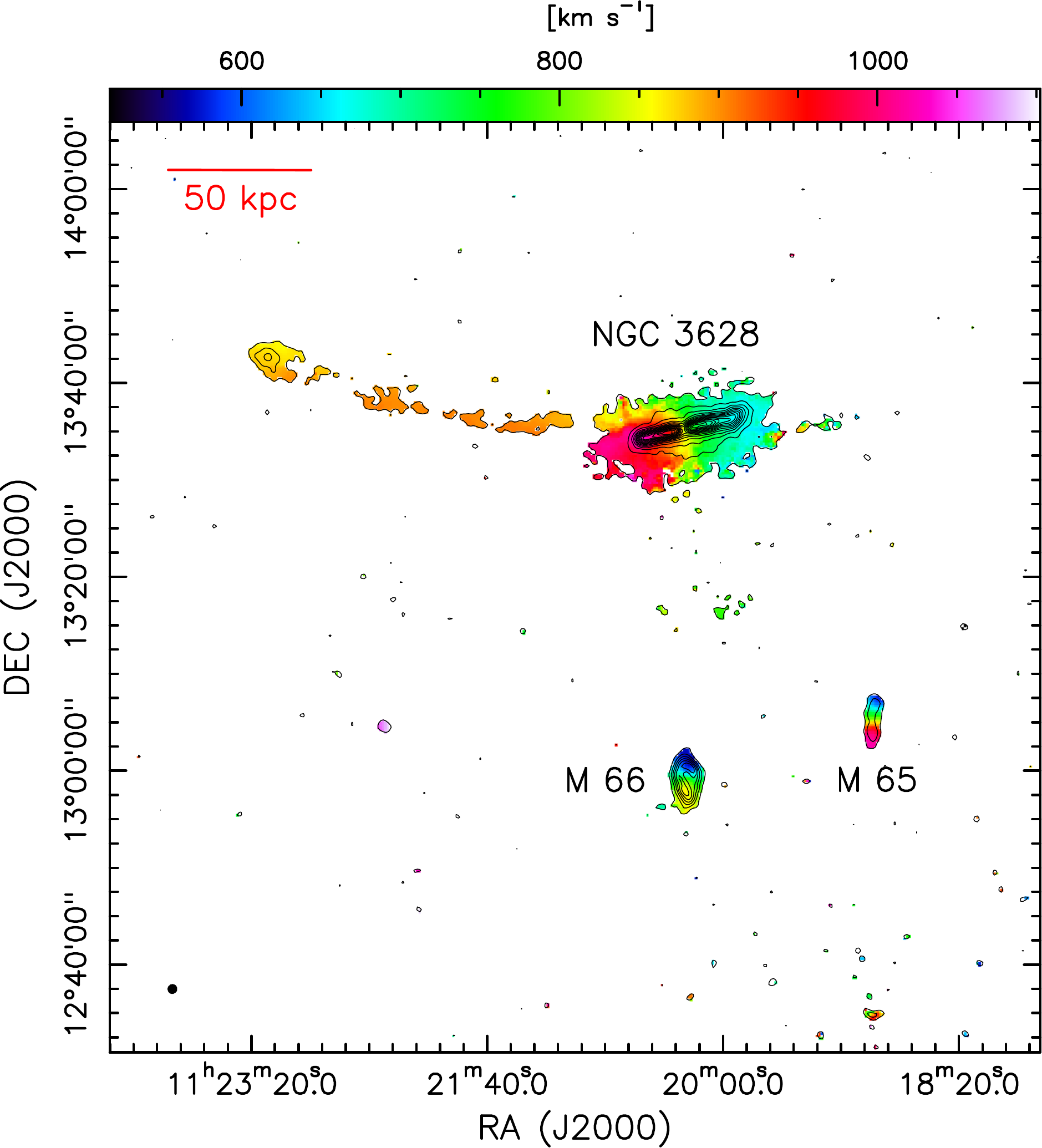}
\caption{The $\HI$ velocity-integrated emission (zeroth moment, left panel) and intensity-weighted velocity (first moment, right panel) maps. The integration range is from  $V_{\rm HEL}$ = 506.7 to 1125.1\,$\kms$ to cover all the $\HI$ features. In both panels the contour levels are set to $(5, 15, 25, 35,...145)\times 0.055$ $\Jypbkms$.
The galaxies NGC 3628, M\,65, and M\,66 are labeled in both panels and a tiny filled ellipse in the lower left shows the beam. The red line in the upper left illustrates the 50\,kpc scale at a distance of 11.3 Mpc. Three blue dashed ellipses in the left panel demonstrate the main $\HI$ emission regions of the three galaxies.}
\label{fig:mom0_2-31}
\end{figure*}

\begin{figure*}
\includegraphics[width=\hsize]{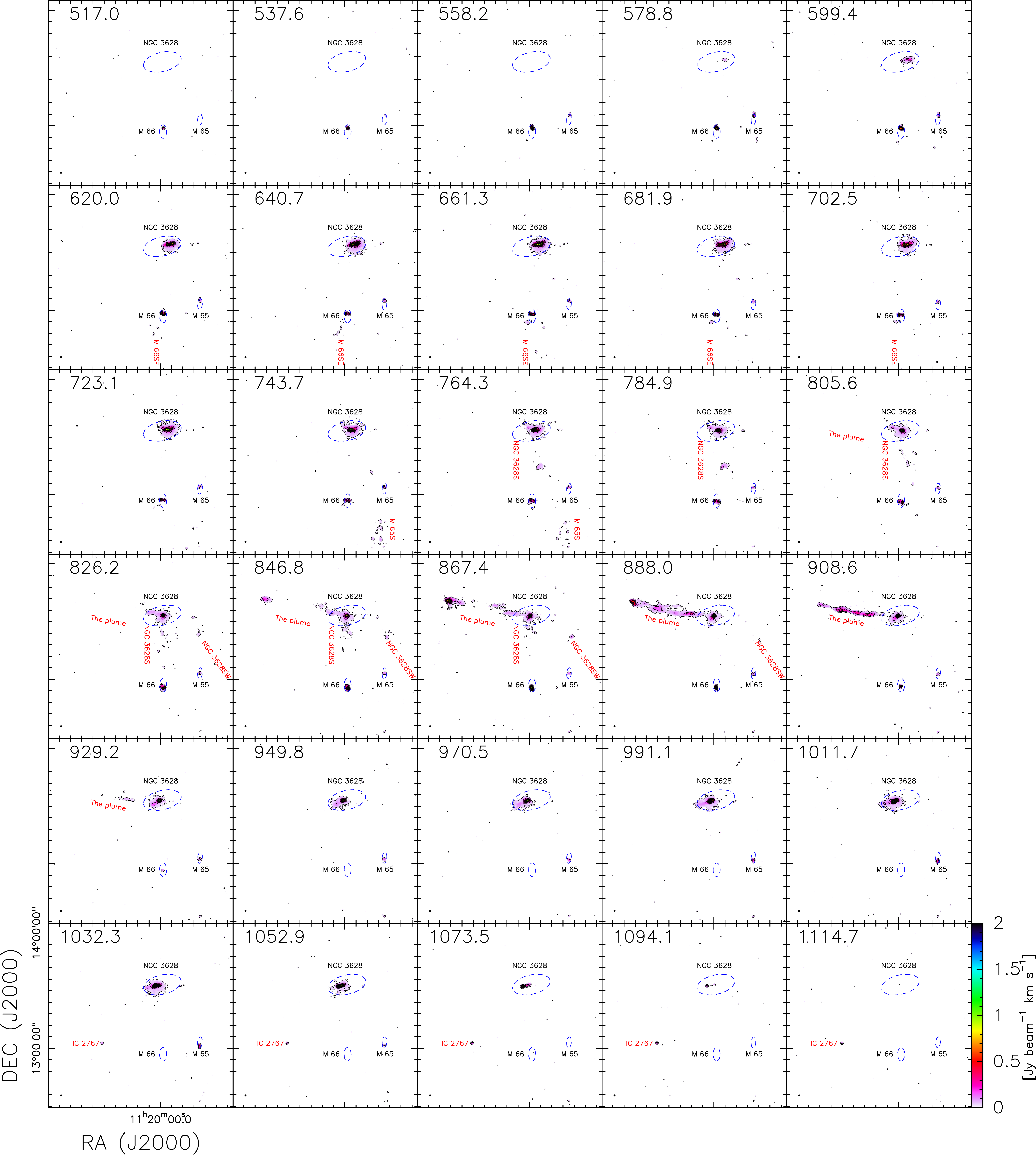}
\caption{Heliocentric velocity channel maps of the $\HI$ emission. The contours in all the panels start at 3$\sigma_{\rm ch}$ and go up in steps of 9$\sigma_{\rm ch}$, where $\sigma_{\rm ch}$ = 0.01 $\Jypbkms$ (see Section \ref{ob}). The galaxies NGC\,3628, M\,65, M\,66 are labeled in each panel in black. M\,66SE (640.7--702.5\,$\kms$), NGC\,3628S (764.3--867.4\,$\kms$), NGC\,3628SW (826.2--888.0\,$\kms$), M\,65S (743.7--764.3\,$\kms$), the plume (805.6--929.2\,$\kms$), and IC2767 (1032.3--1114.7\,$\kms$) are also labeled (in red) in their detected channels. The three blue dashed ellipses demonstrate the locations of the three galaxies belonging to the Leo Triplet.}
\label{fig:chMap}
\end{figure*}

\begin{figure*}
    \centering
    \includegraphics[width=0.45\hsize]{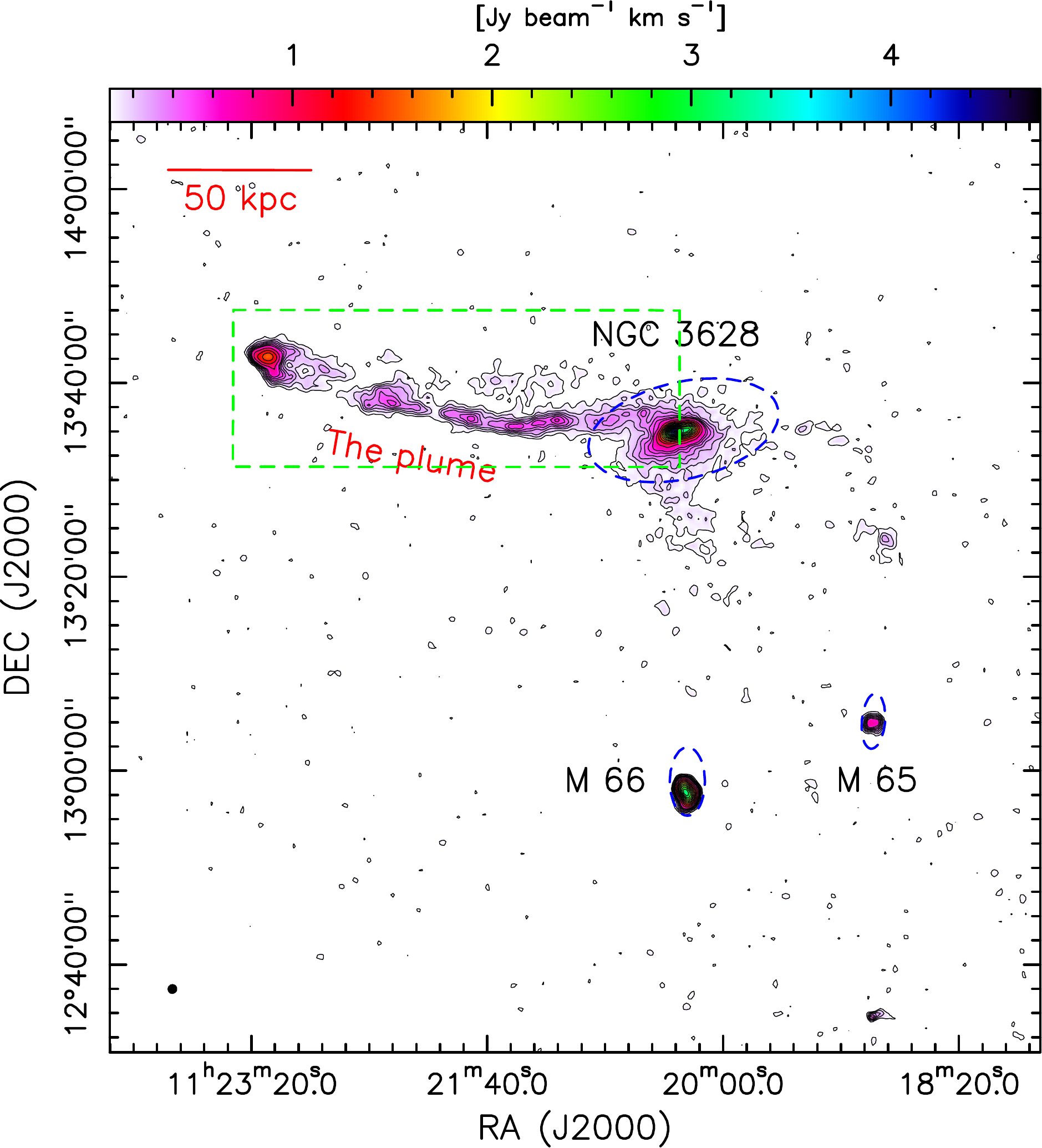}
    \includegraphics[width=0.45\hsize]{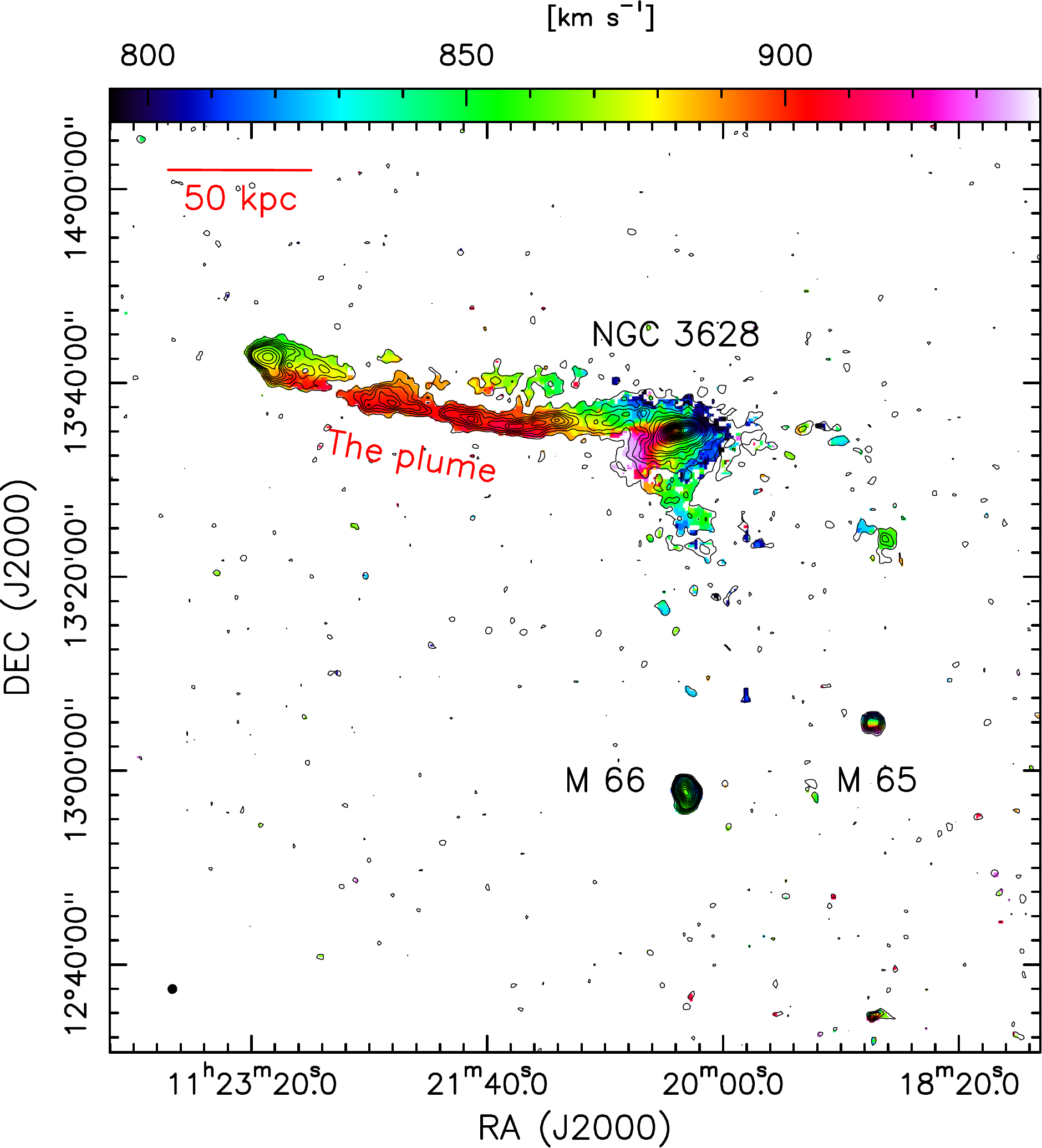}
    \includegraphics[width=0.45\hsize]{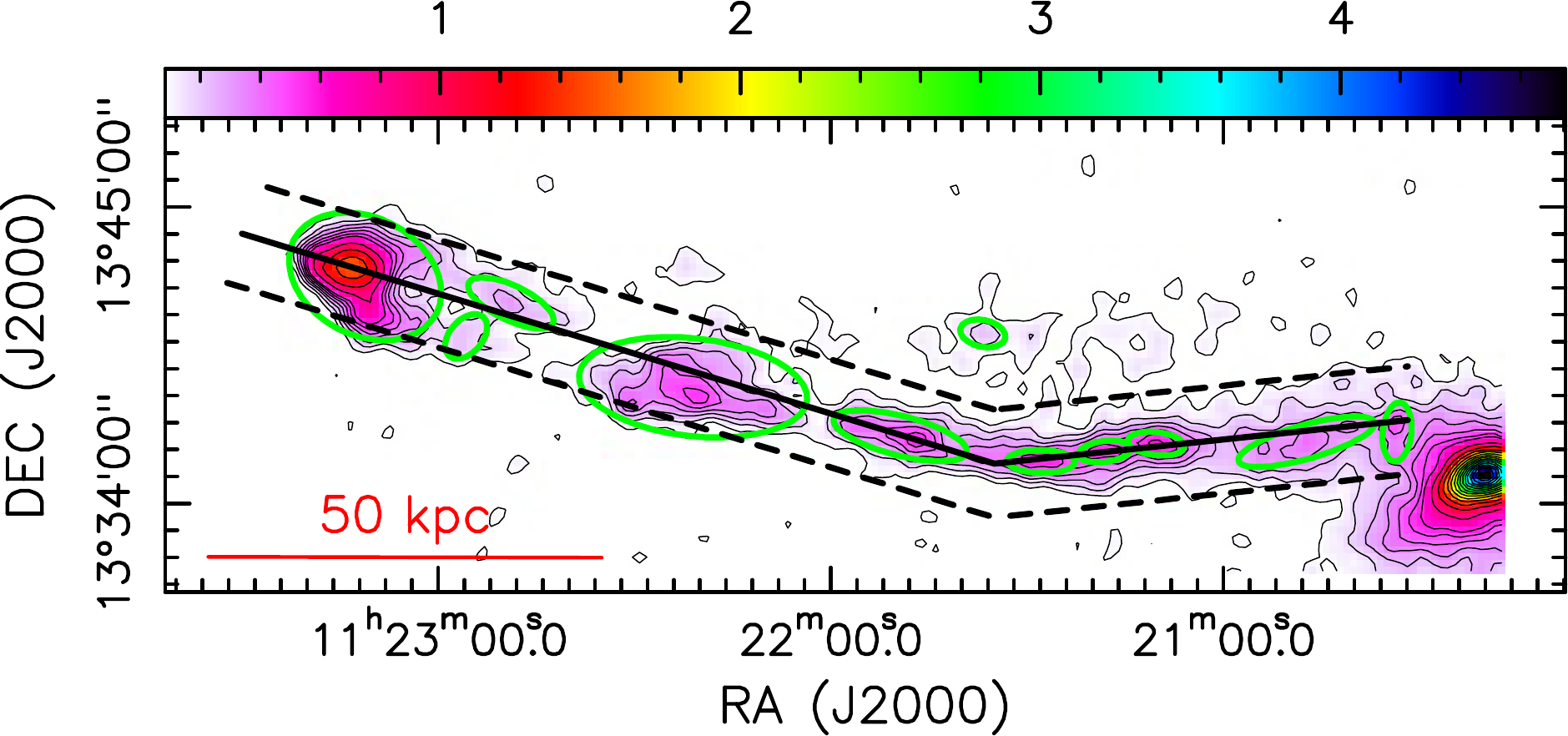}
    \includegraphics[width=0.45\hsize]{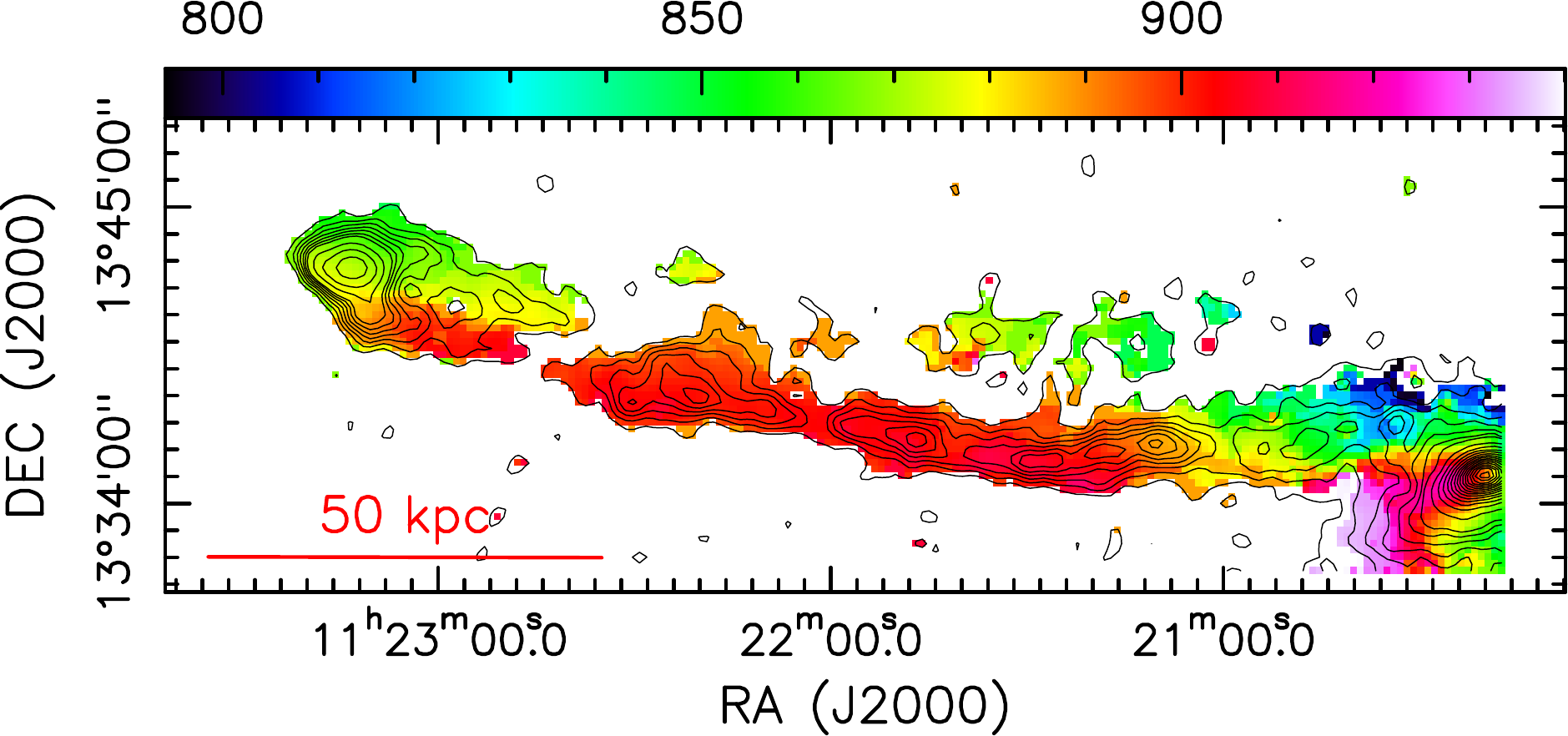}
    \caption{The $\HI$ velocity-integrated emission (zeroth moment) and intensity-weighted velocity (first moment) maps of the plume. Top two panels: In the left panel, contours and also the color image show the zeroth moment map. The integration range is from V$_{\rm HEL}$ = 795.3 to 939.5\,$\kms$. The contour levels are set to (3, 6, 9, 12, 15, 18, 21, 30, 40, 50, 60, ..., 180)\,$\times\,0.027$ $\Jypbkms$.
    The green ellipses demonstrate the clumps detected in the plume. In the right panel, the color image shows the first moment map. The contours overlaid are the same as the ones in the left panel. The galaxies NGC\,3628, M\,65, M\,66, and NGC\,3628E (the plume) are also labeled in both panels and a filled ellipse in the lower left shows the synthesized beam. The red line in the top left illustrates the 50\,kpc scale at a distance of 11.3 Mpc. The three blue dashed ellipses demonstrate the locations of the three galaxies. Bottom two panels: Same as the top two panels but zooming into the region that is illustrated by the green dashed rectangle in the top left panel. The black solid and dashed lines in the lower left panel show the loci and the averaged width of the radial profiles in Fig. \ref{fig:plume-dist}.}
    \label{fig:ngc3628e}
    \end{figure*}

\begin{figure*}
    \centering
    \includegraphics[width=0.45\hsize]{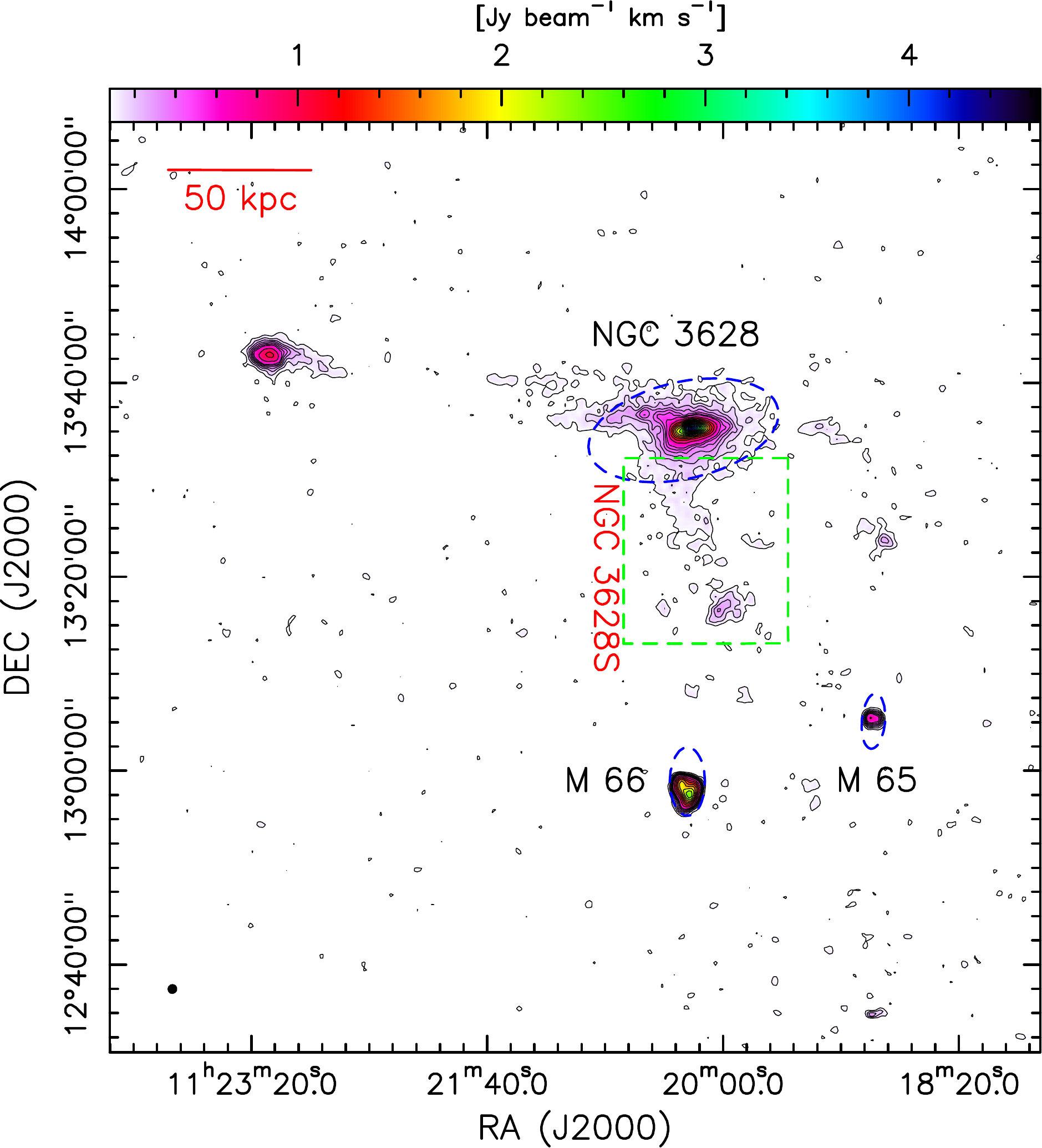}
    \includegraphics[width=0.45\hsize]{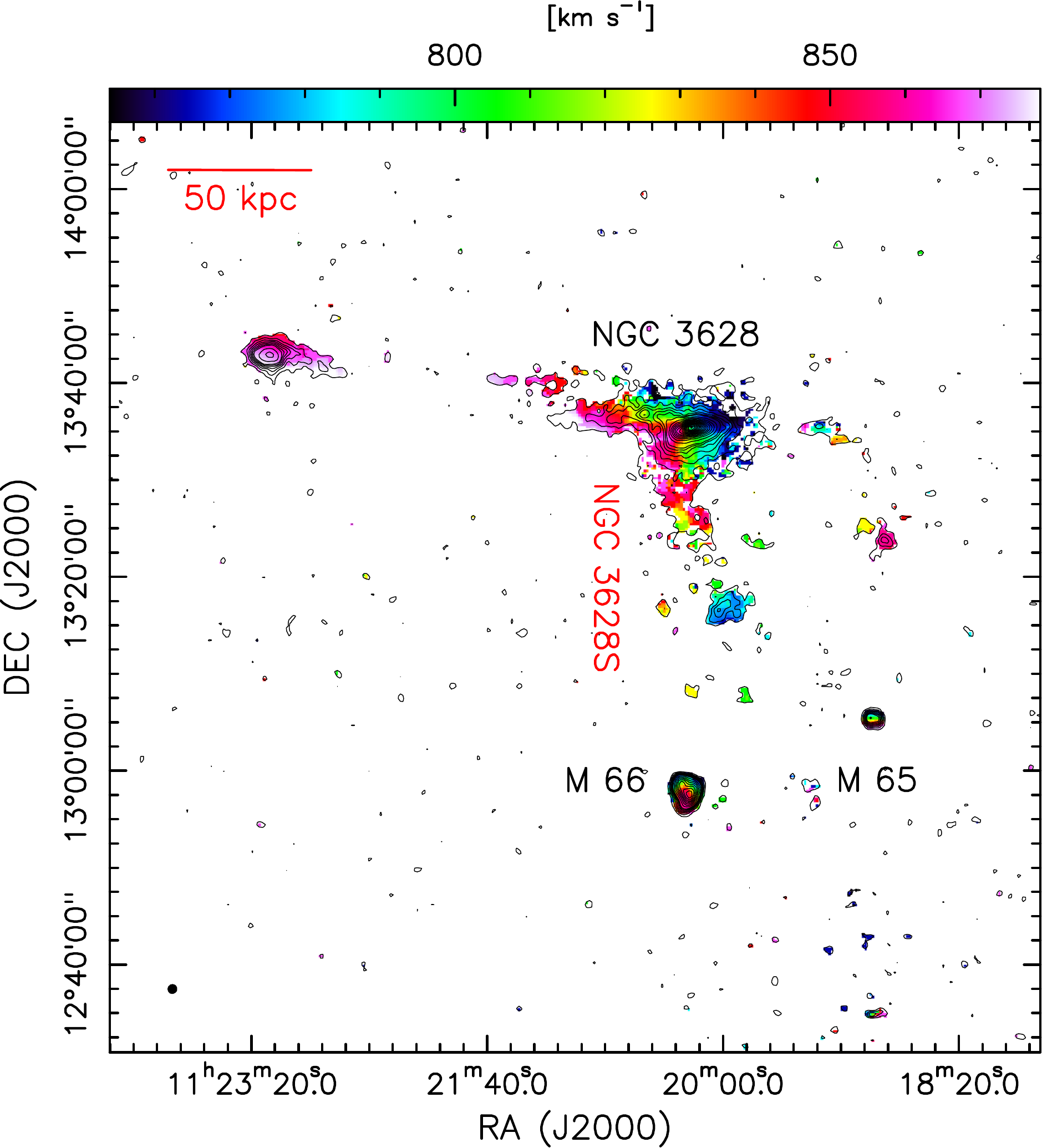}
    \includegraphics[width=0.45\hsize]{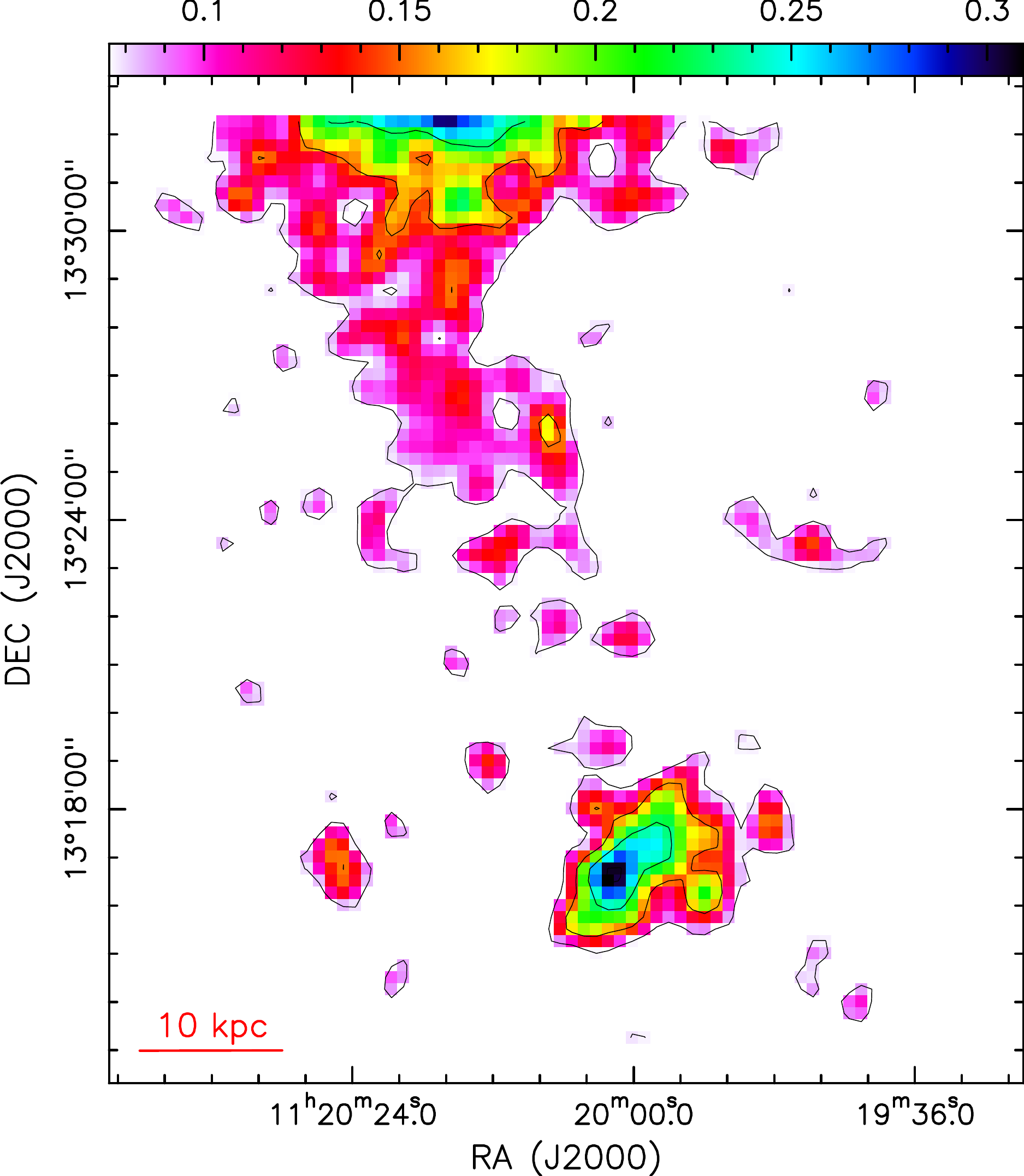}
    \includegraphics[width=0.45\hsize]{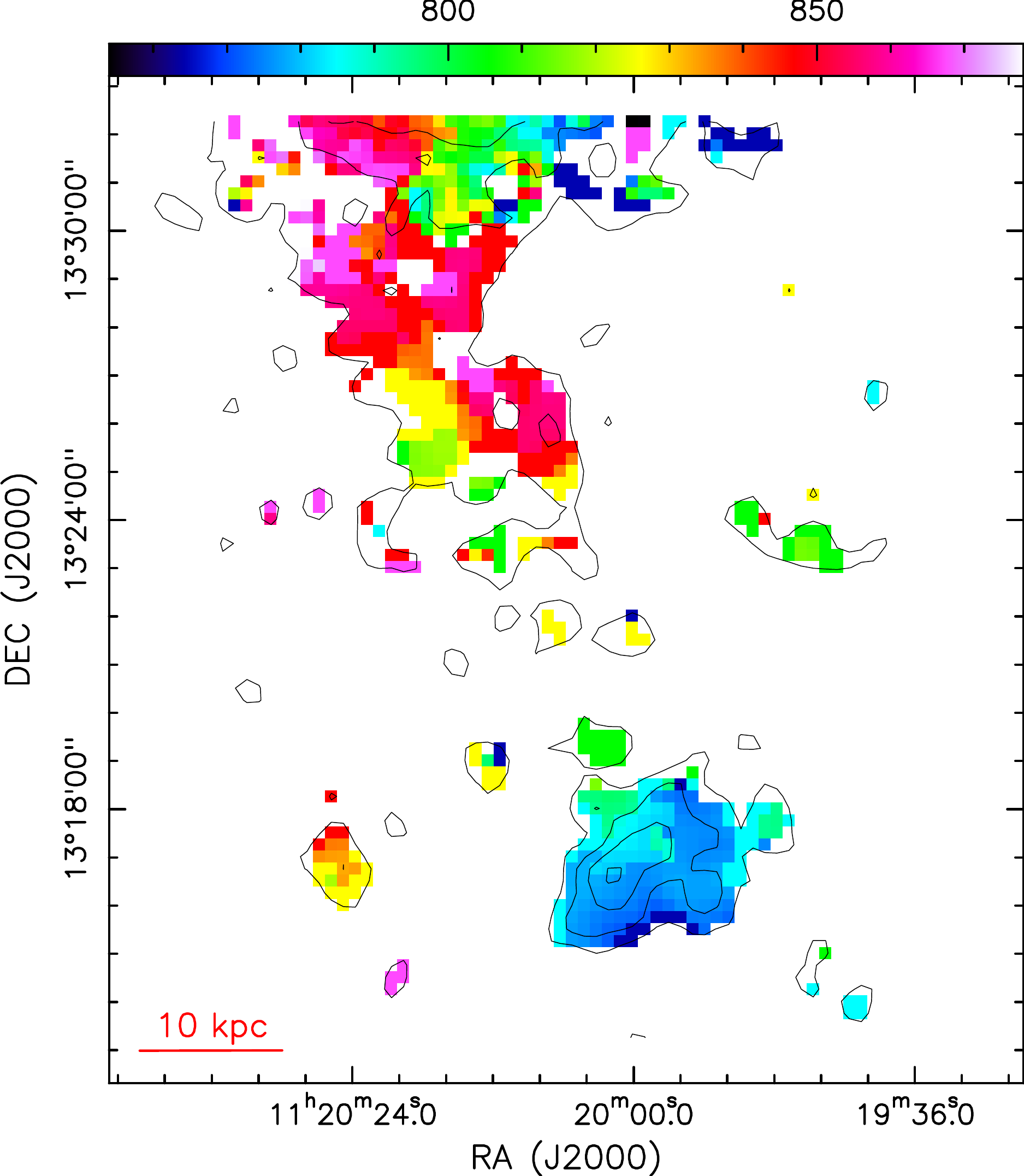}
    \caption{
    Same as Fig. \ref{fig:ngc3628e} but for NGC\,3628S. The velocity range is from 754.0 to 878.7\,$\kms$. The contour levels are set to (3, 6, 9, 12, 15, 18, 21, 30, 40, 50, 60, ..., 180)\,$\times\,0.025$\,$\Jypbkms$.}
    \label{fig:ngc3628s}
    \end{figure*}

\begin{figure*}
    \centering
    \includegraphics[width=0.45\hsize]{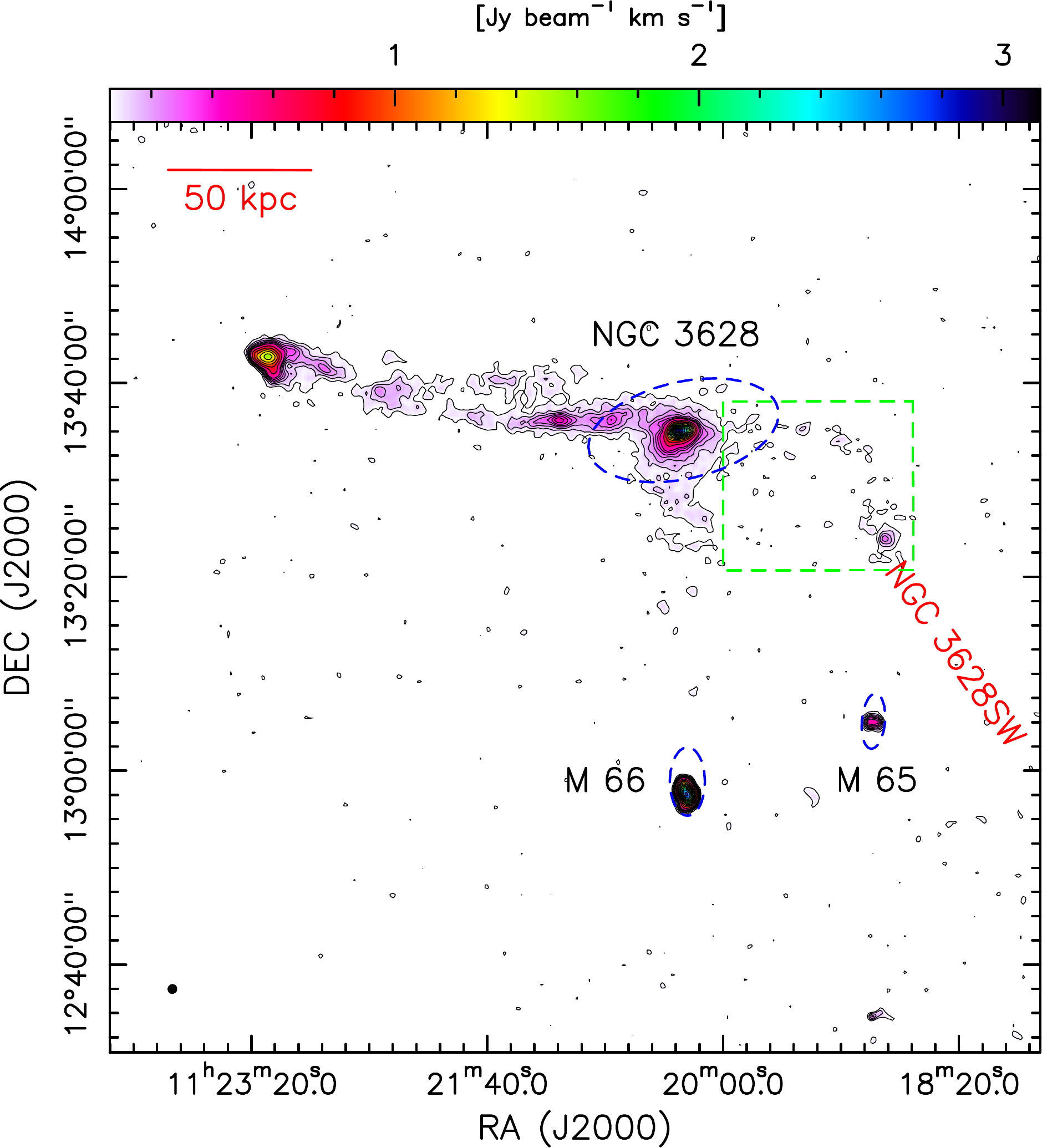}
    \includegraphics[width=0.45\hsize]{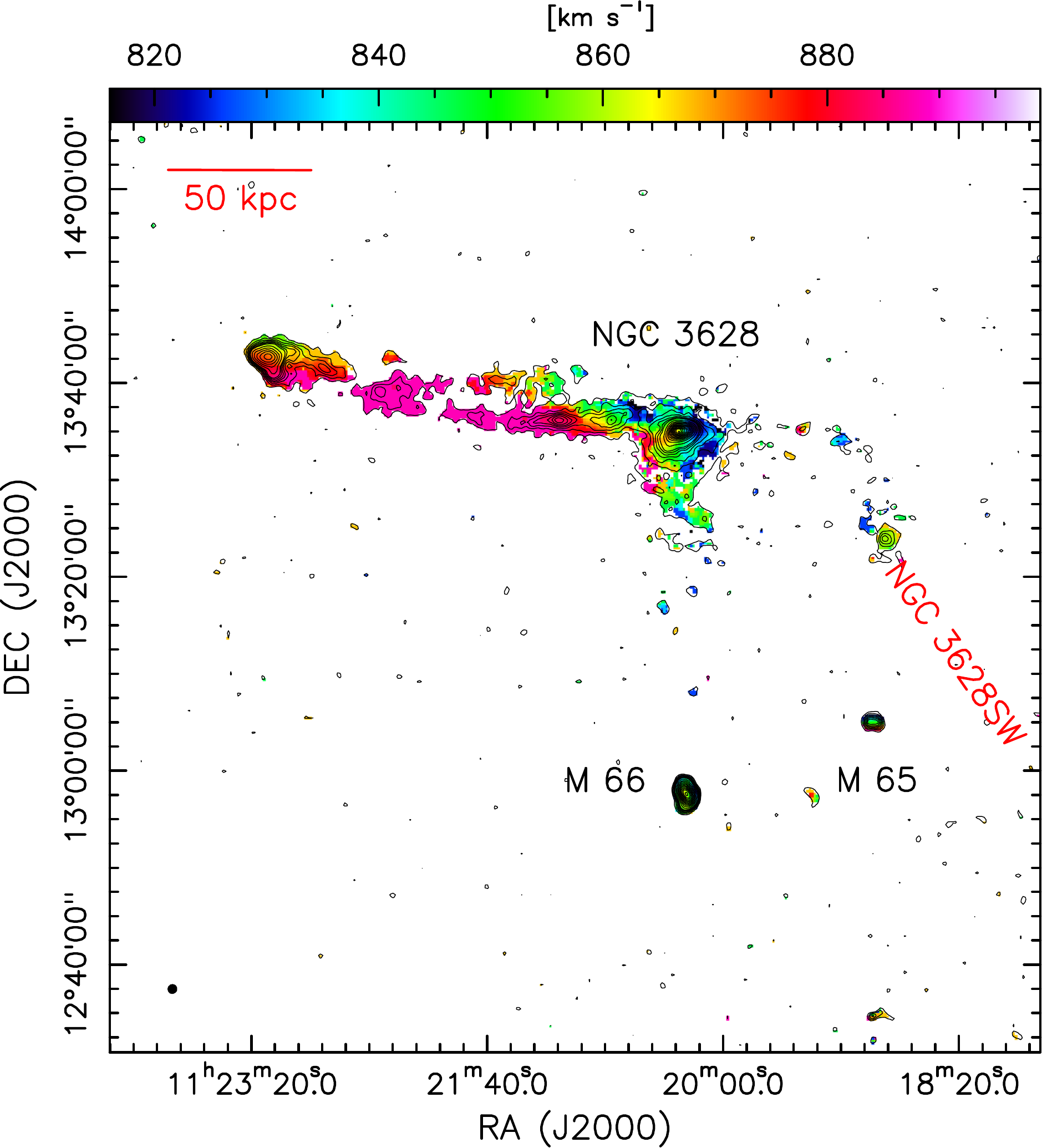}
    \includegraphics[width=0.45\hsize]{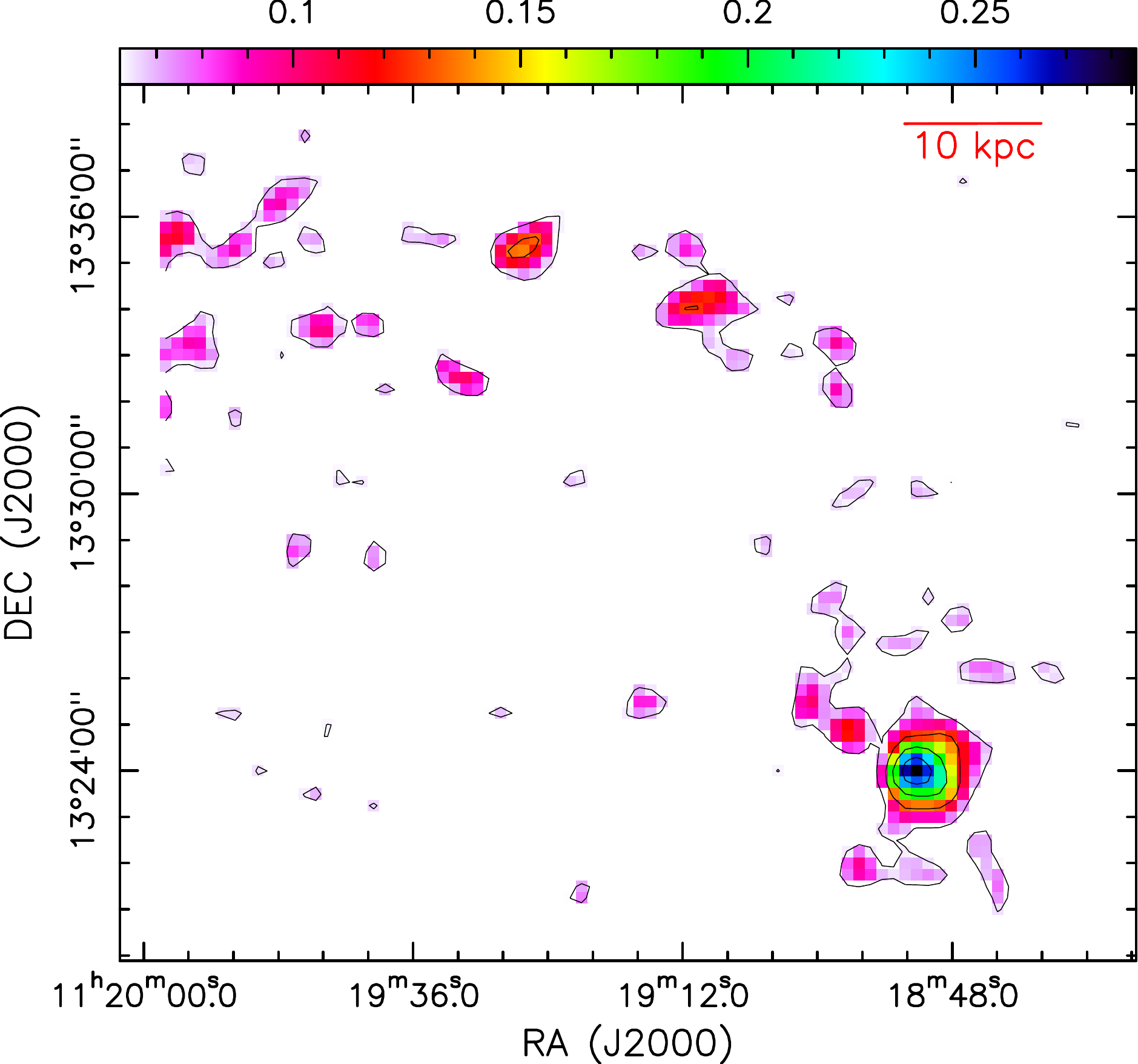}
    \includegraphics[width=0.45\hsize]{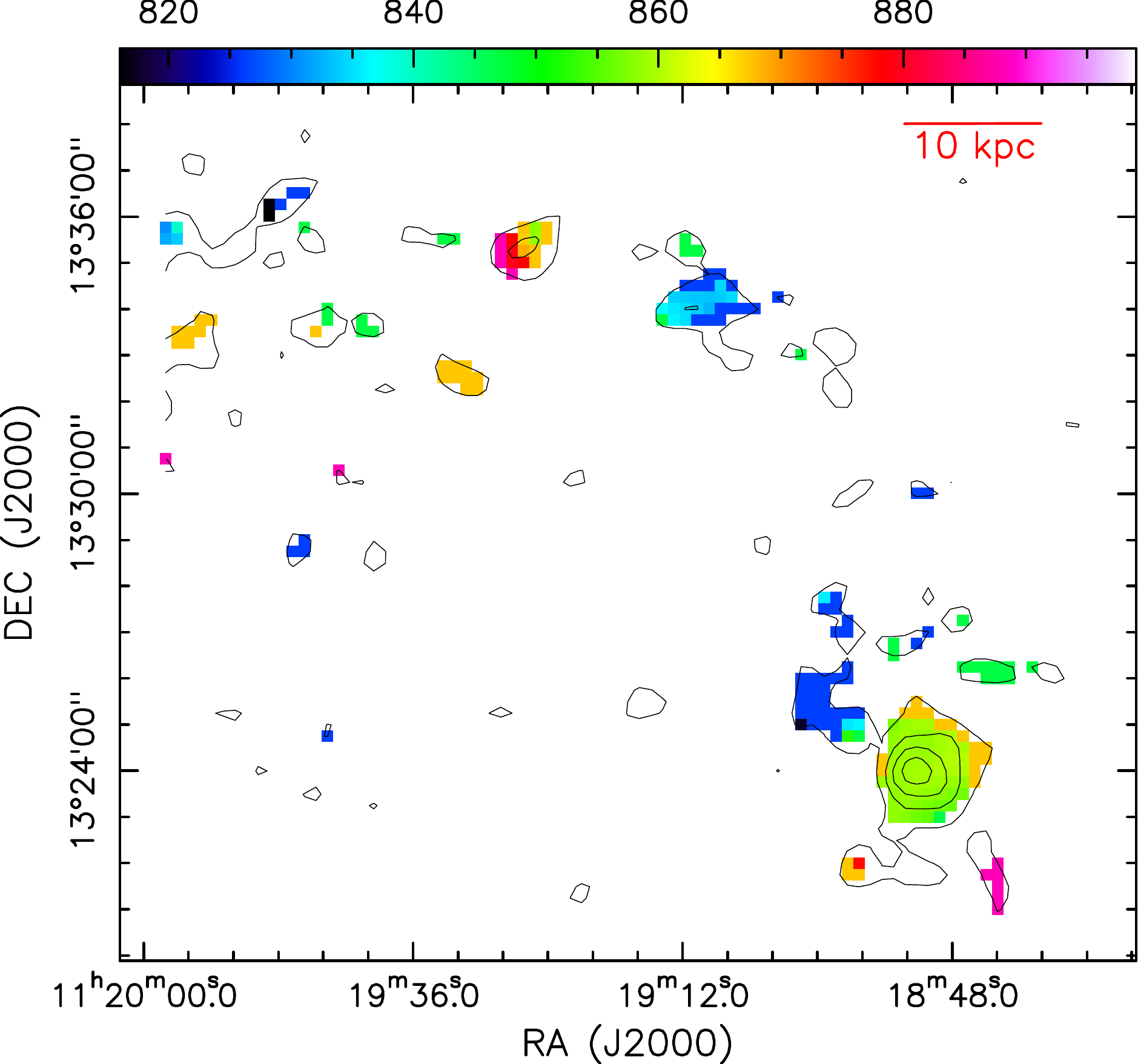}
    \caption{Same as Fig. \ref{fig:ngc3628e} but for NGC\,3628SW. The velocity range is from 815.9 to 898.3\,$\kms$. The contour levels are set to (3, 6, 9, 12, 15, 18, 21, 30, 40, 50, 60, ..., 150)\,$\times\,0.021$\,$\Jypbkms$.}
    \label{fig:ngc3628sw}
    \end{figure*}

\begin{figure*}
    \centering
    \includegraphics[width=0.45\hsize]{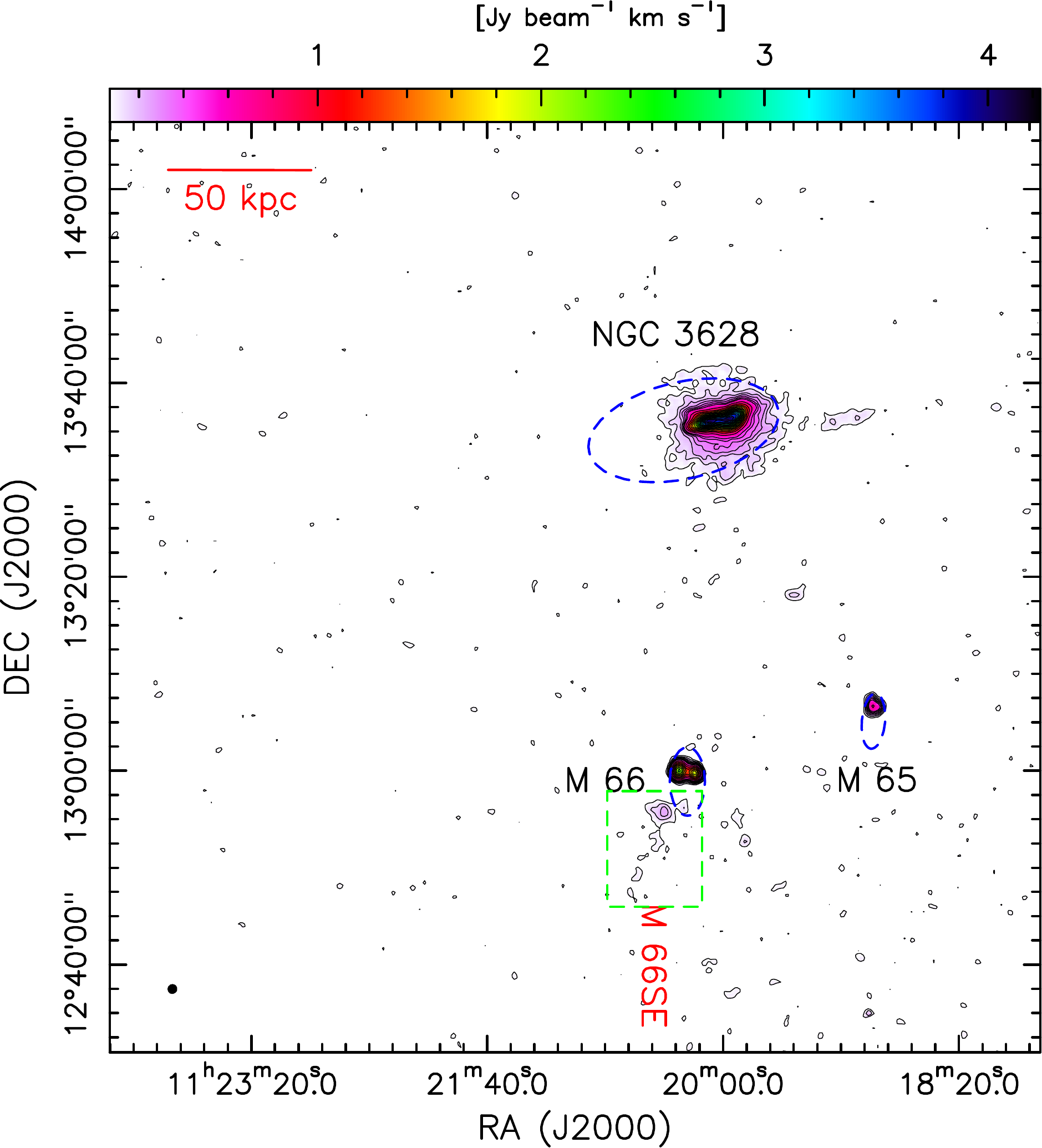}
    \includegraphics[width=0.45\hsize]{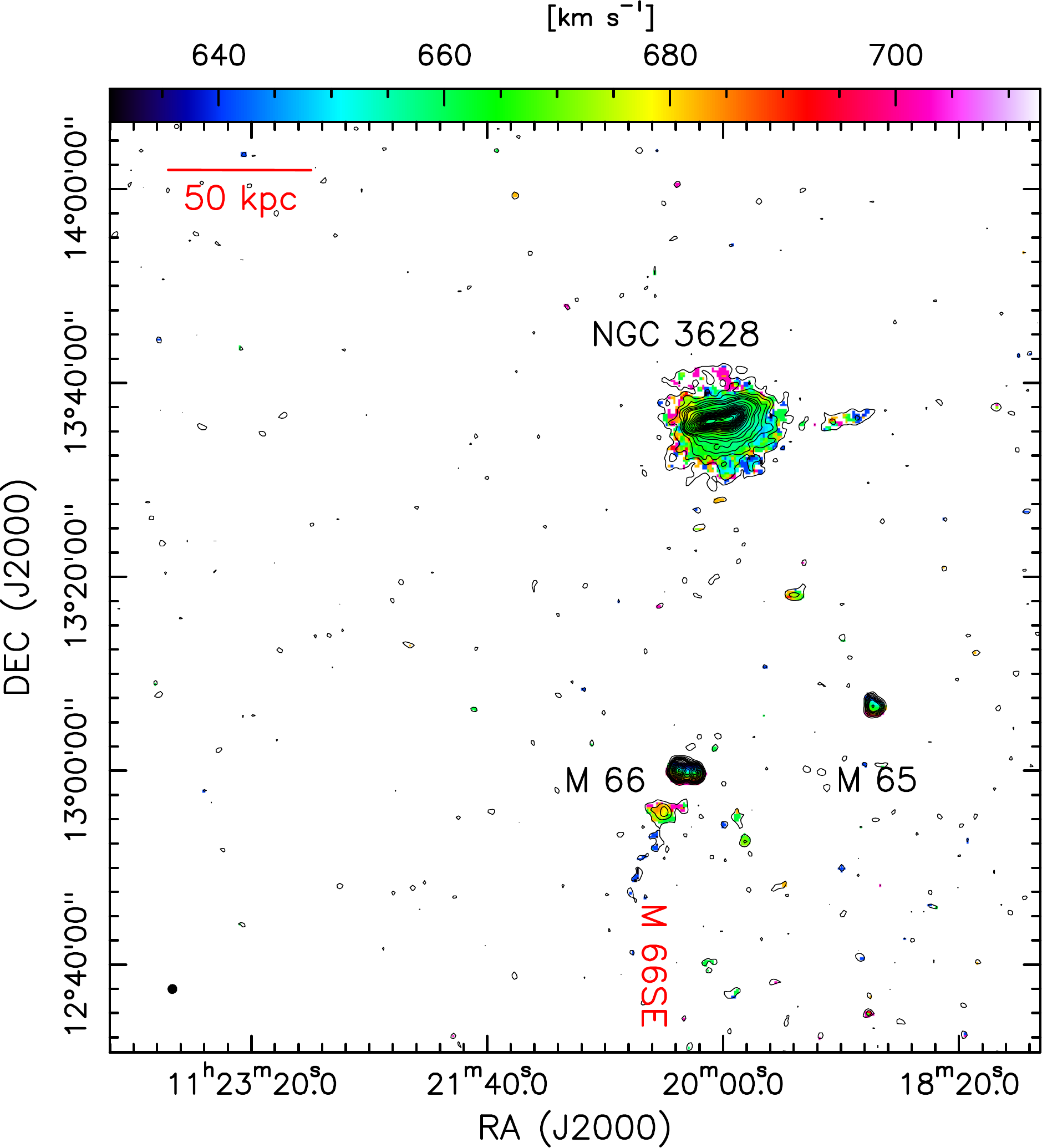}
    \includegraphics[width=0.45\hsize]{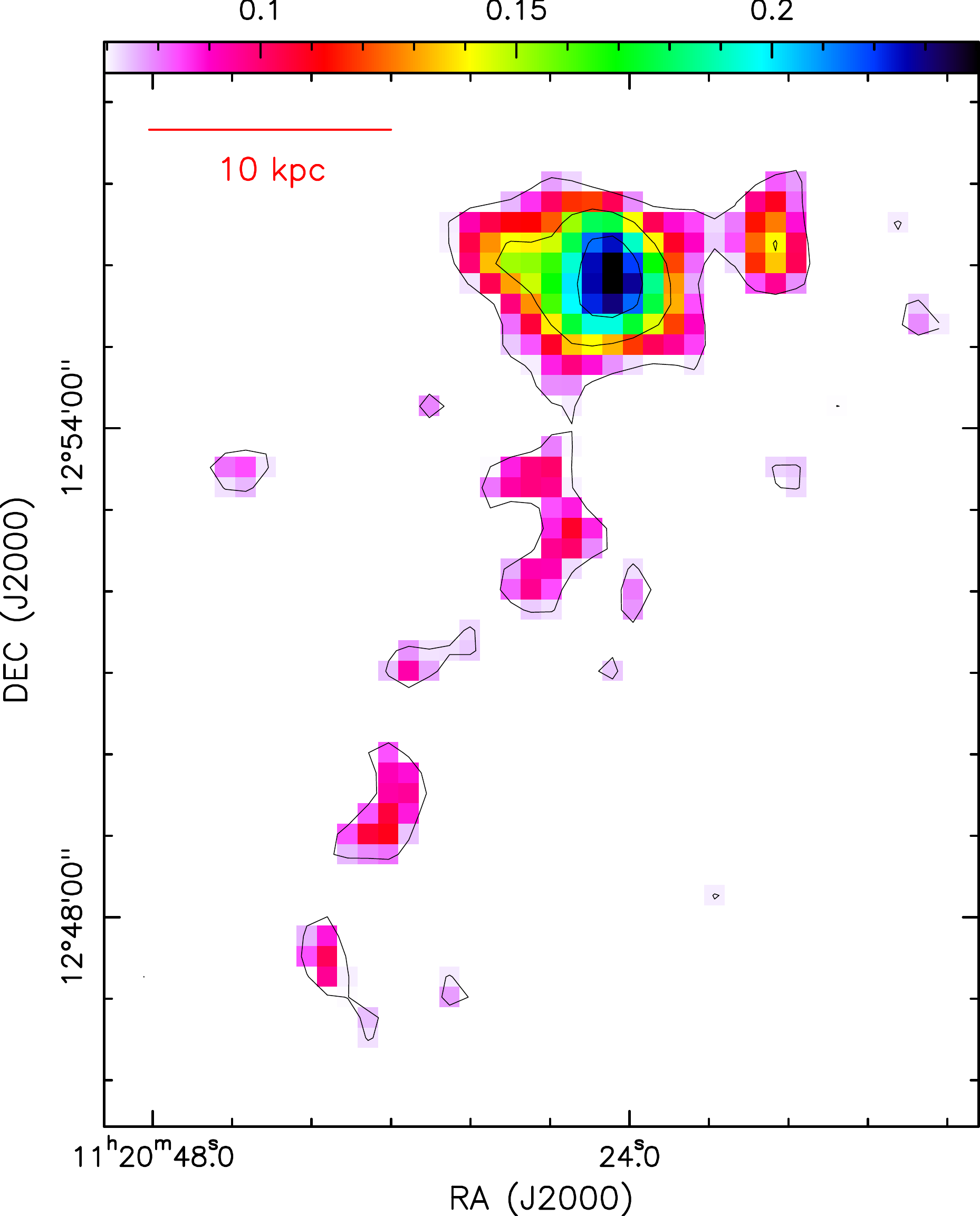}
    \includegraphics[width=0.45\hsize]{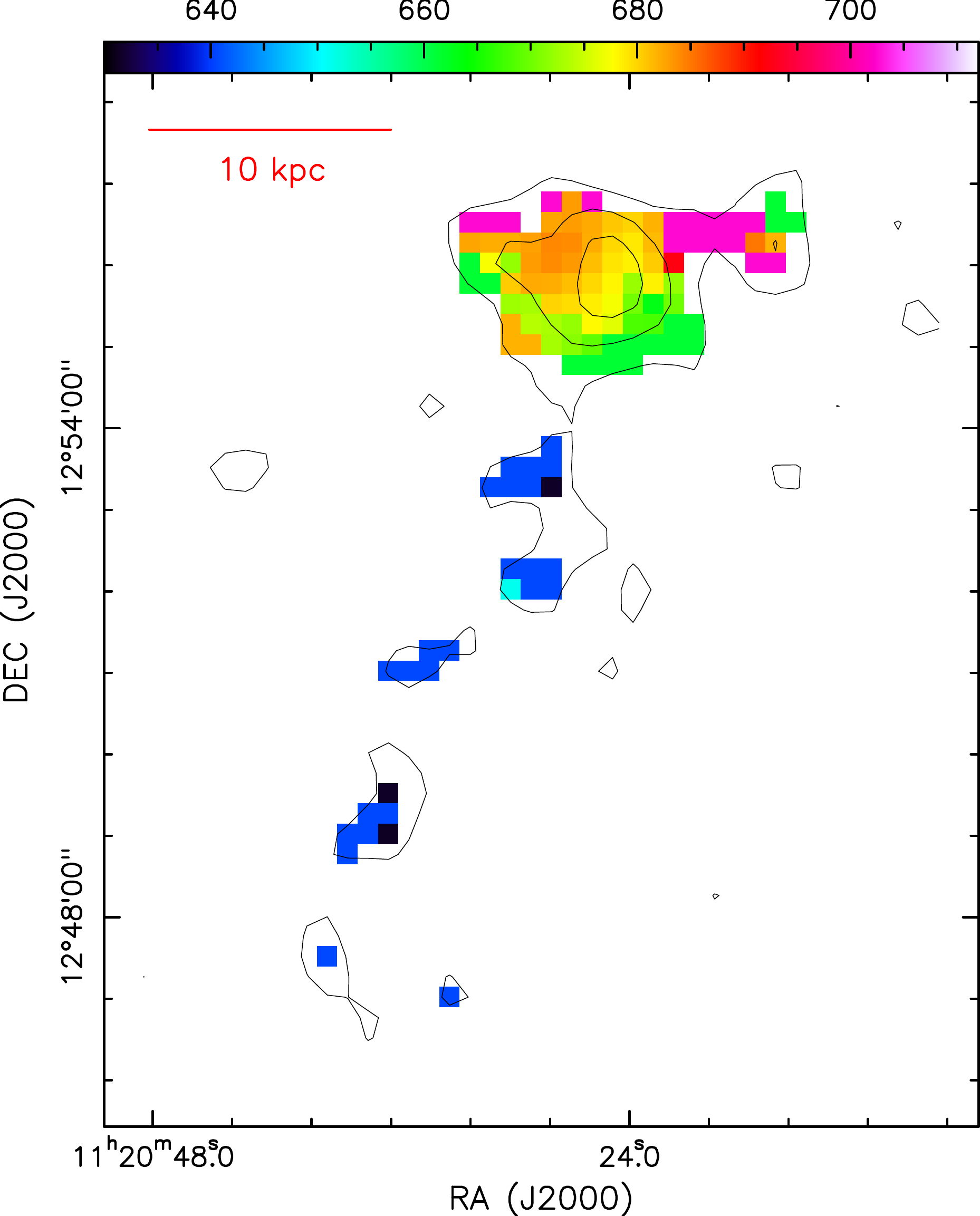}
    \caption{Same as Fig. \ref{fig:ngc3628e} but for M\,66SE. The velocity range is from 630.3 to 712.8\,$\kms$. The contour levels are set to (3, 6, 9, 12, 15, 18, 21, 30, 40, 50, 60, ..., 170)\,$\times\,0.023$\,$\Jypbkms$.}
    \label{fig:m66se}
    \end{figure*}

\begin{figure*}
    \centering
    \includegraphics[width=0.45\hsize]{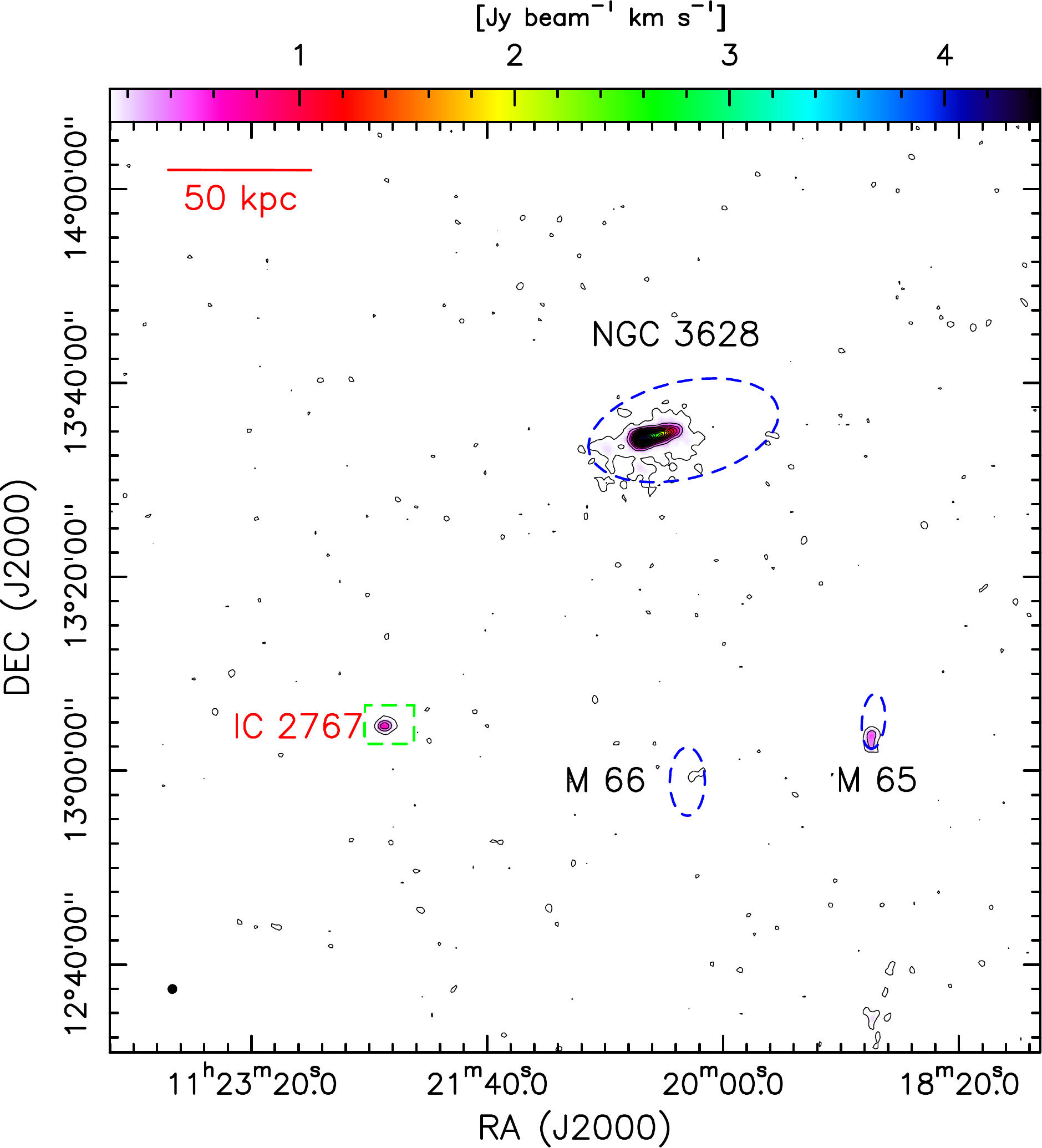}
    \includegraphics[width=0.45\hsize]{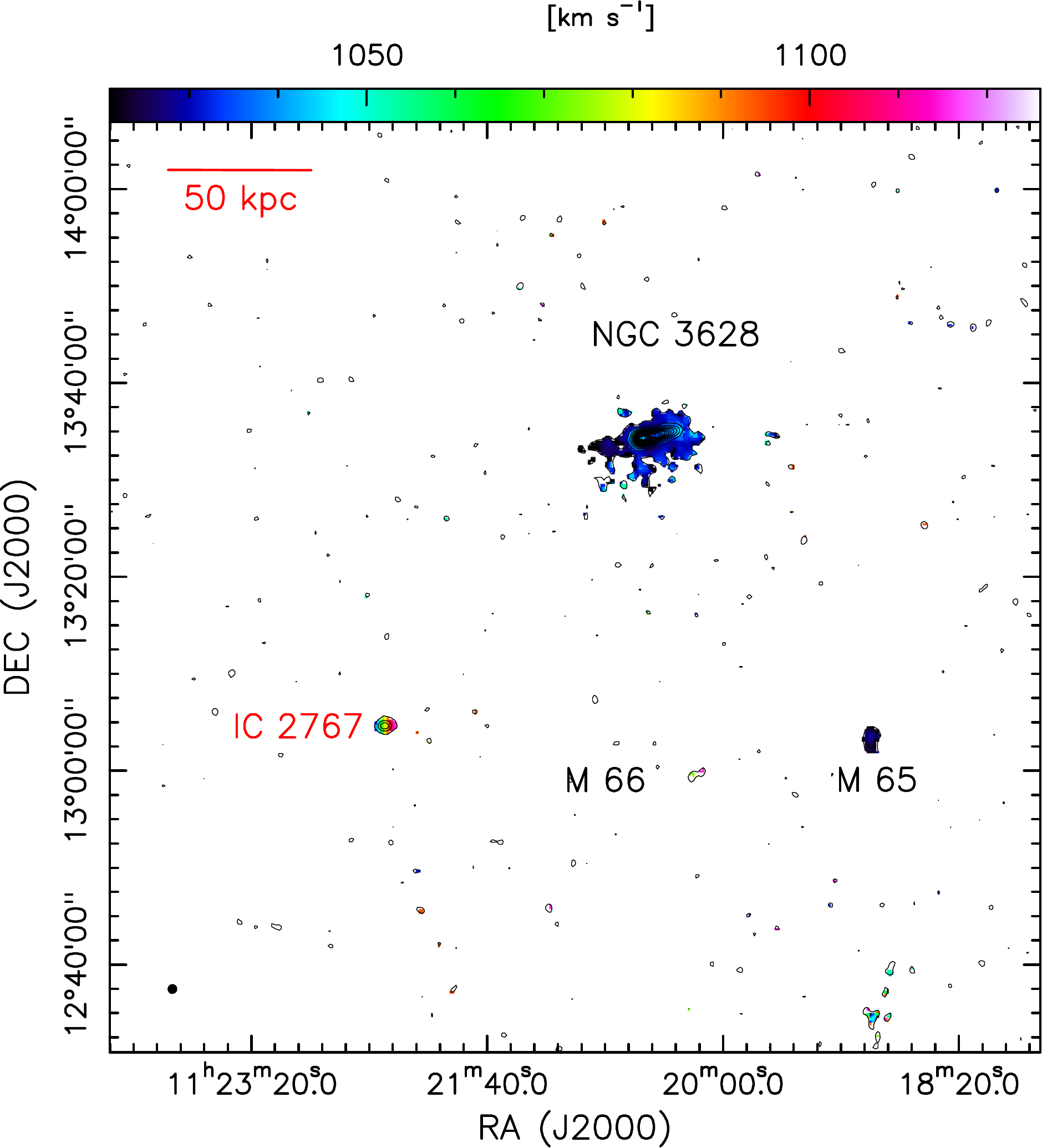}
    \includegraphics[width=0.45\hsize]{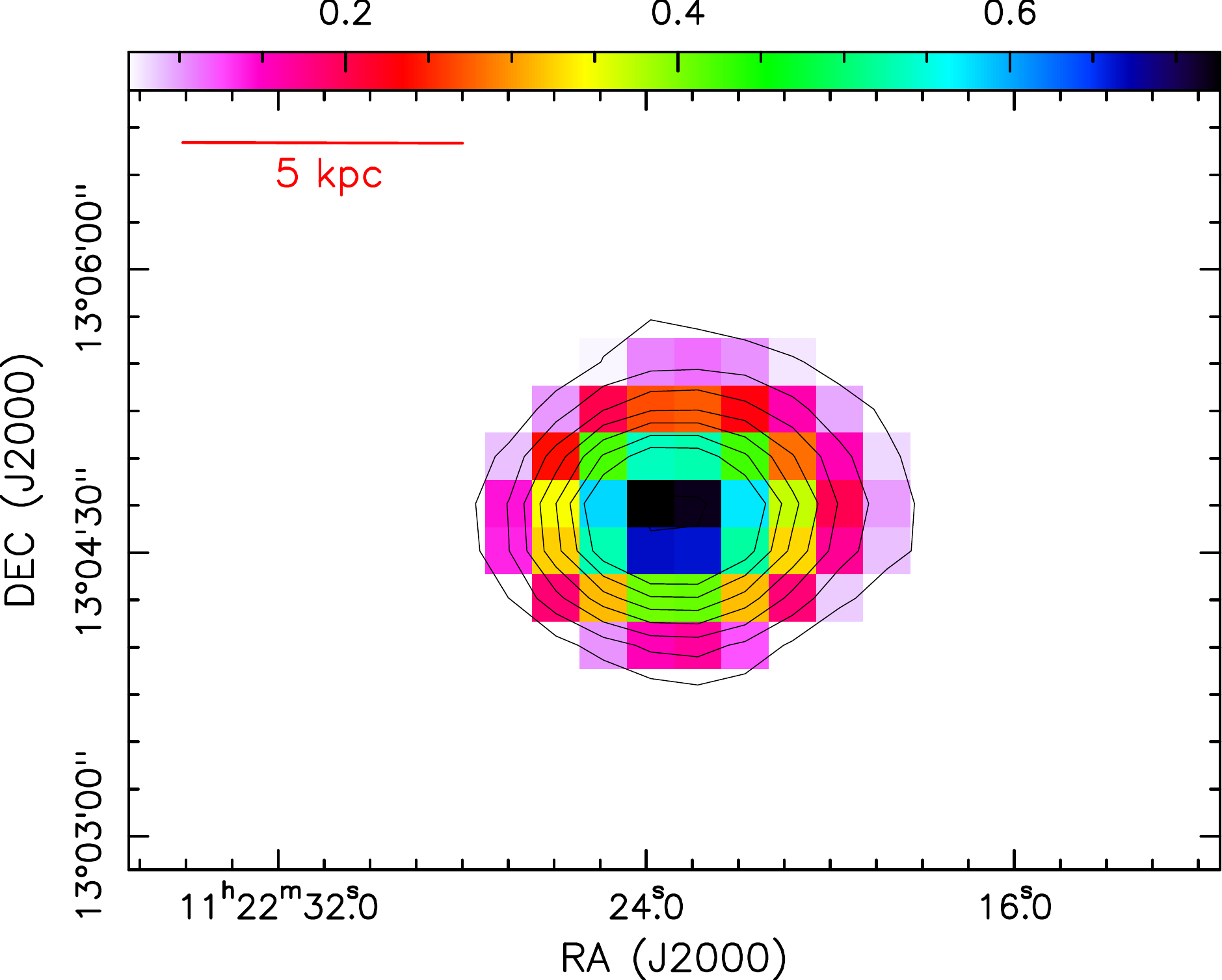}
    \includegraphics[width=0.45\hsize]{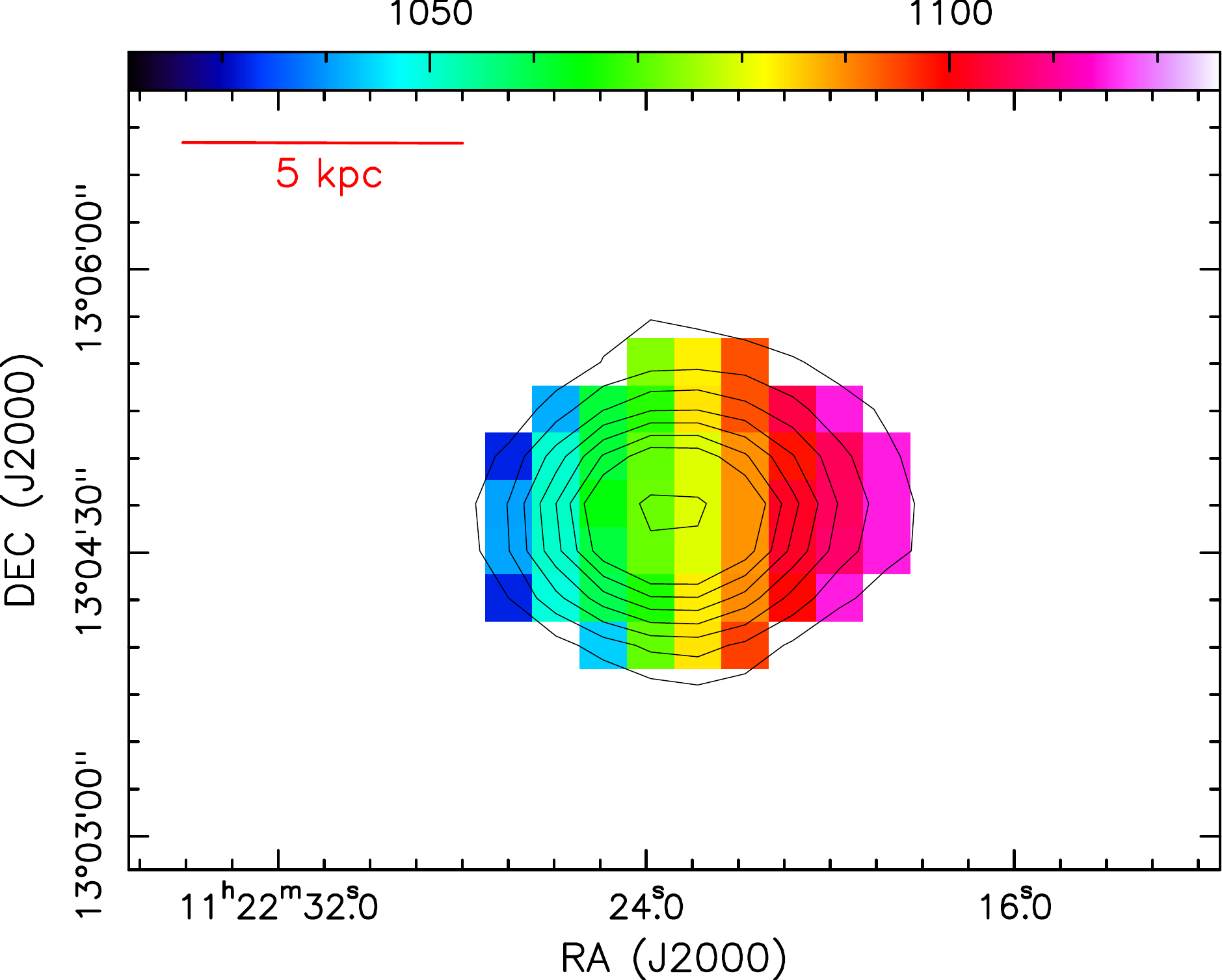}
    \caption{Same as Fig. \ref{fig:ngc3628e} but for IC\,2767. The  velocity range is from 1022.0 to 1125.1\,$\kms$. The contour levels are set to (3, 13, 23, 33, 43, 53, 63, ..., 193)\,$\times\,0.023$ $\Jypbkms$.}
    \label{fig:m66e}
    \end{figure*}

\begin{figure*}
    \centering
    \includegraphics[width=0.45\hsize]{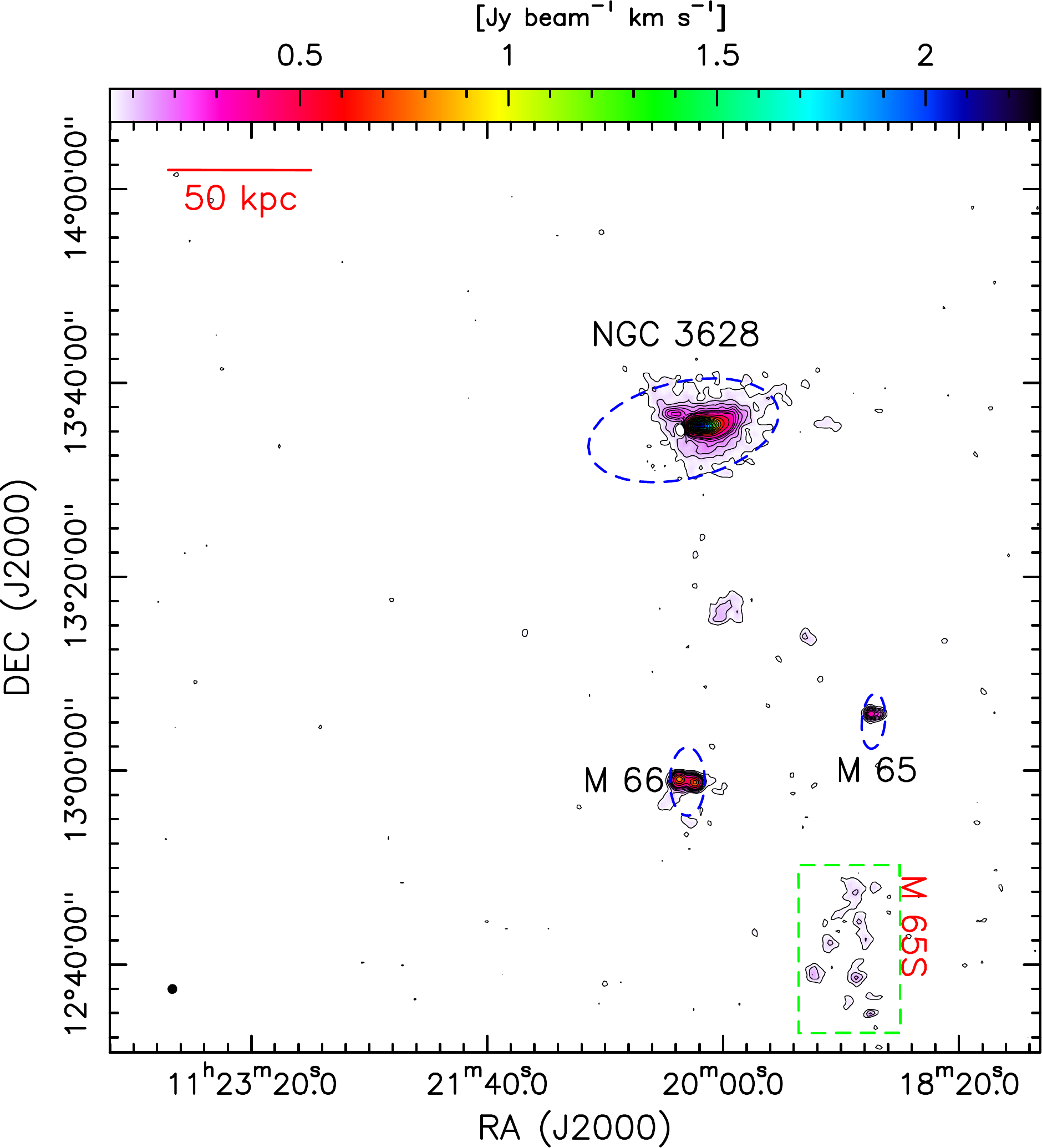}
    \includegraphics[width=0.45\hsize]{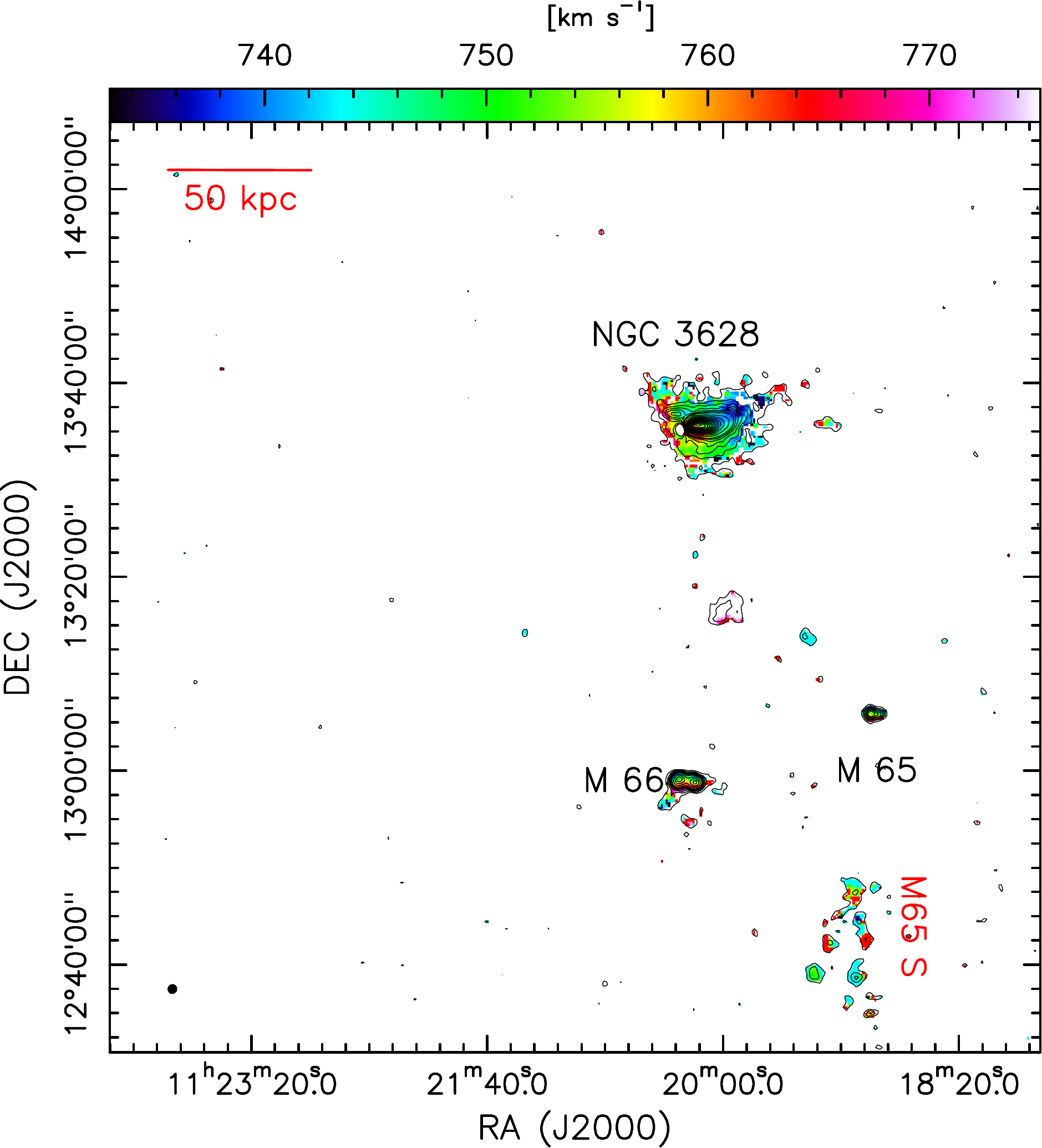}
    \includegraphics[width=0.45\hsize]{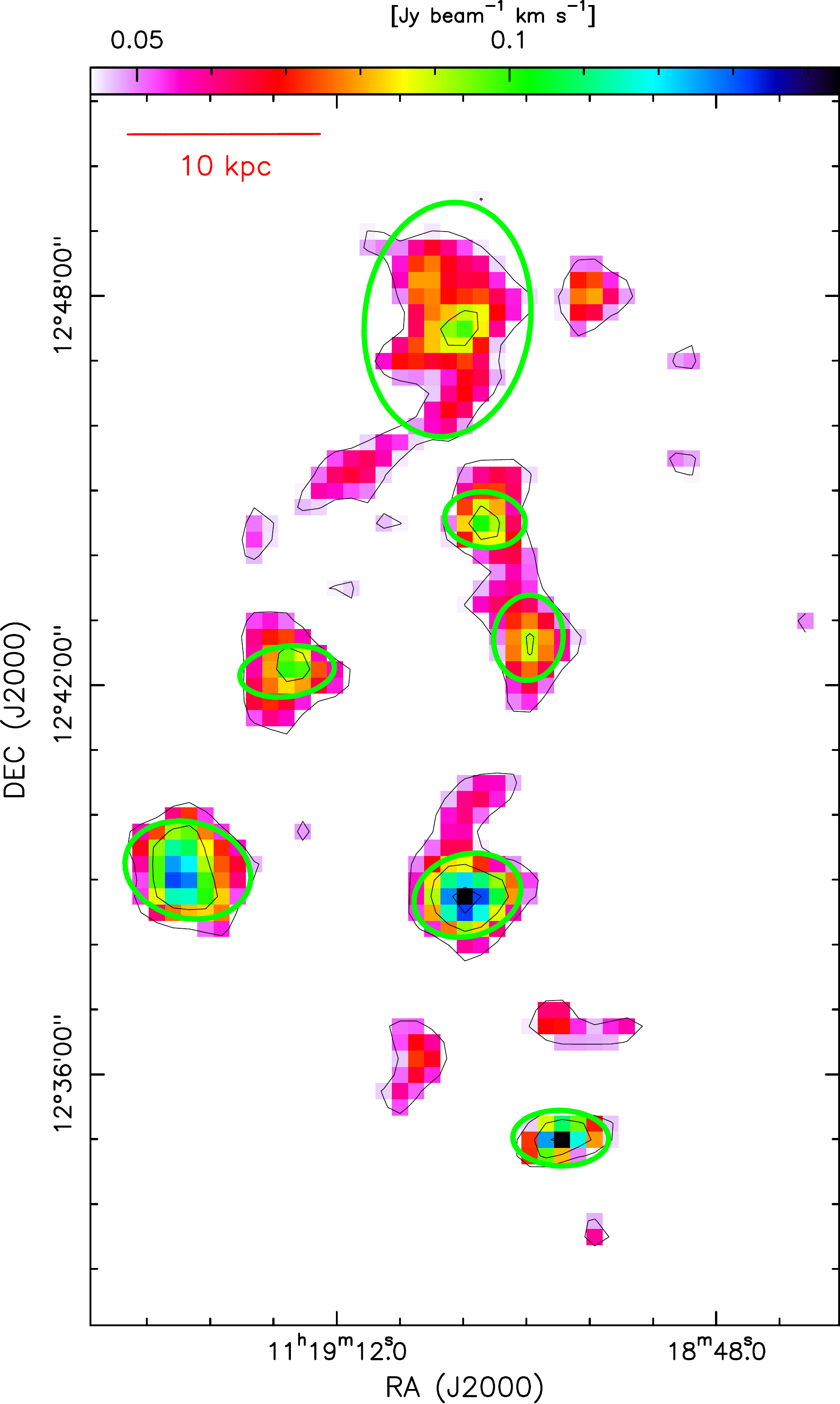}
    \includegraphics[width=0.45\hsize]{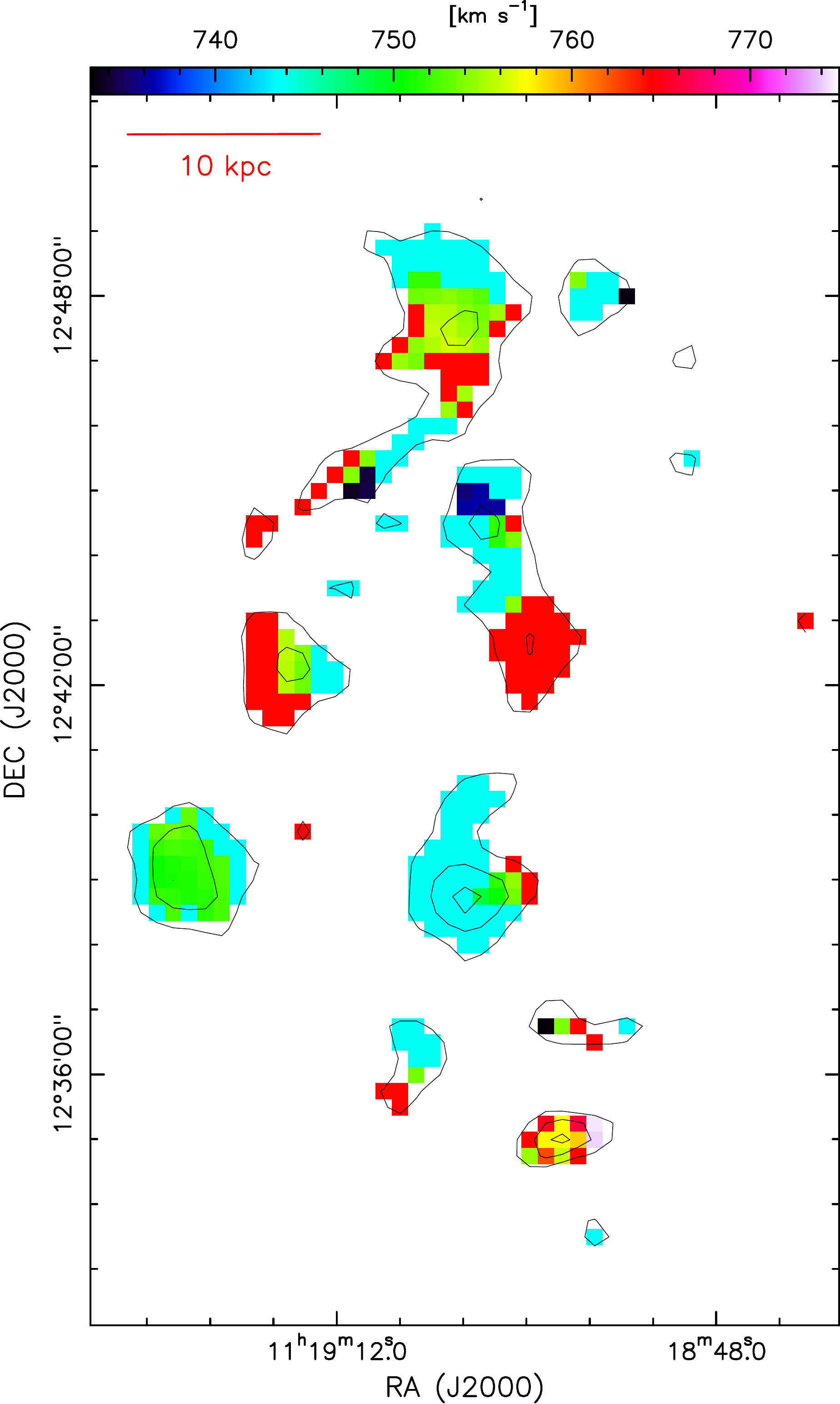}
    \caption{Same as Fig. \ref{fig:ngc3628e} but for M\,65S. The velocity range is from 733.4 to 774.6\,$\kms$. The contour levels are set to (3, 6, 9, 12, 15, 18, 21, 30, 40, 50, ..., 170)\,$\times\,0.015$\,$\Jypbkms$.}
    \label{fig:m65s}
    \end{figure*}

Therefore, the $\HI$ zeroth moment (integrated intensity) and first moment (velocity) maps are presented in Fig. \ref{fig:mom0_2-31} to show the overall distributions of the $\HI$ emission and velocity distributions in the system.
In the left panel, contours and the color image show the zeroth moment map. The integration range is from $V_{\rm HEL}$ = 506.7 to 1125.1\,$\kms$ to cover all the $\HI$ features. The contour levels are set to $(5, 15, 25, 35, 45,...145) \times \sigma$ $\Jypbkms$. The uncertainty of the integrated image $\sigma$ is set to  $\sigma_{\rm ch} \times \sqrt{N_{\rm ch}} = 0.055$\,$\Jypbkms$, where $\sigma_{\rm ch}$  is the typical noise in a single channel (see Section \ref{ob}), and $N_{\rm ch}$ is the number of channels in the integrated velocity range ($N_{\rm ch} = 28$ for Fig. \ref{fig:mom0_2-31}). We also use three blue dashed ellipses in the left panel to cover the main $\HI$ emission regions of the three galaxies. More importantly, the same ellipses are also added to Figs. \ref{fig:chMap} -- \ref{fig:m65s} to indicate the relative locations of the features.
In the right panel of Fig. \ref{fig:mom0_2-31}, the color image shows the first moment map. The contours overlaid show the zeroth moment map in the same way as the left panel.

We can see in  Fig. \ref{fig:mom0_2-31}, NGC\,3628, M\,65, and M\,66 are all detected in the zeroth moment map (left panel), and all show clear velocity gradients in the first moment map (right panel).
However, the weaker features revealed by the right panel of Fig. \ref{fig:op-tpeak} only appear in a relatively narrow velocity range. Due to the large $\sigma$, proportional to $\sqrt{N_{\rm ch}}$, only the  plume can be seen in the vicinity of NGC\,3628 in Fig. \ref{fig:mom0_2-31}.

To further investigate the features along the velocity axis, we present the channel maps in Fig. \ref{fig:chMap}, each one covering 20.7\,$\kms$. The contours in all the panels start at 3$\sigma_{\rm ch}$ and go up in steps of 9$\sigma_{\rm ch}$, where $\sigma_{\rm ch}$ is the typical noise of the integrated intensity in an individual channel (see Section \ref{ob}).
The additional structures with narrow line widths, M\,66SE (640.7--702.5\,$\kms$), NGC\,3628S (764.3--867.4\,$\kms$), NGC\,3628SW (826.2--888.0\,$\kms$), M\,65S (743.7--764.3\,$\kms$), the plume (805.6--929.2\,$\kms$), and IC2767 (1032.3--1114.7\,$\kms$)
are also labeled in Fig. \ref{fig:chMap}\footnote{There is another $\HI$ structure appearing in the west of NGC\,3628  which is very likely caused by a sidelobe (see Appendix \ref{ngc3628w}).}.
The remaining structures are consistent with previous Arecibo $\HI$ observations \citep[$\eg$][]{2009AJ....138..338S} but the angular resolution of our observations is about four times better.

To emphasize these structures and their velocity distribution, we present, in Figs. \ref{fig:ngc3628e}  to \ref{fig:m65s}, their zeroth moment and first moment maps within their specific velocity ranges mentioned above.
In these figures, the top two panels present the full view of the distributions of $\HI$ emission and velocity, while the two lower panels show zoomed images of these structures.
Note that these integrated ranges do not fully cover the $\HI$ emission in the three galaxies, which means only a part of HI emission in each of the galaxies is shown in Figs. \ref{fig:ngc3628e}  to \ref{fig:m66e}.

\textbf{The plume:} Fig. \ref{fig:ngc3628e} presents the zeroth moment (left two panels) and the first moment (right two panels) maps of the plume. In all the panels, the integrated velocity range is from $V_{\rm HEL}$ = 795.3 to 939.5\,$\kms$ to highlight the emission from the plume.
The uncertainty of the velocity-integrated intensity $\sigma = \sigma_{\rm ch} \times \sqrt{N_{\rm ch}}= 0.027\,$ $\Jypbkms$ , where $\sigma_{\rm ch} \approx 0.01$ $\Jypbkms$ (see Sect. \ref{ob}) and $N_{\rm ch} = 7$.

From Fig. \ref{fig:ngc3628e} we can see that there are plenty of condensations along the plume and the strongest one is located at the eastern tip, which was proposed to be a tidal dwarf galaxy by \citet{2014ApJ...786..144N}.
The $\HI$ spectra in the plume are extremely narrow and there is little velocity variation along the plume.
Our Arecibo data have a better velocity resolution of about 1.9\,$\kms$ after smoothing three contiguous velocity channels (see Appendix \ref{arecibo}). As can be seen in the right two panels of Fig. \ref{fig:arecibo}, the narrowest spectra have FWHM line widths of about 20\,$\kms$, evident in the middle of the plume.
These characteristics have also been presented by previous $\HI$ observations \citep[$\eg$][]{1979ApJ...229...83H}.
However, our observations spatially resolve the four $\HI$ clumps revealed by  \citet{2009AJ....138..338S} into about 11 condensations (see Table \ref{tab:clumps} and Fig. \ref{fig:ngc3628e}). The condensations are identified by the astrodendro package\footnote{http://www.dendrograms.org/} \citep{2008ApJ...679.1338R} with an intensity threshold ($min\_value$) of 5$\sigma$ and a size threshold ($min\_npix$) of 10 pixels. The minimum intensity difference ($min\_delta$) to be considered an independent entity (leaf or branch in the package) is set to 1$\sigma$ as suggested by the package.
The 11 condensations along the plume are reminiscent of the chains of tidal dwarfs that were seen in the Dentist's Chair galaxy \citep{2002ApJ...579L..79W}, the Tadpole galaxy \citep{2003ApJ...585..750T}, and in simulations of tidal tails \citep{2007MNRAS.375..805W, 2019RMxAA..55..273N}.

Here we want to emphasize features that have not been noticed before. From the zeroth moment map of Fig. \ref{fig:ngc3628e} we can see that the $\HI$ emission at the eastern tip and middle of the plume shows two branches separated by centrally located weaker $\HI$ emission.
Moreover, from the first moment map we can see that there are clearly two different velocities in the two branches, which are also evident in the Arecibo data in Fig. \ref{fig:arecibo}. From a combined view of the zeroth and first moment maps, we find that the northern branch belongs to a not completely continuous elongated structure with velocities of about 850\,$\kms$, while the stronger southern branch shows velocities closer to 900\,$\kms$. The northern side of the plume exhibits more diffuse gas than the southern one. There is no smooth transition between the two velocity components. Instead, the velocity field looks more like an overlap of two spatially and kinematically distinct gas filaments, which are likely caused by two well-separated arms in the plume drawn out from NGC 3628. This is further discussed in a more comprehensive manner in Section 4.

\begin{table*}
\caption{Condensations in the plume.}
\label{tab:clumps}
\centering
\begin{tabular}{ccccccc}     
\hline\hline
Condensations & R.A. (J2000) & DEC (J2000) & D$_{\rm max} ($\arcsec$)$ & D$_{\rm min} ($\arcsec$)$   & PA  & M$_{\rm HI}$ ( $\times 10^{7}$ M$_{\odot})$  \\
 (1) & (2)  & (3) & (4)   &  (5)  & (6) & (7)  \\
\hline
1   &  11:20:33.5 & 13:36:40  &   66.42$\pm$2.84   &   32.15$\pm$1.37  &     84.55  &       1.23$\pm$0.05 \\
2   &  11:20:47.3 & 13:36:19  &  157.43$\pm$4.55   &   37.91$\pm$1.10  &   $-$164.09  &     3.04$\pm$0.09 \\
3   &  11:21:10.3 & 13:36:15  &   63.61$\pm$1.97   &   25.70$\pm$0.80  &    173.49  &       1.49$\pm$0.05 \\
4   &  11:21:18.2 & 13:35:57  &   59.00$\pm$2.12   &   23.33$\pm$0.84  &   $-$172.58 &      1.17$\pm$0.04 \\
5   &  11:21:27.9 & 13:35:35  &   75.74$\pm$2.28   &   25.74$\pm$0.78  &    178.70  &       1.67$\pm$0.05 \\
6   &  11:21:36.8 & 13:40:20  &   51.05$\pm$4.89   &   30.06$\pm$2.88  &    168.23  &       0.46$\pm$0.04 \\
7   &  11:21:49.4 & 13:36:32  &  152.82$\pm$3.21   &   44.35$\pm$0.93  &    165.84  &       4.46$\pm$0.09 \\
8   &  11:22:21.0 & 13:38:19  &  255.60$\pm$34.15  &  108.76$\pm$14.53 &    173.49  &       14.21$\pm$0.19 \\
9   &  11:22:48.9 & 13:41:27  &  103.93$\pm$4.99   &   38.95$\pm$1.87  &    154.58  &       1.51$\pm$0.07 \\
10  &  11:22:55.7 & 13:40:12  &   57.92$\pm$5.22   &   36.76$\pm$3.31  &     48.69  &       0.58$\pm$0.05 \\
11  &  11:23:11.2 & 13:42:25  &  178.56$\pm$1.14   &  131.04$\pm$0.84  &    152.75  &       27.20$\pm$0.17 \\
\hline
\end{tabular}
\tablefoot{M$_{\rm HI}$ = 2.36 $\times$10$^{5}$ D$^{2}_{\rm Mpc}$$\int$S$_{\nu}$d$\nu$, where D$_{\rm Mpc}$ is 11.3 and S$_{\nu}$d$\nu$ is in $\Jykms$.
The position error in columns (2) and (3) can be  roughly estimated as $0.5\times\theta_{\rm beam}/{\rm SNR}\approx6\arcsec$ with signal-to-noise ratios $\rm SNR\ga$ 5 and beamsize $\theta_{\rm beam}\approx60\arcsec$ \citep{2014ARA&A..52..339R}.}
\end{table*}

\textbf{NGC\,3628S:} Fig. \ref{fig:ngc3628s} presents the zeroth moment (left two panels) and the first moment (right two panels) maps of NGC\,3628S. In all the panels the velocity ranges from V$_{\rm HEL}$ = 754.0 to 878.7\,$\kms$.
The uncertainty of the velocity-integrated intensity $\sigma = \sigma_{\rm ch} \times \sqrt{N_{\rm ch}}$ = 0.025\,$\Jypbkms$, where $\sigma_{\rm ch} \approx 0.01$ $\Jypbkms$ (see Sect. \ref{ob}) and $N_{\rm ch} = 6 $.

From the zeroth moment map we can see that NGC\,3628S extrudes from NGC\,3628 with a uniform intensity of around 3$\sigma$ towards the south. The strongest emission of about 12$\sigma$ is found at the southern tip.
Several pixels with intensities of about 3$\sigma$  further to the east ($\approx$ 5$\arcmin$) are beyond the scope of this work because of their limited size and intensities.
The first moment map shows an overall velocity gradient in a roughly north-south direction along NGC\,3628S, but with large dispersions.
According to the categories in \citet{1972ApJ...178..623T} and assuming that there was a tidal interaction between NGC\,3628 and M\,66, this part should be the `bridge'.
However, this bridge seems to be too prominent with respect to simulations of encounters of galaxies with similar total masses \citep[$\eg$][]{1978AJ.....83..219R}.
According to the model provided by \citet{1972ApJ...178..623T}, the material in the bridge might eventually fall back to the original galaxy, $\ie$ NGC\,3628 in this case.  Again, assuming that this is a bridge caused by tidal interaction, the velocity gradient along NGC\,3628S might be caused by the gas falling back.

\textbf{NGC\,3628SW:} Fig. \ref{fig:ngc3628sw} presents the zeroth moment (left two panels) and the first moment (right two panels) maps of NGC\,3628SW. In all the panels the velocity ranges from V$_{\rm HEL}$ = 815.9 to 898.3\,$\kms$.
The uncertainty of the velocity-integrated intensity $\sigma = \sigma_{\rm ch} \times \sqrt{N_{\rm ch}}$ = 0.021\,$\Jypbkms$, where $\sigma_{\rm ch} \approx 0.01$ $\Jypbkms$ (see Sect. \ref{ob}) and $N_{\rm ch} = 4 $.

From the zeroth moment map of Fig. \ref{fig:ngc3628sw}, we can see that NGC\,3628SW consists of several pieces of small $\HI$ clouds and a strongest condensation reaching a peak at about the 12$\sigma$ level.
These cloudlets seem to form an arc-like structure first oriented west- and then southwards and seem to be connected to NGC\,3628. The strong compact condensation at the end of the arc-like structure
may show a northeast-southwest velocity gradient with blue velocities in the northeast.

\textbf{M66\,SE:} Fig. \ref{fig:m66se} presents the zeroth moment (left two panels) and the first moment (right two panels) maps of M\,66SE. The velocity range is from V$_{\rm HEL}$ = 630.3 to 712.8\,$\kms$ in all the panels.
The uncertainty of the velocity-integrated intensity $\sigma = \sigma_{\rm ch} \times \sqrt{N_{\rm ch}}$ = 0.023\,$\Jypbkms$, where $\sigma_{\rm ch} \approx 0.01$ $\Jypbkms$ (see Sect. \ref{ob}) and $N_{\rm ch} = 5 $.

From the zeroth moment map of Fig. \ref{fig:m66se} we can see that the main part of M\,66SE consists of a strong round  clump and a faint tail extending to the south. On its western side, there are also a few separate pixels with emission of around 3$\sigma$, which are beyond the scope of this paper because of their limited sizes and intensities. From the first moment map we can see that the dense clump presents a south-north velocity gradient with highest velocities in the north. Meanwhile this velocity gradient is reversed compared to the velocity gradient of M\,66 (Fig. \ref{fig:mom0_2-31}). The faint tail in the south of the clump shows velocities of about 640\,$\kms$.
We will further discuss this structure in Section \ref{disscussion}.

\textbf{IC\,2767:} Fig. \ref{fig:m66e} presents the zeroth moment (left two panels) and the first moment (right two panels) maps of IC\,2767. The velocity range is from V$_{\rm HEL}$ = 1022.0 to 1125.1\,$\kms$ in all the panels.
The uncertainty of the velocity-integrated intensity $\sigma = \sigma_{\rm ch} \times \sqrt{N_{\rm ch}}$ = 0.023\,$\Jypbkms$, where $\sigma_{\rm ch} \approx 0.01$ $\Jypbkms$ (see Sect. \ref{ob}) and $N_{\rm ch} = 5 $.

This $\HI$ structure is a compact source with an $\approx$34\,$\kms$ arcmin$^{-1}$ east-west velocity gradient with blue velocities in the east. According to its location and velocity gradient, it is probably associated with the galaxy IC\,2767 as is also suggested by \citet{2009AJ....138..338S}. The systemic velocity  of IC\,2767 is 1080\,$\kms$ \citep{2011AJ....142..170H, 2012A&A...545A..16G}
which is consistent with that of the $\HI$ observations.
The distance of IC 2767 is much larger than that of the Leo Triplet. \citet[][]{2012A&A...545A..16G} and \citet[][]{2011AJ....142..170H} suggested 19.6 and 24 Mpc, respectively, which means IC 2767 is not likely a member of the Leo Triplet.

\textbf{M\,65S:} Fig. \ref{fig:m65s} presents the zeroth moment (left two panels) and the first moment (right two panels) maps of M\,65S. The velocity range is from V$_{\rm HEL}$ = 733.4 to 774.6\,$\kms$ in all the panels.
The uncertainty of the velocity-integrated intensity $\sigma = \sigma_{\rm ch} \times \sqrt{N_{\rm ch}}$ = 0.015\,$\Jypbkms$, where $\sigma_{\rm ch} \approx 0.01$ $\Jypbkms$ (see Sect. \ref{ob}) and $N_{\rm ch} = 2 $.

\citet{2009AJ....138..338S} also detected a diffuse cloud in this region with Arecibo observations. Our observations resolve it into about 7 main clumps (see Table \ref{tab:M65Sclumps} and Fig. \ref{fig:m65s}) with intensities of 3 -- 9$\sigma$, forming an upside-down `Y' morphology.
The condensations are identified by the astrodendro package with $min\_value$ = 5$\sigma$, $min\_npix$ = 5, and $min\_delta$ = 1$\sigma$.
From the first moment map we can see that M\,66S shows an irregular velocity field.
It is helpful to first figure out which galaxy (M\,65 or M\,66) this structure is associated with.
Firstly, M\,65S is nearly equidistant with respect to the two galaxies, but slightly closer to M\,65.
Secondly, the velocities of M\,65S range from 743.7 to 764.3\,$\kms$ and are well separated from those of M\,66SE (640.7-702.5\,$\kms$). Thus M\,65S is not likely a part of M\,66SE. Meanwhile M\,65S is also well separated from the orbit of the tidal interaction model with NGC\,3628 and M\,66 \citep[$\eg$][]{1978AJ.....83..219R}.
Thirdly,  M\,65 is located along the extension of the elongated M\,65S structure.
Based on these admittedly not very strong arguments, it seems plausible that M\,65S is associated with M\,65.

Following this idea,  M\,65 might have been involved in the interaction of this triplet system, either having participated in the direct interaction with (one of) the other two galaxies or representing a captured part of the gas from the relic of the interaction between NGC\,3628 and M\,66.

\begin{table*}
\caption{Condensations in M\,65S.}
\label{tab:M65Sclumps}
\centering
\begin{tabular}{ccccccc}     
\hline\hline
Condensations & R.A. (J2000) & DEC (J2000) & D$_{\rm max} ($\arcsec$)$ & D$_{\rm min} ($\arcsec$)$   & PA  & M$_{\rm HI}$ ( $\times 10^{6}$ M$_{\odot})$ \\
 (1) & (2)  & (3) & (4)   &  (5)  & (6) & (7)  \\
\hline
1 &  11:18:57.9 & 12:35:01  &   44.42$\pm$4.51   &   22.61$\pm$2.29  &   -161.96  &       1.90$\pm$0.19  \\
2 &  11:18:59.9 & 12:42:44  &   39.53$\pm$5.55   &   26.39$\pm$3.70  &    104.46  &       1.40$\pm$0.20  \\
3 &  11:19:02.6 & 12:44:33  &   37.44$\pm$4.84   &   31.25$\pm$4.04  &    100.00  &       1.61$\pm$0.21  \\
4 &  11:19:03.8 & 12:38:46  &   50.29$\pm$3.59   &   42.05$\pm$3.00  &    175.63  &       3.92$\pm$0.28  \\
5 &  11:19:04.9 & 12:47:38  &   76.79$\pm$6.89   &   49.28$\pm$4.42  &    100.45  &       4.17$\pm$0.37  \\
6 &  11:19:15.1 & 12:42:13  &   44.10$\pm$5.22   &   37.44$\pm$4.43  &     74.41  &       2.09$\pm$0.25  \\
7 &  11:19:21.4 & 12:39:10  &   59.04$\pm$4.01   &   47.59$\pm$3.23  &    114.19  &       4.75$\pm$0.32  \\

\hline
\end{tabular}
\tablefoot{M$_{\rm HI}$ = 2.36 $\times$10$^{5}$ D$^{2}_{\rm Mpc}$$\int$S$_{\nu}$d$\nu$, where D$_{\rm Mpc}$ is 11.3 and S$_{\nu}$d$\nu$ is in $\Jykms$.
The position error in columns (2) and (3) can be roughly estimated as $0.5\times\theta_{\rm beam}/\rm SNR\approx6\arcsec$ with signal-to-noise ratios $\rm SNR\ga$ 5 and  beamsize $\theta_{\rm beam}\approx60\arcsec$   \citep{2014ARA&A..52..339R}.}
\end{table*}

\section{Discussion}
\label{disscussion}

In Section \ref{results}, we have presented specific morphologies and kinematics of the appendages in the Leo Triplet. Compared with  previous $\HI$ observations conducted with Arecibo \citep{1979ApJ...229...83H, 2009AJ....138..338S}, our higher resolution observations show unprecedented details of these appendages.
Our observations resolve the previously known four $\HI$ clumps into 11 condensations along the plume. Furthermore, we also demonstrate for the first time that the plume shows not only a kinematical but also a spatial two-branch morphology.
We also reveal a more detailed morphology of the southern bridge of NGC\,3628 (NGC\,3628S) and an overall velocity gradient along NGC\,3628S. This is different from the finding of \citet{1979ApJ...229...83H} who suggested irregular velocities on the basis of lower angular resolution data.
The gas at the southwestern side of NGC\,3628 (NGC\,3628SW) is  seen for the first time in an arc-like structure which seems to be connected with NGC\,3628.
We clearly show a separated clump with a reversed velocity gradient relative to the velocity gradient of M\,66 and a newly detected tail connected with the clump and extending southward (M\,66SE).
We resolve the diffuse gas detected by \citet{2009AJ....138..338S} in the south of M\,65 into 7 clumps with intensities of 3 -- 9$\sigma$ forming an upside-down `Y' morphology (M\,65S), which is likely associated with M\,65.

Below, we shall discuss several interesting features based on the results in Section \ref{results}.

\subsection{Two arms in the plume}
\label{twoArm}
In the pioneering work by \citet{1972ApJ...178..623T}, it has been proven that the far side of the victim galaxy (a point mass surrounded by a disk of test particles---NGC\,3628 in this case) relative to the pericenter of the interaction would form a long and slender `counterarm' (see $\eg$, their Fig. 3 for the motions of the outer disk). The formation of the slender structure is caused by lagging test particles (the backward arm  relative to the victim galaxy), which are accelerated more by the companion galaxy (point mass without disk---M\,66 in this case), overtaking the forward arm forming a relatively narrow counterarm.
However, as has been argued by \citet{1972ApJ...178..623T}, the observed structure of this narrow feature relies on the viewing angle. The counterarm will show a two-arm structure at some angles.

As we have mentioned in Section \ref{results}, a two-branch morphology is observed at the eastern tip and middle of the plume. The northern, more diffuse gas shows a coherent velocity field, and the velocity field of the plume looks more like two overlapping velocity regimes since there is no smooth transition between the two velocities.
Furthermore, we can also find evidence hinting at the existence of two arms in the plume in the 867.4 and 888.0\,$\kms$ maps of Fig. \ref{fig:chMap}. In the 867.4\,$\kms$ map, there is a clear jump in the middle of the plume which shows an elongated morphology. Meanwhile, in the 888.0\,$\kms$ map, we can also see a two branch structure in the middle of the plume.

Compared with the zeroth moment image, the right panel of Fig. \ref{fig:op-tpeak} (the peak intensity image) highlights the $\HI$ emission in the appendages, especially the $\HI$ emission in the plume due to their simple (usually Gaussian) and spectrally narrow profiles. From the right panel of Fig. \ref{fig:op-tpeak}, we can clearly see a two-arm feature in the plume which is in good  agreement with simulations \citep[$\eg$ ][]{1972ApJ...178..623T, 2016MNRAS.463.3637R}.
Note that probably because the western part of the upper branch (especially the part near NGC\,3628) is too weak to show a continuous morphology in our observations there exist some separated regions of enhanced emission along its extension direction.
Just from this projected image, it is hard to deduce the 3D distribution of these two arms.
Combined with the consistent velocity distributions in Fig. \ref{fig:ngc3628e}, it is possible that the lower red arm is the catching-up arm, and the upper green arm is the forward arm, which needs to be verified by additional simulations.

Although the Arecibo $\HI$ observations did not show the two-arm structure due to their lower angular resolution \citep[$\eg$][]{1979ApJ...229...83H, 2009AJ....138..338S},
we note that these data showed already double-peaked spectra in the plume \citep[see the panel b in Fig. 7 of ][]{1979ApJ...229...83H}, which is consistent with our more recently taken Arecibo map.
Two clear velocity regimes are shown in the middle panels of Fig. \ref{fig:arecibo}.
We conclude that we have identified a spatially well-separated two-arm structure in the plume for the first time by our high-resolution $\HI$ observations. The structure can be explained by the tidal interaction model. The detection of the catching-up sub arms in the plume could also be an important constraint on the interaction models.

\begin{figure*}
    \centering
    \includegraphics[width=\hsize]{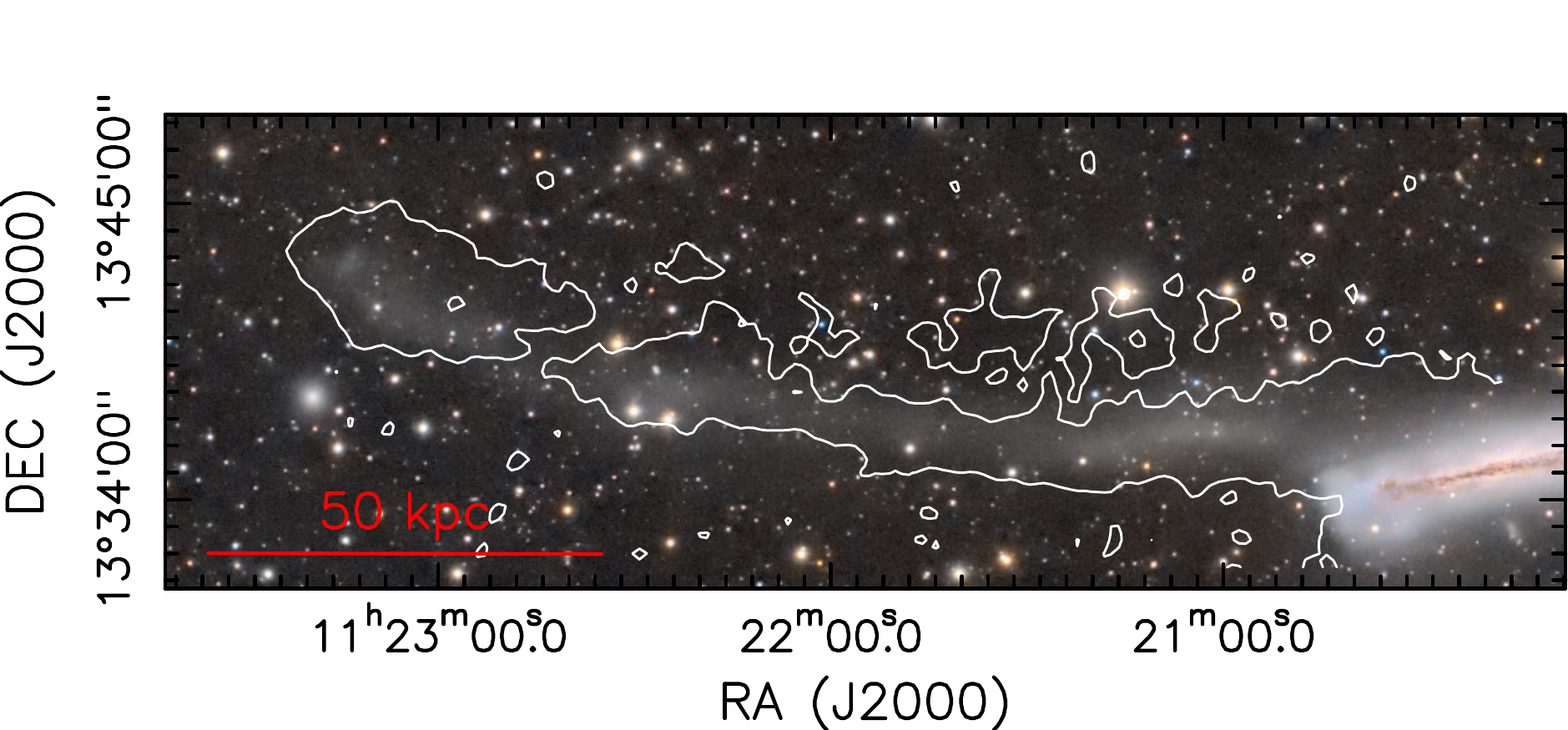}
    \includegraphics[width=\hsize]{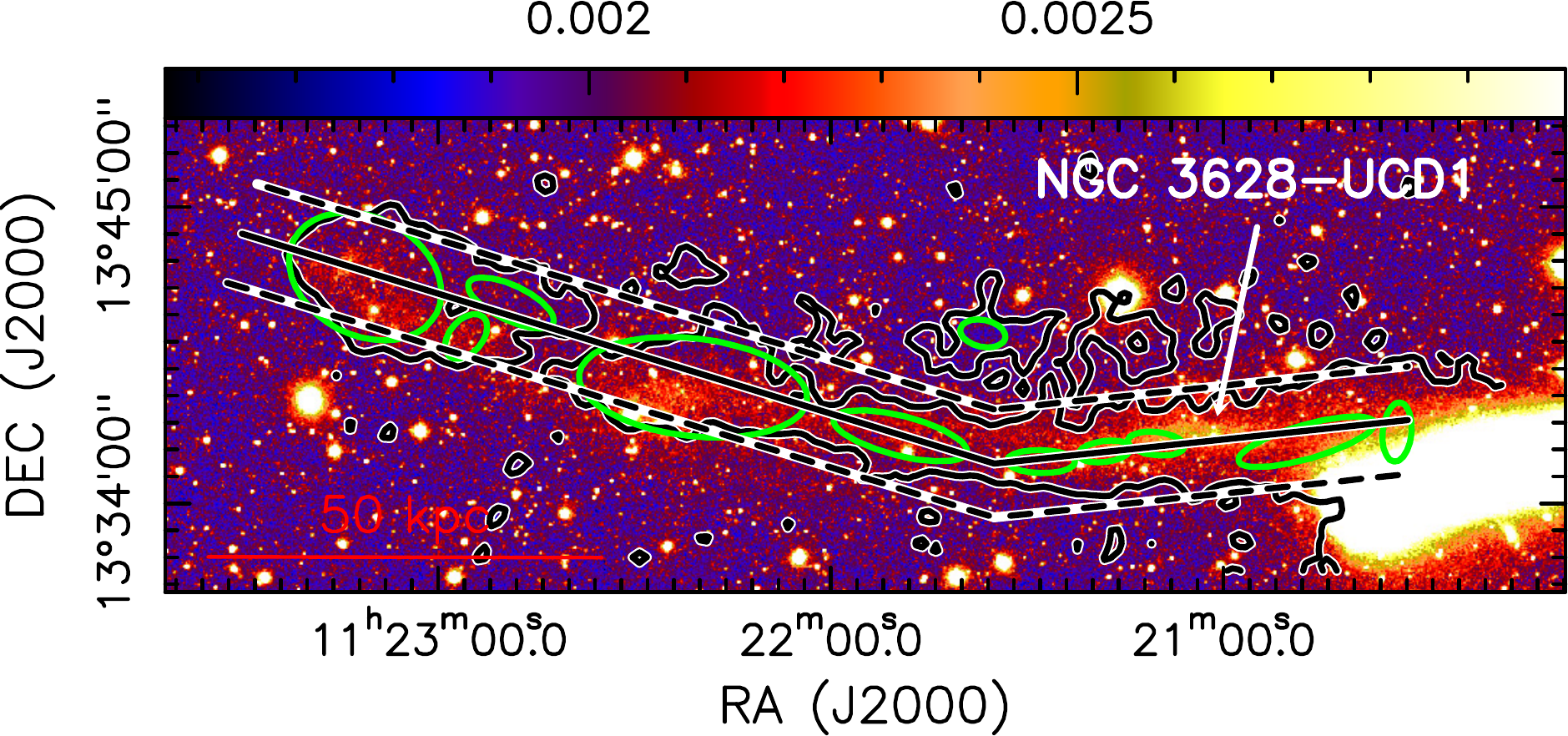}
    \caption{The full-color optical image of the plume (top panel). The luminance-filter optical image of the plume (see Section \ref{op}) excluding the pixels with uncalibrated flux larger than 0.003 (lower panel). The 3-$\sigma$ (3$\times$0.027 $\Jypbkms$) $\HI$ contours in  the two panels are the same as those in the lower panels in Fig \ref{fig:ngc3628e}. The green ellipses demonstrate the condensations identified in the plume. The black solid and dashed lines in the lower panel show the loci and the averaged width for the radial profiles in Fig. \ref{fig:plume-dist}.}
    \label{fig:HI-op}
    \end{figure*}

\begin{figure}
    \centering
    \includegraphics[width=\hsize]{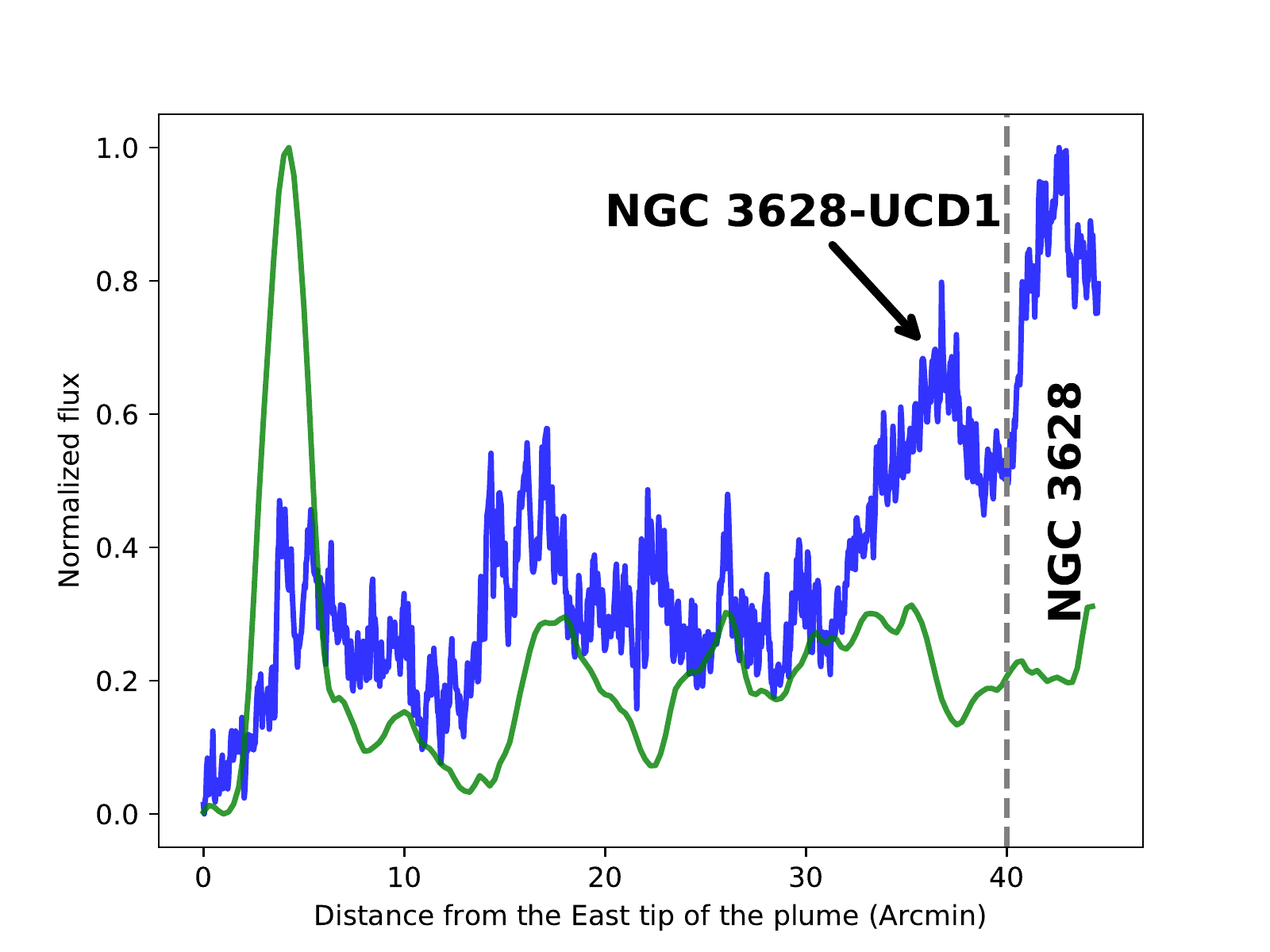}
    \caption{Radial profiles of the $\HI$ (green line) and optical (blue line) fluxes along the plume. The loci are illustrated in the lower left panel of Fig. \ref{fig:ngc3628e} and the lower panel of Fig. \ref{fig:HI-op}. The profiles with x-axis values larger than 40$\arcmin$ (grey dashed line) are likely affected by NGC\,3628.}
    \label{fig:plume-dist}
    \end{figure}

Finally, we compare the $\HI$ plume with its optical counterpart.
The top panel of Fig. \ref{fig:HI-op} presents the full-color optical image of the plume overlaid with the $\HI$ 3-$\sigma$ contours which are the same as those in the lower
panels in Fig. \ref{fig:ngc3628e}.
We can see from this panel that the optical plume is mainly associated with the southern arm of the $\HI$ plume, e.g. the $\HI$ gas with velocities of about 900\,$\kms$ (see  Fig. \ref{fig:ngc3628e}), which is also valid for the eastern tip of the plume.
To further quantitatively explore the distribution of matter, we seek to derive the radial profiles of the $\HI$ and optical fluxes along the plume.
Note that the optical observations derived from the amateur telescope show great details of the plume due to the long time exposure but are not calibrated. Thus here we only compare their relative flux distributions.
The $\HI$ zeroth moment map (the lower left panel of Fig.\ref{fig:ngc3628e}) and the luminance-filter optical image (see Section \ref{op}) are used to extract the $\HI$ and optical fluxes along the plume. To enhance the emission of the plume in the optical image, we exclude the pixels with uncalibrated flux intensities larger than 0.003. The image of the remaining pixels is shown as the color image in the lower panel of Fig. \ref{fig:HI-op}.
The $\HI$ and optical fluxes are derived along the loci shown by the black solid lines in the lower left panel of Fig. \ref{fig:ngc3628e} and the lower panel of Fig. \ref{fig:HI-op} by averaging the pixels within a width of 4$\arcmin$ (dashed lines in these panels). The final flux profiles are then normalized to the 0--1 range and shown in Fig. \ref{fig:plume-dist}.

As we can first see in the lower panel of Fig.\ref{fig:HI-op}, the peaks of the $\HI$ and optical fluxes are generally consistent, except for two of the $\HI$ condensations (the second and sixth ones in Table \ref{tab:clumps}) being likely associated with the northern arm and the optical peak near NGC\,3628-UCD1.
Then we can see in Fig. \ref{fig:plume-dist} that the $\HI$ and optical flux profiles are also roughly consistent with each other, especially the peaks at 5$\arcmin$ and 18$\arcmin$.
The consistency of the optical and $\HI$ peaks has also been reported by \citet{1998AJ....115.2331C}.
However, we can also see in Fig. \ref{fig:plume-dist} that there is a peak in the optical flux profile at about 37$\arcmin$ corresponding to the star cluster NGC\,3628-UCD1. Here the $\HI$ profile shows a minimum. Furthermore, the $\HI$ profile indicates a peak at the eastern tip and then generally shows a flat distribution along the plume. The optical profile presents an overall increasing trend  along the plume from the eastern tip to NGC\,3628.
A more comprehensive study of the plume will be conducted in a future contribution.

\subsection{Two bases of the plume}

We now discuss the connecting part (base) of the plume with NGC\,3628.
From the right panel of Fig. \ref{fig:op-tpeak} we can see that the lower arm of the plume seems to be connected with the western (blueshifted) side of NGC\,3628. Due to the weak emission of the western part of the upper arm, we cannot clearly see the base of the upper arm. However, according to several separated cloudlets along its extension, it seems that the upper arm originates from a position to the west of that of the lower arm.

\citet{1993MNRAS.263.1075W} conducted a study of VLA $\HI$ observations with a resolution of 15$\arcsec$ covering NGC\,3628 and part (out to about 8$\arcmin$ from NGC\,3628) of the plume.
They found a `limb' structure with velocities from 650 to 750\,$\kms$ (most clearly shown at velocities around 691\,$\kms$), which is argued to be associated with the plume. Furthermore, they also found another warped $\HI$ gas component with velocities about 750\,$\kms$, which is not rotating with the disk. \citet{1993MNRAS.263.1075W} suggested that this $\HI$ gas would eventually join the limb, being also a part of the plume.

The base of the plume is more clearly shown in the channel maps in Fig. \ref{fig:chMap}.
Firstly, from the 743.7 and 764.3\,$\kms$ maps in  Fig. \ref{fig:chMap} we can clearly find a protruding structure extending northeastwards. In the following maps with larger velocities, this protruding structure is smoothly connected with the lower arm of the plume. This protrusion may be associated with the warped $\HI$ gas found by \citet{1993MNRAS.263.1075W}. However, \citet{1993MNRAS.263.1075W} did not detect the $\HI$ plume, which is shown in our 784.9 to 929.2\,$\kms$ maps in Fig. \ref{fig:chMap}. This is probably the case because of their lower sensitivity (their rms noise in a 20.6\,$\kms$ wide channel is 0.04 $\Jypbkms$, about four times larger than our $\sigma_{ch}$ $\approx$ 0.01 $\Jypbkms$) and our combined VLA and Arecibo observations are more sensitive to the diffuse $\HI$ gas.
Meanwhile, from the 702.5\,$\kms$ map, we can also see a  protrusion of length $\approx$2$\arcmin$ extending northwards, which is probably associated with the limb structure found by \citet{1993MNRAS.263.1075W}. In the following 723.1\,$\kms$ map it is still seen and there is a new smaller protrusion to its east.

As mentioned in Section \ref{twoArm}, there are two arms in the plume. Meanwhile, the right panel of Fig. \ref{fig:op-tpeak} shows that the upper arm seems to originate further to the west, which means its base should be more blueshifted.
Thus it is plausible that the larger protrusion appearing both in the 702.5 and 723.1\,$\kms$ maps is the base of the fainter upper arm of the plume.
In the  simulation of the interaction between NGC\,3628 and M\,66  by \citet{1978AJ.....83..219R},  the motion of M\,66 was prograde with respect to the revolution of NGC\,3628.
The resulting plume was corotating with NGC\,3628 and located at the backside relative to NGC\,3628 along the line of sight. The plume relative to the bases should show more redshifted velocities \citep[see Fig. 4 in ][]{1978AJ.....83..219R}. This is consistent with the observed plume with velocities of 795.3-939.5\,$\kms$, which is more redshifted than the bases.

In summary, the plume originates from the blueshifted side of NGC\,3628. There seem to exist two bases, associated with the two arms in the plume.
The velocities of the bases of the upper and lower arms are about 700 and 750\,$\kms$, respectively.

\subsection{The peculiar velocity clump and its tail}

\citet{1979ApJ...229...83H} detected an $\HI$ bump in the southeast of M\,66 and found it to not corotate with M\,66.
Combining the $\HI$ data from \citet{1979ApJ...229...83H} and CO (1-0) data, \citet{1993ApJ...418..100Z} further confirmed the existence of a `peculiar velocity' clump in the southeast of M\,66. Through the analysis of the velocities of the spectra, they concluded that this seems to be a non-corotating clump that was pulled out of the inner spiral arms due to the close encounter of NGC\,3628 and M\,66.
\citet{1993ApJ...418..100Z} also argued  that the peculiar velocity clump is located at a high latitude, outside the plane of the disk of M\,66.

The peculiar velocity clump identified by \citet{1979ApJ...229...83H} and \citet{1993ApJ...418..100Z} is associated with the dominant clump observed by us in M\,66SE.
As we have mentioned in Section \ref{results}, the clump has a nearly  reversed velocity gradient compared to that of M\,66 which is consistent with the peculiar velocity found by \citet{1993ApJ...418..100Z}. In addition, we also detected a tail connected with the clump and extending to the south for the first time. As can be seen from the first moment map of Fig. \ref{fig:m66se},
the velocities of the tail and the clump can be roughly defined by a consistent velocity gradient.

In previous observations, the southeastern part of M\,66 has shown some different features compared to the observations in the northwestern part. For example, significant changes in $^{12}$CO (2-1) to $^{13}$CO (2-1) line ratios over small scales are found in the southeastern region, the CO line widths seem to be larger than those of the northwestern part, and the lines are characterized by a double peak \cite[$\eg$][]{1996A&A...306..721R, 2011ASPC..446..111D}.
Moreover,
assuming that there has been an interaction between NGC\,3628 and M\,66 as suggested by \citet{1978AJ.....83..219R},
M\,66SE is situated just on the orbit proposed by \citet{1978AJ.....83..219R} at the forward direction of M\,66.
In the tidal interaction models of \citet{1972ApJ...178..623T}, there should be some material captured by the companion (M\,66) from the victim galaxy (NGC\,3628) at the location where M\,66SE is situated.
This scenario is clearly demonstrated in the simulations of \citet{2011A&A...532A.118P} who showed that accreted tidal debris can form counter-orbiting satellite galaxies.

By combining the velocity pattern of M\,66SE and the tidal model suggested by \citet{1978AJ.....83..219R} and with M\,66SE supposedly representing gas captured from NGC\,3628 during the interaction, this gas may be falling back onto M\,66 giving rise to the overall velocity gradient of the clump and tail and resulting in the features previously observed in the southeastern part of M\,66,  $\eg$ large line widths and double peaks.

\subsection{The origin of the appendages}

The last item to be discussed is the origin of the appendage structures and a brief comparison of our observations with the dynamical models in the literature.

From an overall perspective, tidal interactions between NGC\,3628 and the other member galaxies of the Leo Triplet are probably necessary to explain the rich appendage structures found in this system.
However, the previous interaction models between NGC\,3628 and M\,66 (M\,65 is typically excluded in these models) seem to be insufficient to fully explain our observations, $\eg$ the prominent bridge (NGC\,3628S) and M\,65S.

Regarding the plume, its morphologies and kinematics have been basically reproduced by the tidal dynamical simulations between NGC\,3628 and M\,66 \citep{1978AJ.....83..219R}.
The newly detected two-arm structure in the plume and M\,66SE can also be explained by the tidal models, but further simulations would be required in order to demonstrate this in more detail.
As we have mentioned in the introduction, the alternative scenario of a minor merger is also proposed to explain the plume \citep[][]{2015ApJ...812L..10J}.
Specific simulations are required to compare these scenarios
%
and to clarify whether the plume is a result of a minor merger between NGC\,3628 and a dwarf galaxy or the by-product of the interaction between NGC\,3628 and the other member galaxies (M\,65 and/or M\,66).

\section{Summary}
We have conducted detailed morphological and kinematic studies of the appendages of the Leo Triplet ($\eg$ NGC\,3628, M\,65/NGC\,3623, and M\,66/NGC\,3627) by using high resolution and sensitivity fully-sampled observations of the neutral hydrogen in this system. The main conclusions are:

\begin{enumerate}
  \item We find six appendage structures besides the three galaxies, $\ie$ M\,66SE ($V_{\rm HEL}$ = 640.7--702.5\,$\kms$), NGC\,3628S (764.3--867.4\,$\kms$), NGC\,3628SW (826.2--888.0\,$\kms$), M\,65S (743.7--764.3\,$\kms$),  the plume (805.6--929.2\,$\kms$), and  IC\,2767 (1032.3--1114.7\,$\kms$).
      The morphologies and kinematics of these appendages can be used to constrain future kinematic models.
  \item We detect a two-arm structure in the plume for the first time that can be explained, at least approximately, by the tidal interaction model. The optical counterpart of the plume is mainly associated with the southern arm in the $\HI$ plume, e.g. $\HI$ gas with velocities around 900\,$\kms$. The peaks of the normalized optical and $\HI$ flux profiles are approximately consistent. However, there is a significant discrepancy at the location of the star cluster NGC\,3628-UCD1. Furthermore, the $\HI$ profile shows a roughly flat distribution along the plume from the eastern tip to NGC\,3628. Instead, the optical profile presents an overall increasing trend.
  \item 
        There seem to be two bases associated with the two arms in the plume.
        The base of the lower arm connected with the western (blueshifted) side of NGC\,3628 shows a velocity of about 750\,$\kms$.
        The base of the upper arm originates from a position to the west of that of the lower arm, showing a velocity about 700\,$\kms$.
        The two bases are likely associated with the `limb' structure and the warped $\HI$ gas component found by \citet{1993MNRAS.263.1075W}.
  \item We find an $\HI$ clump with a reversed velocity gradient relative to the velocity gradient of M\,66 and a tail extending to the south in the southeast of M\,66 (M\,66SE). By combining the velocity pattern of the clump and tail and the interaction model suggested by \citet{1978AJ.....83..219R}, we conclude that M\,66SE may represent gas captured from NGC\,3628. Meanwhile it is falling into M\,66, resulting in the features previously observed in the southeastern part of M\,66,  $\eg$ large line widths and spectral double peaks.

  \item An upside-down `Y' shaped $\HI$ gas component (M\,65S) is detected in the south of M\,65 and is probably associated with M\,65. This suggests that M\,65 might have been involved in the interaction of this triplet system,
      either having participated in the direct interaction with (one of) the other two galaxies or representing a captured part of the gas from the relic of the interaction between NGC\,3628 and M\,66.

\end{enumerate}

\begin{acknowledgements}
We thank an anonymous referee for useful suggestions improving the paper.
GW thanks Dustin Lang and Dalei Li for helpful discussions and providing the DESI data.
This work was funded by the CAS "Light of West China" Program under grant No. 2018-XBQNXZ-B-024 and the National Natural Science foundation of China under grant Nos. 11433008, 12173075, 12103082, and 11973076.
DMD acknowledges financial support from the Talentia Senior Program (through the incentive ASE-136) from Secretar\'\i a General de  Universidades, Investigaci\'{o}n y Tecnolog\'\i a, de la Junta de Andaluc\'\i a.
DMD acknowledges funding from the State Agency for Research of the Spanish MCIU through the ``Center of Excellence Severo Ochoa'' award to  the Instituto de Astrof{\'i}sica de Andaluc{\'i}a (SEV-2017-0709).
DMD acknowledges funding from the State Agency for Research of the Spanish MCI(Project PDI2020-114581GB-C21/ AEI / 10.13039/501100011033).

This research made use of astrodendro, a Python package to compute dendrograms of Astronomical data (http://www.dendrograms.org/).
The National Radio Astronomy Observatory is a facility of the National Science Foundation operated under cooperative agreement by Associated Universities, Inc.
The Arecibo Observatory is a facility of the National Science Foundation operated under cooperative agreement by the University of Central Florida and in alliance with Universidad Ana G. Mendez, and Yang Enterprises, Inc.

\end{acknowledgements}

%
%

\begin{appendix}

\section{NGC\,3628W}
\label{ngc3628w}

\begin{figure*}[b]
    \centering
    \includegraphics[width=0.45\hsize]{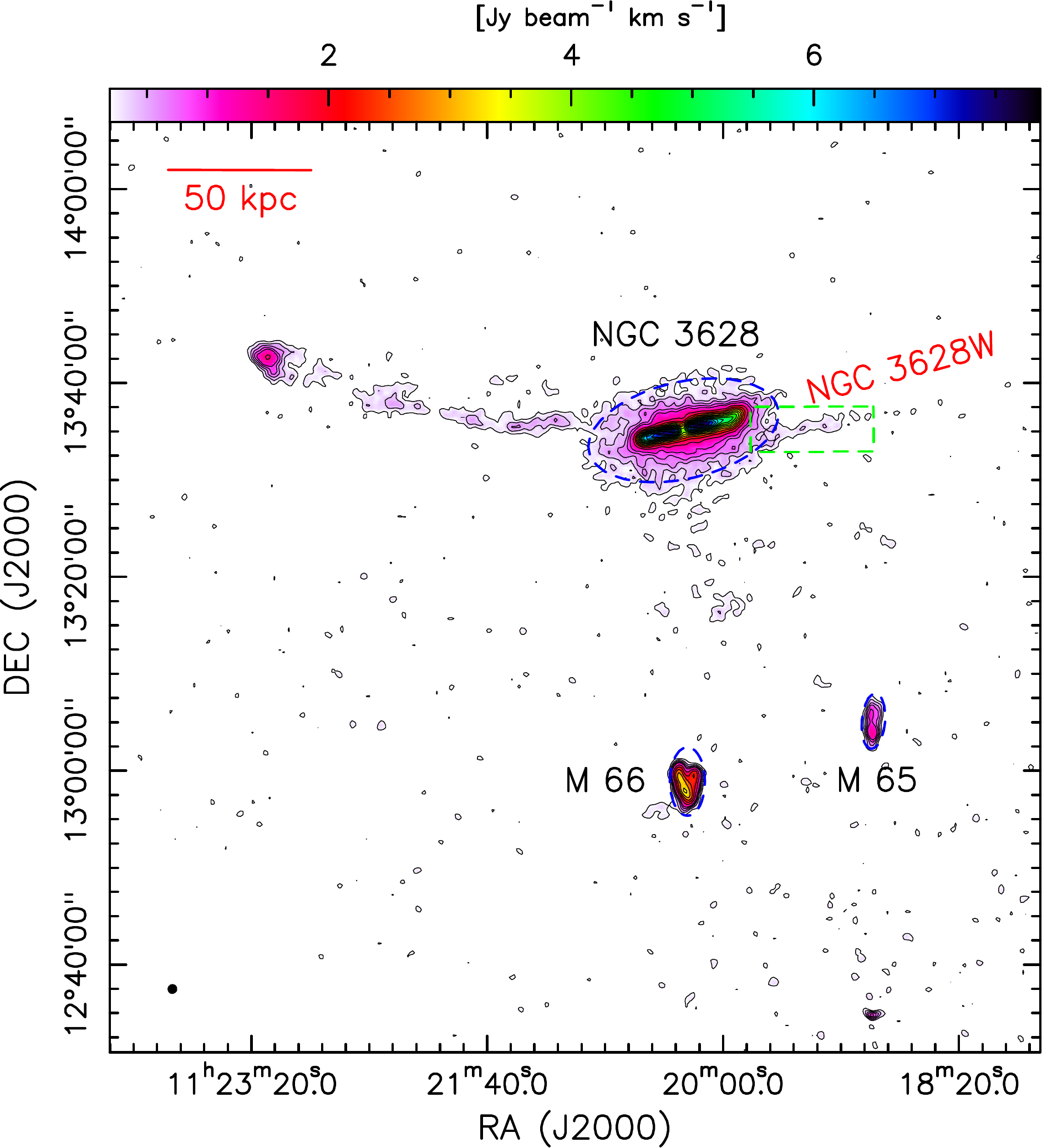}
    \includegraphics[width=0.45\hsize]{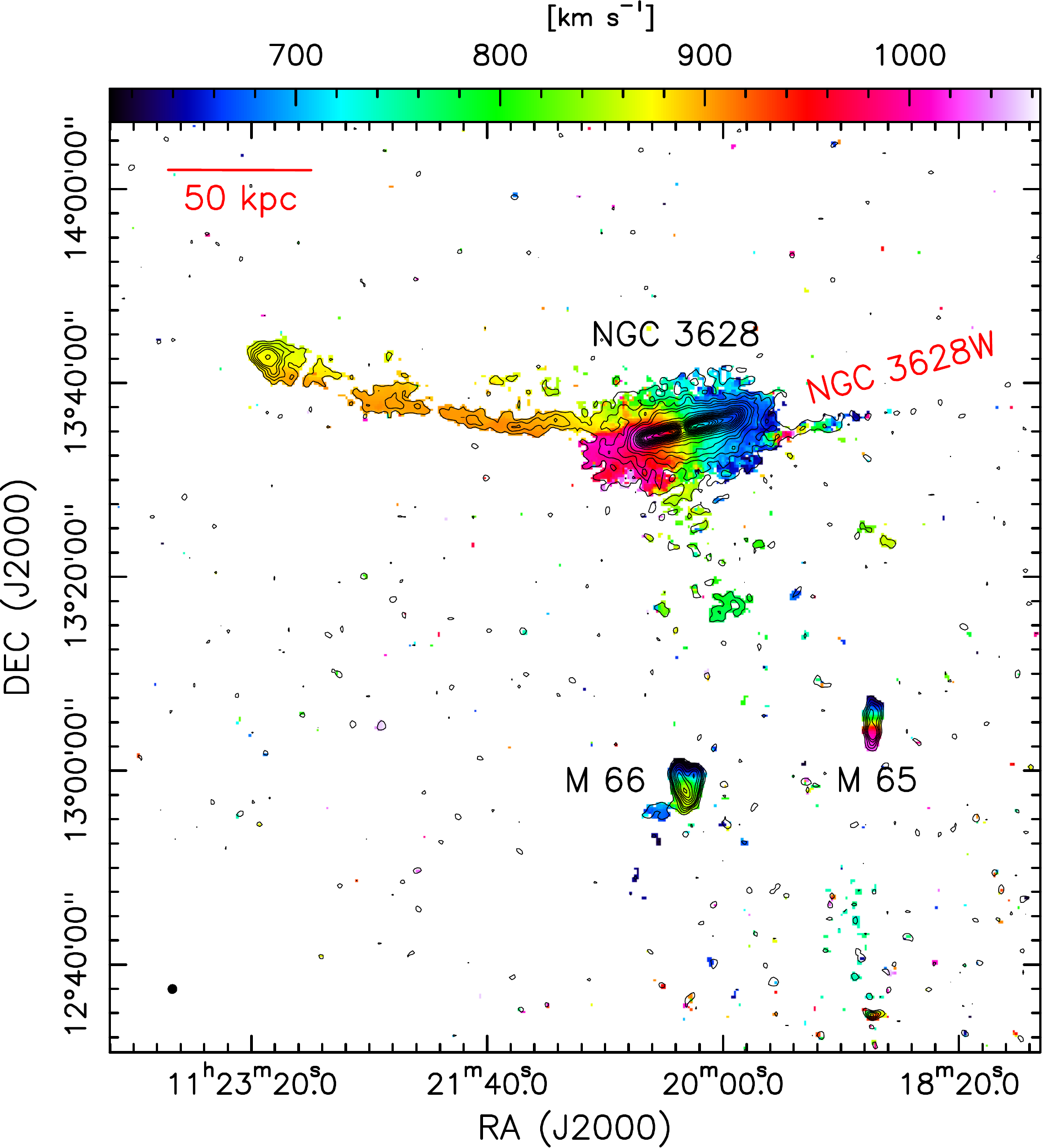}
    \includegraphics[width=0.45\hsize]{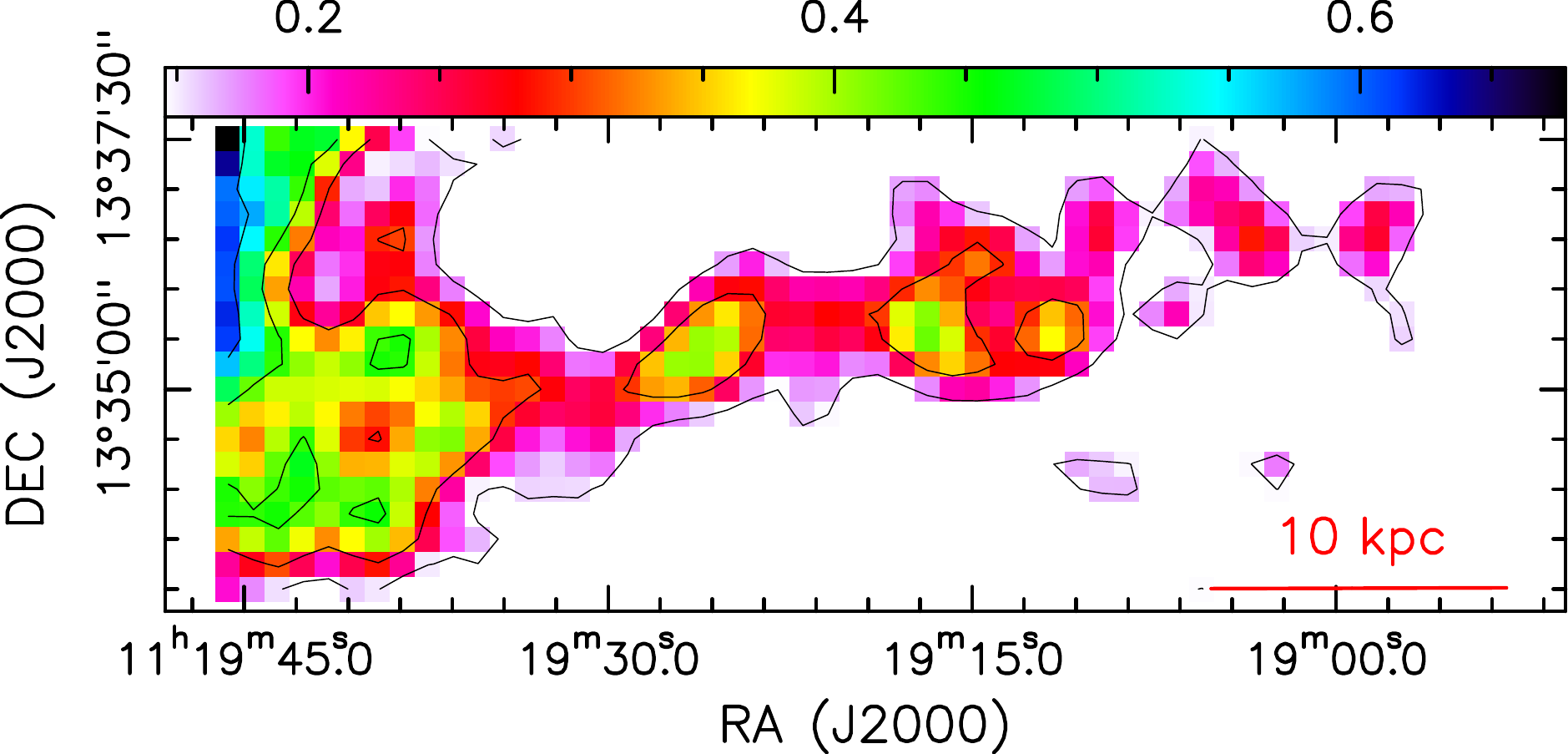}
    \includegraphics[width=0.45\hsize]{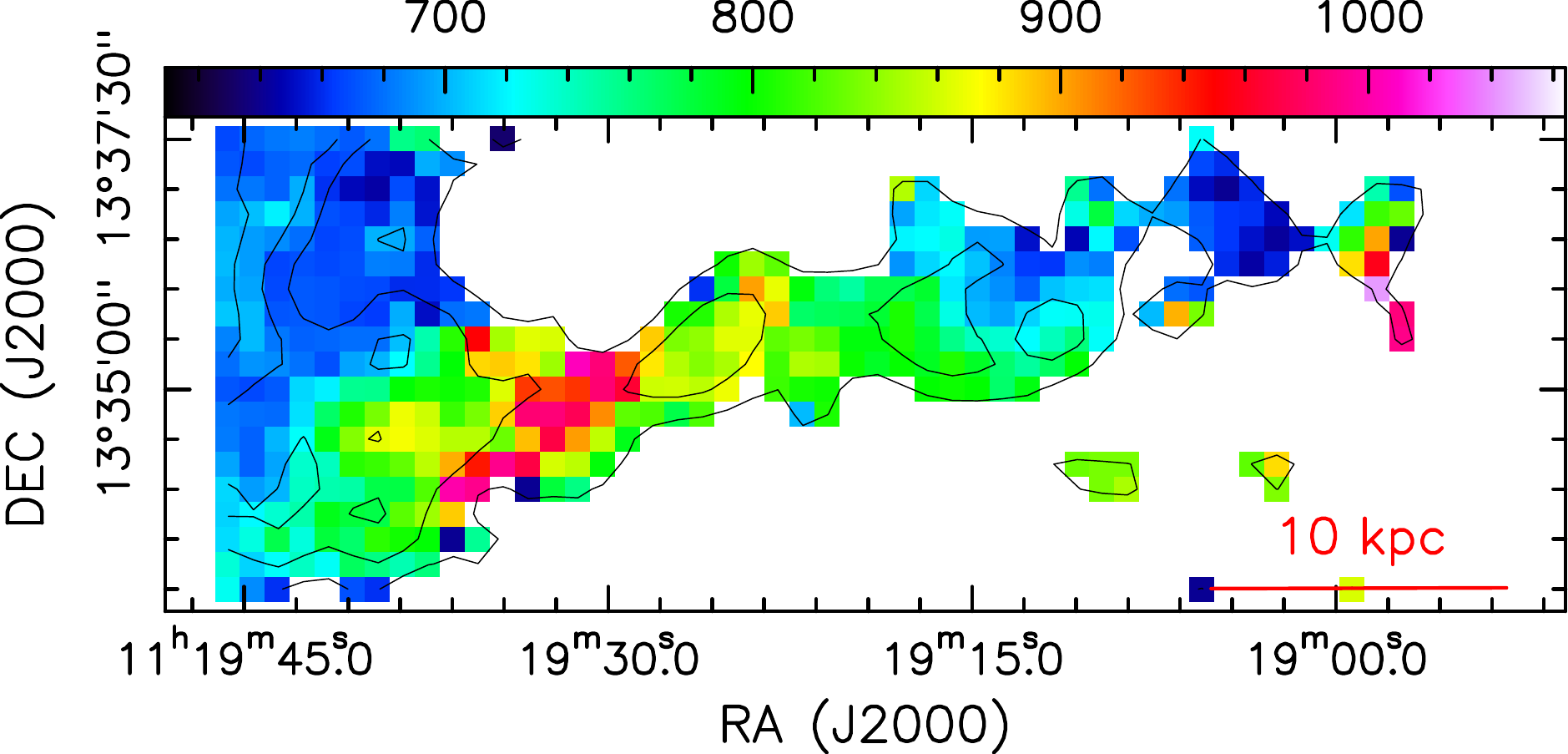}
    \caption{Same as Fig. \ref{fig:ngc3628e} but for NGC\,3628W. The velocity range is from 609.8 to 1063.2\,$\kms$ in both panels. The contour levels are set to (3, 6, 9, 12, 15, 18, 21, 30, 40, 50, ..., 160)\,$\times\,0.048$\,$\Jypbkms$.}
    \label{fig:ngc3628w}
    \end{figure*}

\textbf{NGC\,3628W:} Fig. \ref{fig:ngc3628w} presents the first moment (left panel) and the second moment (right panel) maps of NGC\,3628W. The velocity range is from $V_{\rm HEL}$ = 609.8 to 1063.2\,$\kms$ in both panels. The uncertainty of the velocity-integrated intensity $\sigma = \sigma_{\rm ch} \times \sqrt{N_{\rm ch}}$ = 0.048\,$\Jypbkms$, where $\sigma_{\rm ch} \approx 0.01$ $\Jypbkms$ (see Sect. \ref{ob}) and $N_{\rm ch} = 22 $.

As we can see in Fig. \ref{fig:ngc3628w}, the morphology and velocity patterns of this structure are very similar to the strong central part of NGC\,3628, $\ie$ they show a similar elongated morphology and velocity gradient. This feature is only present in the VLA data but is absent in the Arecibo data. Thus it is likely caused by the VLA sidelobes. It was also not seen in any other study in spite of its strength. Thus, we conclude that NGC\,3628W is very likely a spurious feature. As a consequence, it is not part of our analysis.

\section{The Arecibo Observations}
\label{arecibo}

\begin{figure*}[t]
    \centering
    \includegraphics[width=0.32\hsize]{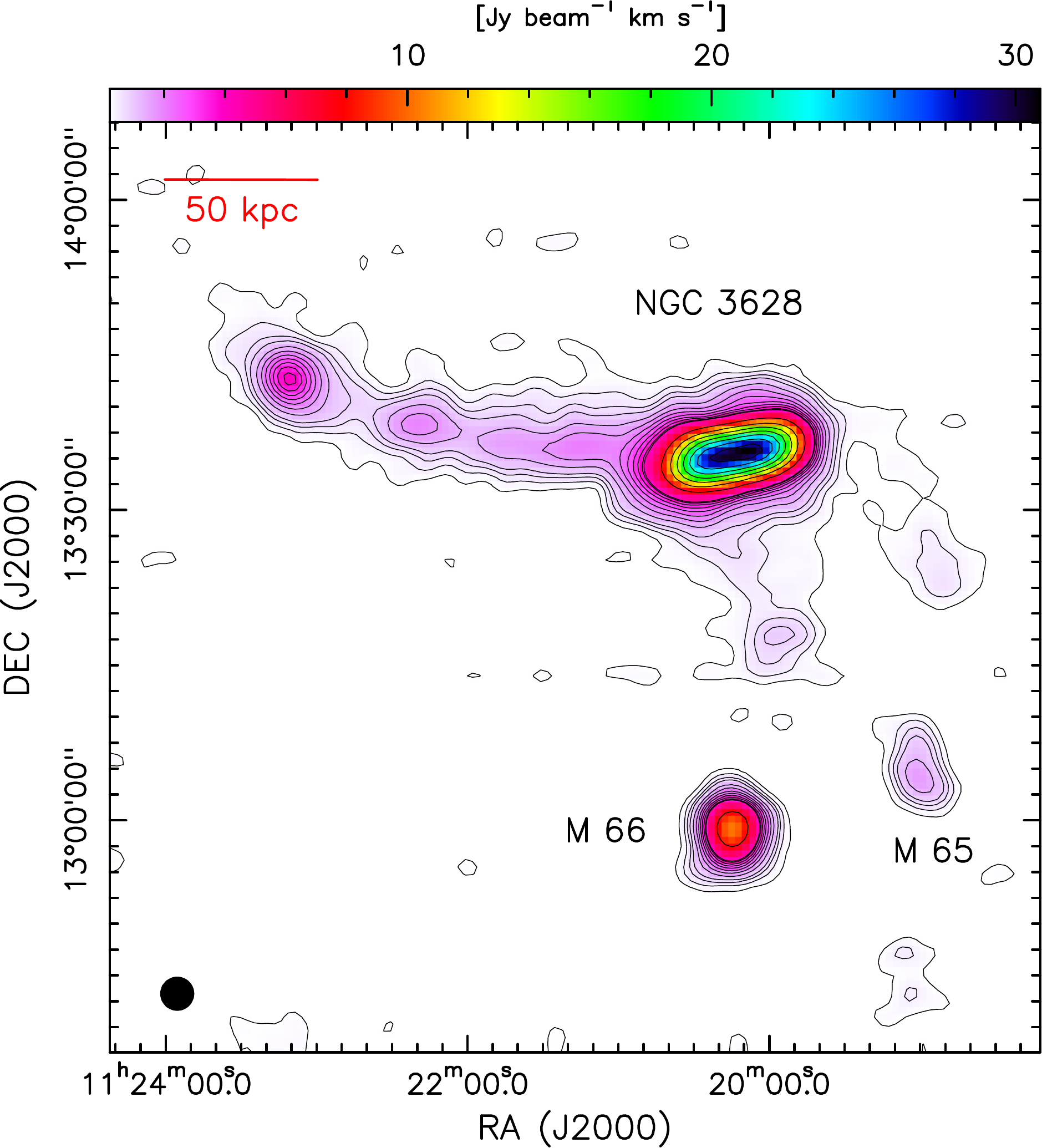}
    \includegraphics[width=0.32\hsize]{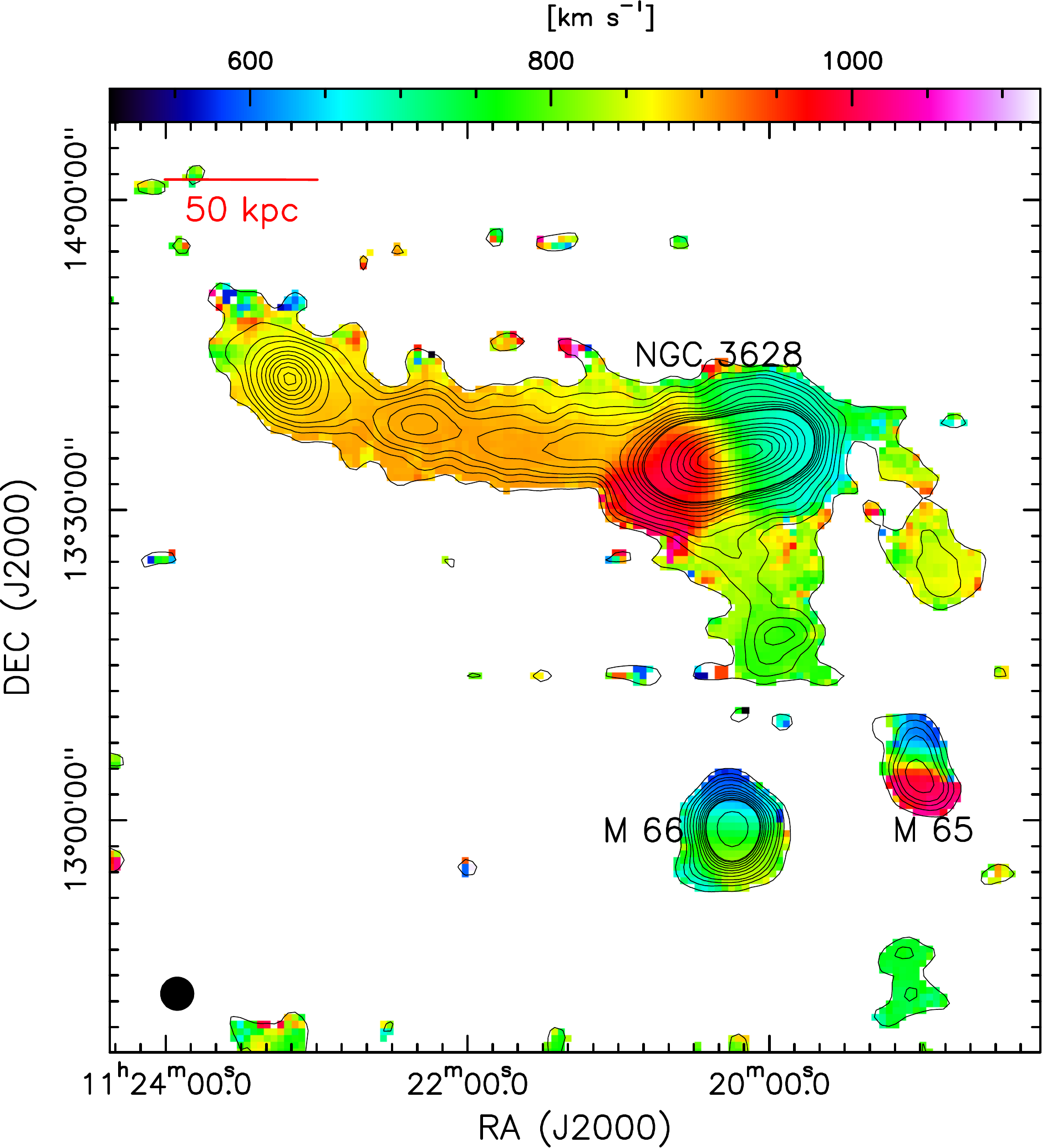}
    \includegraphics[width=0.32\hsize]{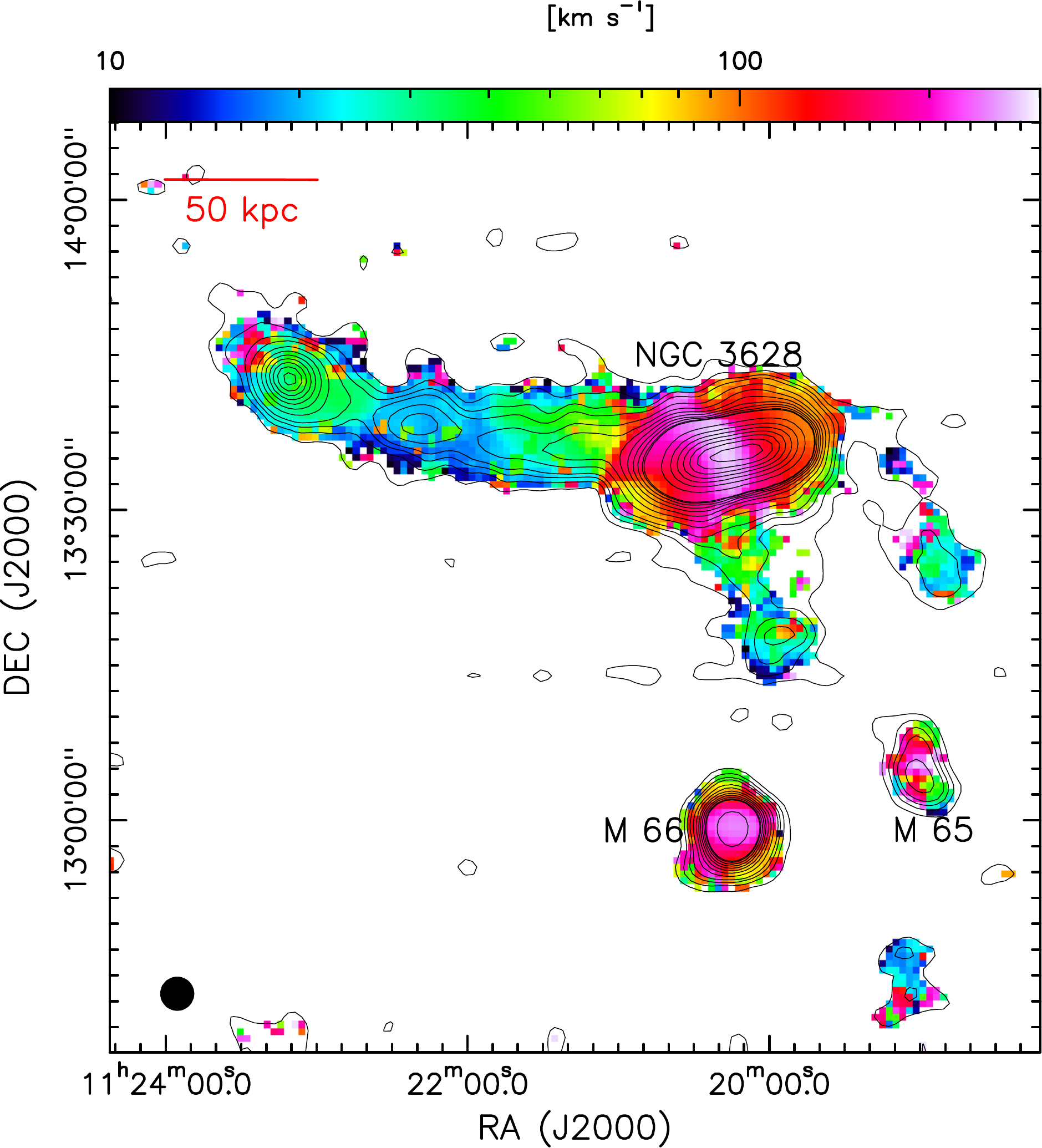}
    \includegraphics[width=0.32\hsize]{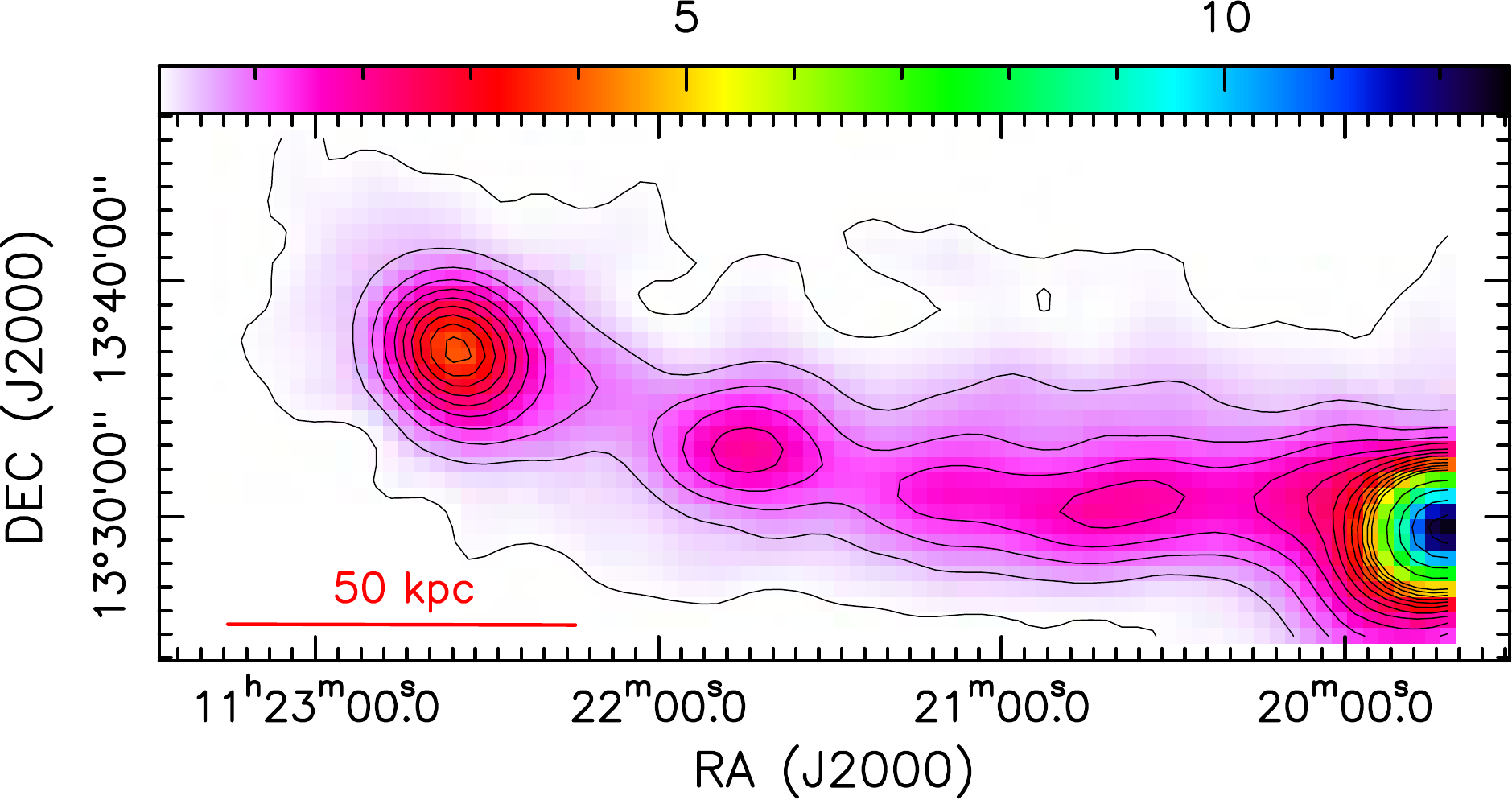}
    \includegraphics[width=0.32\hsize]{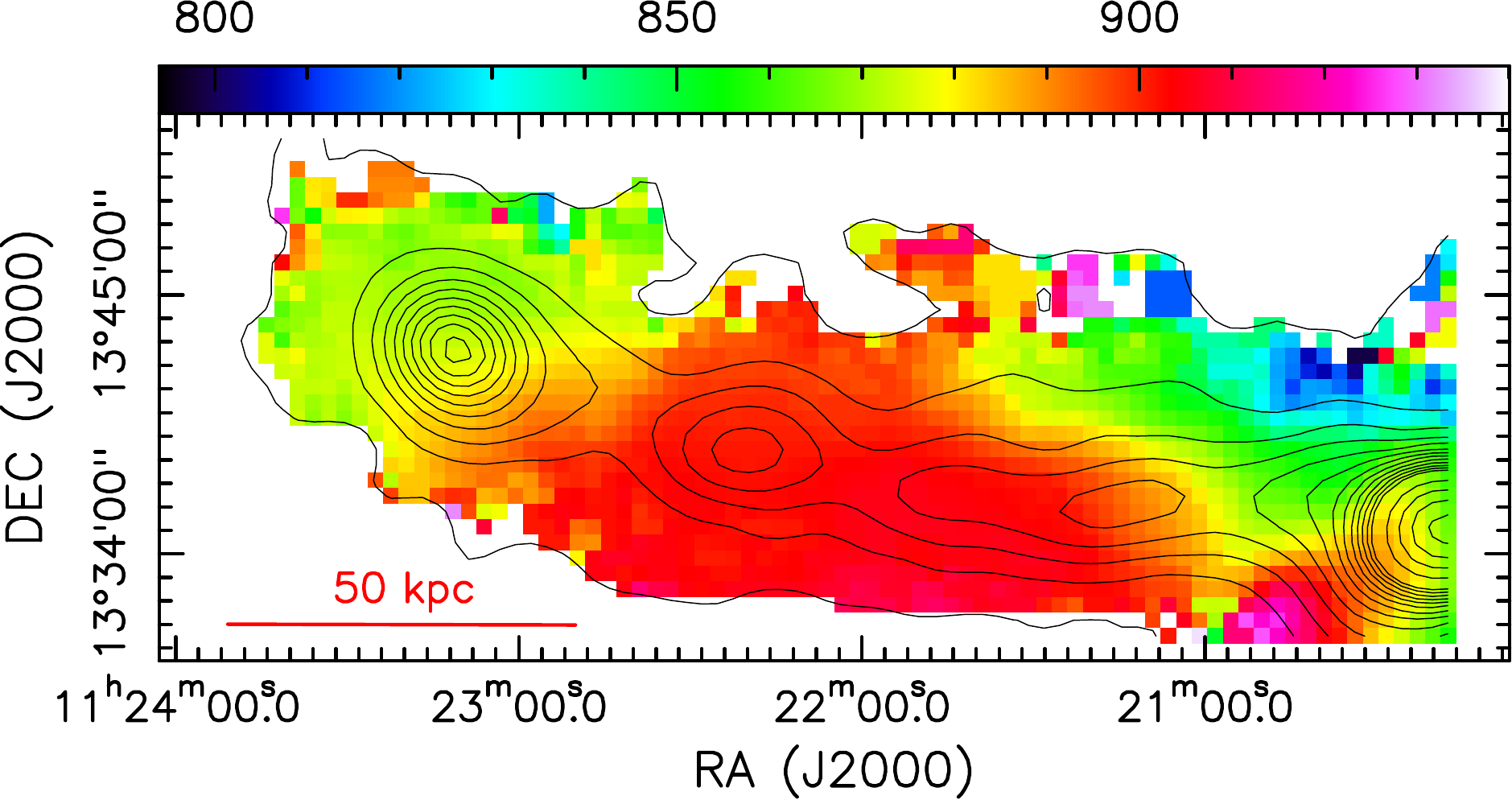}
    \includegraphics[width=0.32\hsize]{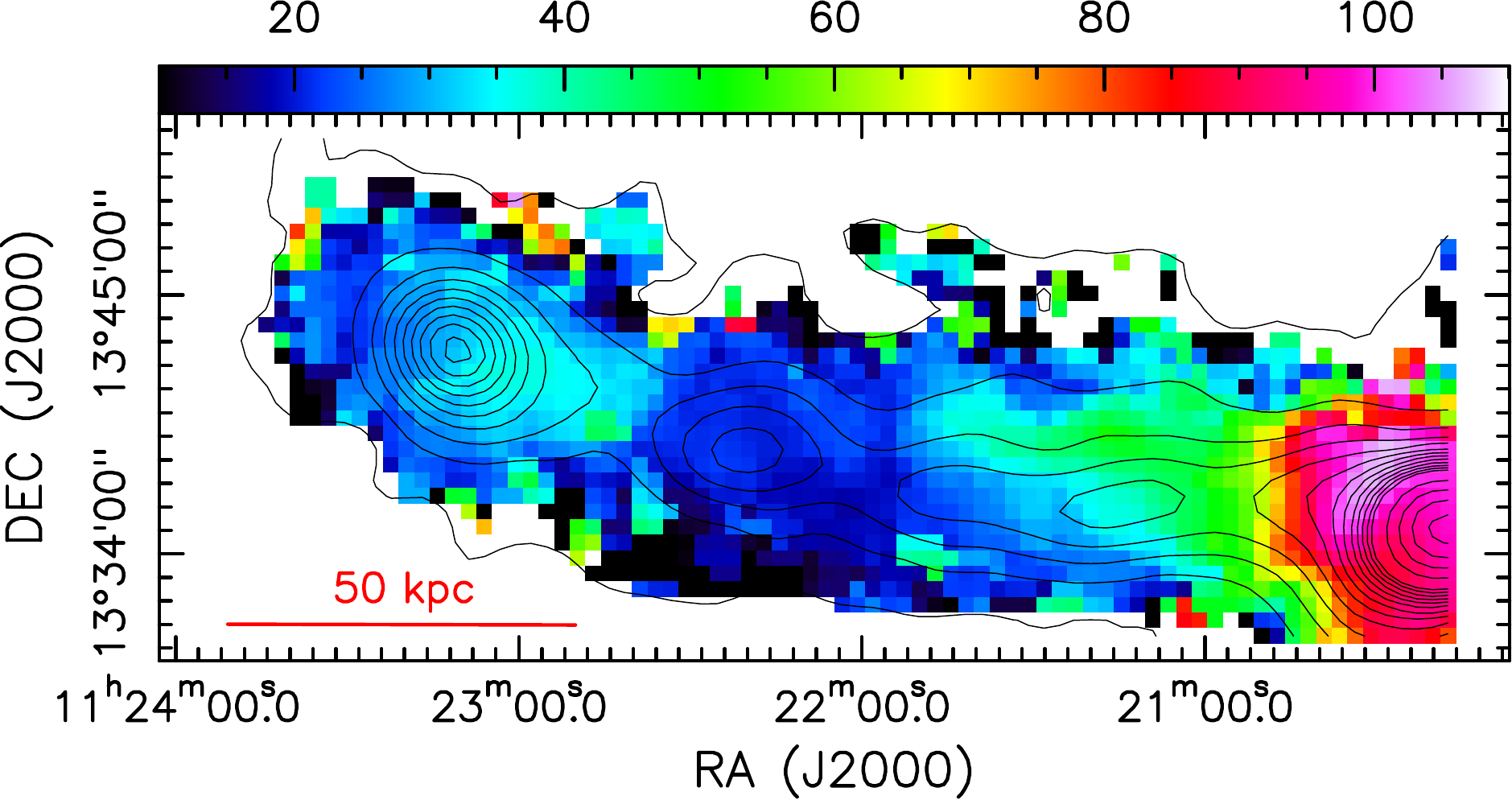}
    \caption{HI data exclusively from the Arecibo telescope with 3$\ffam$3 angular resolution.
    Top three panels: The $\HI$ velocity-integrated emission (zeroth moment),  intensity-weighted velocity (first moment), and intensity-weighted velocity dispersion (second moment) maps of the Leo Triplet. In the left panel, contours and also the color image show the zeroth moment map. The integration range is from V$_{\rm HEL}$ = 506.7 to 1125.1\,$\kms$. The contour levels are set to (3, 6, 9, 12, 18, 24, 30, 36, 42, 48, 54, 100, 150, 200, ..., 600)\,$\times\,0.07$ $\Jypbkms$.
    In the middle and right panels, the color image shows the first and second moment maps. The contours overlaid are the same as the ones in the left panel. The galaxies NGC\,3628, M\,65, and M\,66 are also labeled in both panels and a filled ellipse in the lower left shows the synthesized beam.
    Bottom three panels: Same as the top panels, but exclusively for the plume. The integration range is from V$_{\rm HEL}$ = 795.3 to 939.5\,$\kms$. The contour levels are set to (3, 15, 27, 39, 51, 63, 75, 87, 99, 111, 133, 200, 250, 300, ..., 600)\,$\times\,0.03$ $\Jypbkms$.
    The red line in each panel illustrates the 50\,kpc scale at a distance of 11.3 Mpc.}
    \label{fig:arecibo}
    \end{figure*}

We first masked the strong spikes in the spectra caused by radio frequency interference (RFI) and smoothed three contiguous velocity channels to enhance the S/N ratio of the individual channels in the Arecibo data. The velocity resolution and the noise level of the smoothed data turn into 1.9\,$\kms$ and 2.0 m$\Jypb$, respectively. Our noise level is slightly better than that in \citet{2009AJ....138..338S} ($\approx$ 4 m$\Jypb$ scaled to 1.9\,$\kms$).
Figure \ref{fig:arecibo} shows the zeroth moment (left panels), first moment (middle panels), and second moment (right panels) of the Leo Triplet (top three panels) and the plume (lower three panels).
The velocity ranges of the top and lower panels are from V$_{\rm HEL}$ = 506.7 to 1125.1\,$\kms$ and 795.3 to 939.5\,$\kms$, respectively, which are the same as those for Fig. \ref{fig:mom0_2-31} and \ref{fig:ngc3628e}.
The uncertainties of the upper and lower zeroth moment maps are 0.07\,$\Jypbkms$ and  0.03\,$\Jypbkms$.
A threshold of 3 $\sigma_{\rm ch}$, where $\sigma_{\rm ch}$ is the noise level in a channel, is used to derive the first and second moment maps to avoid emphasizing contributions from spectral noise and the RFI.


We can see in the top left panel of Fig. \ref{fig:arecibo} that our Arecibo map is consistent with that presented by \citet{2009AJ....138..338S}.
We note that an about $10\arcmin$-long spur, reported by \citet{2009AJ....138..338S} in the north of NGC\,3628 extending further north, is not evident in our map. We checked our data and found a very narrow and faint feature with velocities of about 950\,$\kms$, which is similar to that reported by \citet{2009AJ....138..338S}. But this feature is not discussed in our paper due to its poor SNR. Meanwhile, M\,65S has also been detected by \citet{2009AJ....138..338S} but this feature extends further north in their map. These discrepancies might be partially caused by larger noise levels at the edges of our Arecibo images.
From the middle panels, we can see that the Arecibo data show quite a similar velocity pattern as that seen in Figs. \ref{fig:mom0_2-31} and \ref{fig:ngc3628e}. In the plume, there are also two velocity regimes with similar spatial distributions as in Fig. \ref{fig:ngc3628e}.
%
Since the Arecibo data have a better velocity resolution than the combined data, we further provide the second moment maps in Fig. \ref{fig:arecibo}. From the two maps, we can see that the narrowest spectra are located in the middle of the plume, showing linewidths of about 20\,$\kms$, which is consistent with the observations in \citet{1979ApJ...229...83H} and the velocity resolution of our combined Arecibo and VLA data.


\end{appendix}

\end{document}